\documentstyle [12pt,epsf] {article}

\newcommand{\nwc}{\newcommand}
\nwc{\cl}  {\clubsuit}

\nwc{\hyp} {\hyphenation}
\nwc{\be}  {\begin{equation}}
\nwc{\ee}  {\end{equation}}
\nwc{\ba}  {\begin{array}}
\nwc{\ea}  {\end{array}}
\nwc{\bdm} {\begin{displaymath}}
\nwc{\edm} {\end{displaymath}}
\nwc{\bea} {\be\ba{rcl}}
\nwc{\eea} {\ea\ee}
\nwc{\ben} {\begin{eqnarray}}
\nwc{\een} {\end{eqnarray}}
\nwc{\bda} {\bdm\ba{lcl}}
\nwc{\eda} {\ea\edm}
\nwc{\bc}  {\begin{center}}
\nwc{\ec}  {\end{center}}
\nwc{\ds}  {\displaystyle}
\nwc{\bmat}{\left(\ba}
\nwc{\emat}{\ea\right)}
\nwc{\non} {\nonumber}
\nwc{\bib} {\bibitem}
\nwc{\lra} {\longrightarrow}
\nwc{\Llra}{\Longleftrightarrow}
\nwc{\ra}  {\rightarrow}
\nwc{\Ra}  {\Rightarrow}
\nwc{\lmt} {\longmapsto}
\nwc{\prl} {\partial}
\nwc{\iy}  {\infty}
\nwc{\ol}  {\overline}
\nwc{\hm}  {\hspace{3mm}}
\nwc{\lf}  {\left}
\nwc{\ri}  {\right}
\nwc{\lm}  {\limits}
\nwc{\lb}  {\lbrack}
\nwc{\rb}  {\rbrack}
\nwc{\ov}  {\over}
\nwc{\pr}  {\prime}
\nwc{\nnn} {\nonumber \vspace{.2cm} \\ }
\nwc{\Sc}  {{\cal S}}
\nwc{\Lc}  {{\cal L}}
\nwc{\Rc}  {{\cal R}}
\nwc{\Dc}  {{\cal D}}
\nwc{\Oc}  {{\cal O}}
\nwc{\Cc}  {{\cal C}}
\nwc{\Pc}  {{\cal P}}
\nwc{\Mc}  {{\cal M}}
\nwc{\Ec}  {{\cal E}}
\nwc{\Fc}  {{\cal F}}
\nwc{\Hc}  {{\cal H}}
\nwc{\Kc}  {{\cal K}}
\nwc{\Xc}  {{\cal X}}
\nwc{\Gc}  {{\cal G}}
\nwc{\Zc}  {{\cal Z}}
\nwc{\Nc}  {{\cal N}}
\nwc{\fca} {{\cal f}}
\nwc{\xc}  {{\cal x}}
\nwc{\Ac}  {{\cal A}}
\nwc{\Bc}  {{\cal B}}
\nwc{\Uc}  {{\cal U}}
\nwc{\Vc}  {{\cal V}}
\nwc{\Th} {\Theta}
\nwc{\th} {\theta}
\nwc{\vth} {\vartheta}
\nwc{\eps}{\epsilon}
\nwc{\si} {\sigma}
\nwc{\Gm} {\Gamma}
\nwc{\gm} {\gamma}
\nwc{\bt} {\beta}
\nwc{\La} {\Lambda}
\nwc{\la} {\lambda}
\nwc{\om} {\omega}
\nwc{\Om} {\Omega}
\nwc{\dt} {\delta}
\nwc{\Si} {\Sigma}
\nwc{\Dt} {\Delta}
\nwc{\al} {\alpha}
\nwc{\vph}{\varphi}
\nwc{\zt} {\zeta}
\def\Tr{\mathop{\rm Tr}}

\def\VEV#1{\left\langle #1\right\rangle}

\def\abs#1{\left| #1\right|}
\def\pr#1{#1^\prime}

\nwc{\Id}  {{\bf 1}}
\nwc{\diag} {{\rm diag}}
\nwc{\inv}  {{\rm inv}}
\nwc{\mod}  {{\rm mod}}
\nwc{\hal} {\frac{1}{2}}
\nwc{\tpi}  {2\pi i}

\def\slash#1{#1\!\!\!/\!\,\,}

\def\ijmpa#1{Int. J. Mod. Phys. {\bf A#1}}

\def\npb#1{Nucl. Phys. {\bf B#1}}
\def\nc#1{Nuovo Cim. {\bf #1}}

\def\plb#1{Phys. Lett. {\bf #1B}}
\def\pr#1{Phys. Rev. {\bf #1}}

\def\prc#1{Phys. Rep. {\bf C#1}}
\def\prd#1{Phys. Rev. {\bf D#1 }}
\def\prle#1{Phys. Rev. Lett. {\bf #1}}

\def\zpc#1{Z. Phys. {\bf C#1}}

\def\eV {\,{\rm  eV}}     
\def\keV {\,{\rm  keV}}     
\def\MeV {\,{\rm  MeV}}
\def\GeV {\,{\rm  GeV}}

\def \lta {\mathrel{\vcenter
     {\hbox{$<$}\nointerlineskip\hbox{$\sim$}}}}
\def \gta {\mathrel{\vcenter
     {\hbox{$>$}\nointerlineskip\hbox{$\sim$}}}} 
\newsavebox{\nnin} \sbox{\nnin}{$\hspace{1mm}\in\kern -.8em /
                   \hspace{1mm}$}

\newcommand{\sub}{\subset}
\newsavebox{\nnsub} \sbox{\nnsub}{$\hspace{1mm}\sub\kern -.9em /
            \hspace{1mm}$}

\def\KK{{\rm I\kern -.2em  K}}
\def\NN{{\rm I\kern -.16em N}}
\def\RR{{\rm I\kern -.2em  R}}
\def\ZZ{Z \kern -.43em Z}
\def\QQ{{\rm \kern .25em
             \vrule height1.4ex depth-.12ex width.06em\kern-.31em Q}}
\def\CC{{\rm \kern .25em
             \vrule height1.4ex depth-.12ex width.06em\kern-.31em C}}
\def\ZZZ{Z\kern -0.31em Z}

\nwc{\olbt}  {\ol{\bt}}
\nwc{\olgm}  {\ol{\gm}}
\nwc{\olh}   {\ol{h}}
\nwc{\olla}  {\ol{\la}}
\nwc{\olm}   {\ol{m}}
\nwc{\olq}   {\ol{q}}
\nwc{\olmu}  {\ol{\mu}}
\nwc{\olnu}  {\ol{\nu}}
\nwc{\olpsi} {\ol{\psi}}
\nwc{\olsi}  {\ol{\sigma}}
\nwc{\olu}   {\ol{u}}
\nwc{\olzt}  {\ol{\zt}}
\nwc{\olK}   {\ol{K}}
\nwc{\olf}   {\ol{f}}
\nwc{\olchi} {\ol{\chi}}
\nwc{\vp}    {\varphi}
\nwc{\vpd}   {\vp^\dagger}
\nwc{\Phid}  {\Phi^\dagger}
\nwc{\Ud}    {U^\dagger}
\nwc{\prlt}  {\frac{\prl}{\prl t}}
\nwc{\ttau}  {\tilde{\tau}}
\nwc{\tP}    {\tilde{P}}
\nwc{\tU}    {\tilde{U}}
\nwc{\teps}  {\tilde{\eps}}
\nwc{\tla}   {\tilde{\la}}
\nwc{\tit}   {\tilde{t}}
\nwc{\tPhi}  {{\tilde{\Phi}}}
\nwc{\teta}  {{\tilde{\eta}}}
\nwc{\trho}  {{\tilde{\rho}}}
\nwc{\tsi}   {{\tilde{\si}}}
\nwc{\tV}    {{\tilde{V}}}
\nwc{\ham}   {\widehat{m}}
\nwc{\hagm}  {\widehat{\gm}}
\nwc{\iddq}  {\int\frac{d^dq}{(2\pi)^d}}
\nwc{\iddp}  {\int\frac{d^dp}{(2\pi)^d}}
\nwc{\iddQ}  {\int\frac{d^dQ}{(2\pi)^d}}
\nwc{\prpr}  {\prime\prime}
\nwc{\etap}   {{\eta^\prime}}

\nwc{\cst}     {SU_L(3)\times SU_R(3)}
\nwc{\csN}     {SU_L(N)\times SU_R(N)}
\nwc{\vac}     {{\rm vac}}

\newcommand{\sect}[1]{ \section{#1} \setcounter{equation}{0} }

\begin{document}

\begin{titlepage}

  \title{Effective linear meson model\thanks{Supported by the Deutsche
      Forschungsgemeinschaft}}

\author{{\sc D.--U. Jungnickel\thanks{Email: 
D.Jungnickel@thphys.uni-heidelberg.de}} \\
 \\ and \\ \\
{\sc C. Wetterich\thanks{Email: C.Wetterich@thphys.uni-heidelberg.de}} 
\\ \\ \\
{\em Institut f\"ur Theoretische Physik} \\
{\em Universit\"at Heidelberg} \\
{\em Philosophenweg 16} \\
{\em 69120 Heidelberg, Germany}}

\date{June 1996}
\maketitle

\begin{picture}(5,2.5)(-350,-450)
\put(12,-115){HD--THEP--96--19}
\end{picture}

\thispagestyle{empty}

\begin{abstract}
  The effective action of the linear meson model generates the mesonic
  $n$--point functions with all quantum effects included. Based on
  chiral symmetry and a systematic quark mass expansion we derive
  relations between meson masses and decay constants. The model
  ``predicts'' values for $f_\eta$ and $f_{\eta^\prime}$ which are
  compatible with observation. This involves a large momentum
  dependent $\eta$--$\eta^\prime$ mixing angle which is different for
  the on--shell decays of the $\eta$ and the $\eta^\prime$. We also
  present relations for the masses of the $0^{++}$ octet. The
  parameters of the linear meson model are computed and related to
  cubic and quartic couplings among pseudoscalar and scalar mesons. We
  also discuss extensions for vector and axialvector fields. In a good
  approximation the exchange of these fields is responsible for the
  important nonminimal kinetic terms and the $\eta$--$\eta^\prime$
  mixing encountered in the linear meson model.
\end{abstract}

\end{titlepage}

\sect{Introduction}
\label{Introduction}

In quantum field theory all effects of quantum fluctuations are
incorporated in the effective action $\Gamma$, the generating
functional of one--particle irreducible Green functions.  From the
knowledge of these amplitudes the information about particle masses
and decay rates, scattering cross sections, etc.~can be extracted in a
straightforward manner. In particular, the effective action for the
mesons in the lowest mass pseudoscalar octet contains all information
on the physics involving only $\pi^\pm$, $\pi^0$, $K^\pm$, $K^0$,
$\ol{K}^0$ and $\eta$. We emphasize that all quantum fluctuation
effects are already included in the effective action and no further
integration over fluctuations has to be performed\footnote{``Effective
  actions'' are also often used in a different context where only some
  degrees of freedom are integrated out whereas fluctuations of the
  remaining degrees of freedom still need to be computed. This is,
  e.g., the typical setting of chiral perturbation theory and differs
  {}from our approach.}!  Without any further input the effective action
can be viewed simply as a coherent description of the information
gathered by other means about scattering amplitudes, decay rates, etc.
In a very general context it contains already predictive power
following from constraints which describe the analyticity properties
of the momentum dependence of Green functions or general features like
convexity.  Furthermore, all exact symmetry relations are
automatically embodied in the symmetries of $\Gamma$ or the related
Ward identities.

Our aim is to find relations among the $n$--point functions described
by $\Gamma$ which go beyond exact symmetry properties and general
constraints. This allows to establish relations among physical
quantities and to make predictions. (For strong interactions these
``predictions'' are more often ``postdictions'', but they permit an
understanding of already measured quantities.) In the case of meson
physics the ultimate goal is a computation of the effective mesonic
action $\Gamma$ from basic QCD, involving as free parameters only
$\alpha_s(M_Z)$ and the current quark masses. We will be concerned
here with more modest partial answers which follow from a few simple
assumptions about the general properties of the mesonic effective
action.

A lot of information can be extracted from approximate chiral
$SU_L(3)\times SU_R(3)$ symmetry. In the absence of current quark
masses for the up, down and strange quark this is an exact symmetry of
the QCD Lagrangian which is believed to be broken spontaneously by the
chiral condensate to the vector subgroup $SU_V(3)$.  Considering the
explicit symmetry breaking by the quark masses $m_u$, $m_d$ and $m_s$
as a small effect and expanding the Green functions in powers of these
masses gives rise to the very successful chiral perturbation theory
\cite{Wei79-1,GL82-1} in the context of the nonlinear sigma model for
the lowest $0^{-+}$ octet. In the present paper this approach is
extended to a linear meson or sigma model \cite{GML60-1} including
also fields for the $\eta^\prime$, the lowest lying scalar $0^{++}$
octet and a scalar singlet. (The latter is often called ``$\sigma$
particle'', and we use in this work the terms ``linear meson model''
and ``linear sigma model'' synonymously.) Together with the $0^{-+}$
octet these fields are combined into a complex $3\times3$ matrix
$\Phi$ which transforms as a linear $(\ol{\bf 3},{\bf 3})$
representation with respect to $SU_L(3)\times SU_R(3)$. The
corresponding mesons can be interpreted as quark--antiquark bound
states $\ol{q}_L q_R$ \cite{Bij95-1}.  In the absence of quark masses
spontaneous chiral symmetry breaking arises through a nonvanishing
vacuum expectation value of the scalar singlet described by the real
part of $\Tr\Phi$. Nonvanishing quark masses also enforce nonzero
expectation values of the diagonal part of the scalar octet.  In this
context the possible information from approximate chiral symmetry
breaking is twofold: First, there are a few simple linear relations
which can be understood easily on the basis of representation theory.
A typical example is a Gell-Mann--Okubo type mass relation for the
particles in the $0^{++}$ octet. This kind of relation could equally
well be understood in the context of an extended nonlinear model
including the $\eta^\prime$ and the scalar octet. Beyond this, the
linear meson model may imply further constraints for the free
parameters remaining in a nonlinear model to a given order in the
quark mass expansion.  This type of constraint arises typically from
nonlinearities in the map from the linear to the nonlinear sigma model
and is difficult to classify by representation theory. We observe that
the effective action of the nonlinear sigma model for the pseudoscalar
octet is completely contained in the effective action for the linear
meson model, once restricted to the appropriate degrees of freedom.
One may therefore hope to extract some information on those parameters
of the nonlinear sigma model which appear in higher orders in the
quark mass expansion.

The predictive power of the linear  model is greatly enhanced if
approximate chiral symmetry is combined with additional assumptions:
\begin{description}
\item[(i) The derivative expansion] assumes for the inverse propagator
  that the deviation from a momentum dependence $\sim q^2+m^2$ is only
  a small effect. This should hold in a range of $q^2$ in the vicinity
  of the zero at $q^2=-m^2$. It amounts to neglecting terms in the
  effective action which contain more than two derivatives or treating
  the deviations of the inverse propagators from $q^2+m^2$ as small
  corrections in a systematic way. For a determination of masses,
  mixing angles or decay widths only Green functions with on--shell
  external momenta are of interest. Hence, it is natural to expand the
  proper vertices around external momenta corresponding to
  appropriately chosen ``average masses'' for each $SU_V(3)$
  multiplet. For many observables this leads to a derivative expansion
  which formally corresponds to an expansion in powers of quark
  masses. It should be stressed, though, that non--analyticities due
  to multi--particle thresholds clearly restrict the range of validity
  of this expansion.
\item[(ii) The expansion in the chiral condensate] assumes that the
  typical mass scale relevant for spontaneous chiral symmetry breaking
  is small as compared to the typical strong interaction mass scales,
  as, for instance, the string tension or glueball masses. The
  discussion of the relevant scales is somewhat subtle (cf.~section
  \ref{ExpansionInTheChiralCondensate}). It is often sufficient to
  assume that the scale of spontaneous chiral symmetry breaking is not
  large as compared to other strong interaction scales. In the present
  paper we do not exploit explicitly this expansion but rather use it
  in order to establish reasonable ranges for some parameters. We
  emphasize in this context that the present paper makes no polynomial
  expansion of the effective action around $\Phi=0$. We will instead
  expand in the difference $\Phi-\VEV{\Phi}$ with $\VEV{\Phi}$ the
  expectation value of $\Phi$ in the presence of spontaneous chiral
  symmetry breaking and equal quark masses. The latter is justified by
  the observation that a given order in the quark mass expansion only
  involves a maximal power of $\Phi-\VEV{\Phi}$.  Within the expansion
  around $\VEV{\Phi}$ the chiral condensate
  $\sigma_0\sim\Tr\VEV{\Phi}$ appears as a free parameter. It should
  be noted that the validity of a polynomial expansion around $\Phi=0$
  would automatically generate a systematic expansion in powers of
  $\sigma_0$. It is, however, not necessary for this purpose.
\end{description}

Both, the derivative expansion and the expansion in $\sigma_0$ can
also be motivated by the observation that a classical linear sigma
model is in the class of renormalizable theories and remains there if
it is coupled to quarks. (We neglect here the large--distance
nonlocalities in the effective quark interactions reflecting
confinement induced by the gluonic degrees of freedom.) Quantum
fluctuations have then the tendency to induce a flow of the effective
couplings towards the Gaussian fixed point (triviality) in the
vicinity of which the derivative and polynomial expansions become
valid\footnote{Nonlocal quark interactions related to confinement
  counteract this tendency for very low momentum scales. They
  influence the mesons only indirectly and are suppressed by a
  nonvanishing constituent quark mass.}. Because of nonvanishing meson
masses the running extends, however, at most over a range somewhere
inbetween the $\GeV$ scale below which the mesons form as quark bound
states \cite{EW94-1} and approximately $100\MeV$ where the pion mass
acts an an infrared cutoff (with graduation because of the rich scalar
mass spectrum).  Nevertheless, the renormalization effects may be
substantial due to the existence of strong effective couplings
\cite{JW95-1} such that the general form of the effective action may
already be influenced by the vicinity of the Gaussian fixed point. We
do not expect that the derivative expansion or the expansion in the
chiral condensate $\sigma_0$ converge very fast under all
circumstances. The associated dimensionless expansion parameters are
simply not very small. This also holds for the expansion in the
strange quark mass in contrast to an expansion in $m_u$ and $m_d$. We
will discuss these issues in detail and find that the ``rate of
convergence'' depends quite significantly on the physical quantity
considered. As a general rule, the convergence is much better for the
flavored mesons than for the non--flavored ones.
\begin{description}
\item[(iii) The leading mixing approximation] attributes the dominant
  deviation from the low order results of the quark mass or the
  derivative expansion to a mixing of states. A prominent example is
  the $\eta$--$\eta^\prime$ mixing which indeed turns out to be
  responsible for the comparatively slow convergence of the
  straightforward quark mass expansion in this sector. Another
  important feature in this context is the ``partial Higgs effect''
  which describes the mixing with the $0^{-+}$ states contained in the
  divergence $\prl_\mu\rho_A^\mu$ of the axialvector fields.
\end{description}

The smallest common denominator of all these considerations and the
minimal starting point for any systematic study of the linear meson
model assumes an effective action consisting of the most general
effective potential for $\Phi$ and the most general kinetic term
involving two derivatives. By this we mean that all invariants
consistent with $SU_L(3)\times SU_R(3)$ symmetry have to be included
which contribute to a given order in the quark mass expansion. It is
crucial in this respect that the most general kinetic term in the
effective Lagrangian is not simply
$Z_\vph\Tr\prl_\mu\Phi^\dagger\prl^\mu\Phi$. There are other important
invariants involving two derivatives as, for example, a term 
\begin{eqnarray}
  \label{Int0}
  &&\ds{\frac{1}{8}X_\vp^-\Big\{\Tr\left(
 \Phid\prl_\mu\Phi-\prl_\mu\Phid\Phi\right)
 \left(\Phid\prl^\mu\Phi-\prl^\mu\Phid\Phi\right)}\nnn
 && \ds{\hspace{0.8cm} +
 \Tr\left(\Phi\prl_\mu\Phid-\prl_\mu\Phi\Phid\right)
 \left(\Phi\prl^\mu\Phid-\prl^\mu\Phi\Phid\right)
 \Big\}}
\end{eqnarray}
After chiral symmetry breaking this term induces different wave
function renormalizations for the pseudoscalar and the scalar octets,
a momentum dependent mixing of $\eta$ and $\eta^\prime$ and similar
effects which all turn out to be quantitatively important! We
emphasize that the symmetry breaking effects in the wave function
renormalizations $Z_i$ for the various mesons (which are described by
the kinetic terms) are as important for an understanding of the meson
mass spectrum as the ``unrenormalized mass terms'' $\ol{M}_i^2$
(described by the effective potential). With effective inverse
propagators $\sim Z_i q^2+\ol{M}_i^2$ the physical masses are given as
$M_i=\ol{M}_i Z_i^{-1/2}$. A study of the mass splitting between the
scalar and the pseudoscalar octet involves the effect of chiral
symmetry breaking on $\ol{M}_i^2$ and $Z_i$. We therefore investigate
the kinetic terms in the same way as the effective potential. This
explains most of the differences of our results with earlier
investigations \cite{PW84-1}--\cite{Pis95-1} where chiral symmetry
breaking in the kinetic terms was neglected.

The present paper is devoted to a systematic study of the effective
action of the linear meson model based on the considerations discussed
above. For the pseudoscalar sector we take as phenomenological input
$M_{\pi^\pm}$, $M_{K^\pm}$, $M_{K^0}$, $M_{\eta^\prime}$, $f_\pi$ and
$f_{K^\pm}$.  Using this we compute partial decay rates for the
$\pi^0$, $\eta$ and $\eta^\prime$ into two photons as parameterized by
$f_{\pi^0}$, $f_\eta$ and $f_{\eta^\prime}$ as well as other
quantities of interest.  The perhaps most striking outcome is that
$M_\eta$ as well as the decay constants $f_\eta$ and $f_{\eta^\prime}$
are determined essentially as functions of only one additional
parameter, with a rather weak dependence on the other couplings
present in the linear sigma model.  Fixing this parameter by the
measured value $M_\eta=547.5\MeV$ we predict to first order in the
quark mass expansion and first order in the derivative expansion
\begin{eqnarray}
  \label{Int1}
  \ds{\frac{f_{\pi^0}}{f_\pi}} &\simeq& \ds{
  1.00}\nnn
  \ds{\frac{f_\eta}{f_\pi}} &\simeq& \ds{
  1.23}\nnn
  \ds{\frac{f_{\eta^\prime}}{f_\pi}} &\simeq& \ds{
  0.91}\; .
\end{eqnarray}
Taking into account the theoretical uncertainties these results are in
satisfactory agreement with the experimental observations
$(f_{\pi^0}/f_\pi)^{\rm exp}=1.00\pm0.04$, $(f_\eta/f_\pi)^{\rm
  exp}=1.06\pm0.05$ and $(f_{\eta^\prime}/f_\pi)^{\rm
  exp}=0.81\pm0.02$. In view of the lowest order result for vanishing
quark masses, $(f_\eta/f_\pi)^{(0)}=\sqrt{3}$,
$(f_{\eta^\prime}/f_\pi)^{(0)}=\sqrt{3/8}$ this is rather remarkable.
The values of $M_\eta$, $f_\eta$ and $f_{\eta^\prime}$ for different
parameters of the model can be found in sections
\ref{MassRelaionsAndCouplingConstantsToQuadraticOrder},
\ref{DecayConstantsOfEtaAndEtap} and \ref{Results}. One can get an
idea about the ``robustness'' of the estimate (\ref{Int1}) from the
figures and tables of these sections. We also discuss the masses of
the mesons in the lowest lying $0^{++}$ octet (section
\ref{MassRelationsForTheScalarOctet}). We find that the scalar partner
of the $\eta$ has a typical mass of $(1300-1400)\MeV$ and should be
associated with the resonance $f_0(1300)$ \cite{PDG94-1}.
Large mixing effects with two--kaon or four--quark states are
characteristic for the isotriplet $a_0(980)$.  The resonance
$a_0(980)$ may actually be dominantly a two--kaon state and in this
case the model suggests a further isotriplet resonance with a mass
around $1300\MeV$. It may be identified with the reported resonance
$a_0(1320)$ \cite{PDG94-2}.

Four main lines enter our systematic analysis:

{\bf (1)} The relations between the pseudoscalar meson mass
differences within a given multiplet and the differences in decay
constants $f_K-f_\pi$ or $f_{K^\pm}-f_{K^0}$ involve the couplings
between two pseudoscalar octets and one or more scalar octets. Up to a
wave function renormalization the differences of decay constants
correspond to the expectation values $\VEV{h}$ of the diagonal fields
in the scalar octet. Lowest order mass differences follow from the
cubic couplings $\sim\Tr(m^2 h)$ once the expectation value of $h$ is
inserted.  (Here $m$ denotes the traceless hermitean matrix of
pseudoscalar octet fields and $h$ that of the scalar octet.)
Similarly higher order corrections arise from quartic couplings as
$\Tr(m h m h)$ and so on. The quark mass expansion is closely related
to an expansion in powers of the $SU_V(3)$--breaking expectation value
$\VEV{h}$. This mechanism is described in detail in sections
\ref{ScalarMesonMassesToLinearOrder} and
\ref{PseudoScalarMesonMassesToQuadraticOrder}. As a byproduct of our
analysis one also gains information on the cubic and quartic couplings
involving pseudoscalars and scalars which are relevant for the decay
of a scalar into two pseudoscalars etc.

{\bf (2)} We formulate the quark mass expansion as a power series in
the parameters $\Delta_u$, $\Delta_d$, $\Delta_s$ which measure the
deviation of the scalar expectation values from their values for zero
quark masses. In particular,
$\Delta_s-\frac{1}{2}(\Delta_u+\Delta_d)\sim f_K-f_\pi$ corresponds to
the $SU_V(3)$--breaking induced by the mass of the strange quark, and
$\Delta_u-\Delta_d\sim f_{K^\pm}-f_{K^0}$ measures the amount of
isospin breaking. In our analysis we actually never need to determine
the current quark masses. Our ``predictions'' involve directly the
relations between meson masses and decay constants. Within the
language of a general $SU_V(3)$ symmetric model for $m$, $h$, etc.~the
essential ingredient is the determination of the cubic and quartic
couplings between scalars and pseudoscalars as well as the wave
function renormalizations from the couplings of the linear meson
model. This is done in section \ref{LinearSigmaModel}.

A systematic expression of the pseudoscalar octet mass splitting to
order $\Delta$ needs the identification of those terms of the
effective potential which contribute to this order whereas the quark
mass corrections to the kinetic term can be neglected.  Similarly, an
estimate to order $\Delta^2$ (corresponding to second order in the
quark mass expansion) involves the effective potential contribution to
order $\Delta^2$, corrections to the kinetic terms to order $\Delta$
and a lowest order estimate of the terms involving four derivatives.
We find that the apparent convergence of the expansion in $\Delta$ is
quite satisfactory for the flavored pseudoscalars $\pi^\pm$, $K^\pm$,
$K^0$, $\ol{K}^0$ and the $\pi^0$.  On the other hand, the formal
series in $\Delta$ does not converge very well in the
$\eta$--$\eta^\prime$ sector if the singlet mass term generated by the
chiral anomaly is, as usual, considered as a quantity $\sim\Oc(1)$.
The reason are the relatively large mixing effects which are formally
of the order $\Delta$. This is combined with the observation that a
zeroth order mass term for the $\eta^\prime$ (without mixing) is
considerably smaller than the physical $\eta^\prime$ mass and actually
not so much larger than the zeroth order mass of the $\eta$. We wish to
stress that the apparent convergence of the $\Delta$--expansion in the
$\eta$--$\eta^\prime$ sector improves substantially if one includes
systematically all effects to a given order in $\Delta$ for all
elements of the $2\times2$ matrix which describes the inverse
propagator of the $\eta$--$\eta^\prime$ system.  After diagonalization
this procedure amounts for the mass eigenvalues to a partial
resummation of terms which are formally of higher order in $\Delta$.
We believe that these convergence properties of the quark mass
expansion are quite general: The series converges very well only in
situations without large effects from mixing of states. If mixing is
important, a good convergence can only be obtained if the
$\Delta$--corrections are retained for all elements in the relevant
matrix. We encounter a very similar situation if we want to interpret
the $a_0(980)$ as the isotriplet member of the scalar octet described
by $h$.

For a computation of differences in decay constants like $f_K-f_\pi$
the lowest order in the quark mass expansion requires corrections of
order $\Delta$ both for the effective potential and the kinetic terms.
This implies that all relations between meson mass differences and
decay constants need to lowest nontrivial order also the kinetic terms
to order $\Delta$. The only exceptions are the Gell-Mann--Okubo mass
relations to linear order in $\Delta$. They do not involve the decay
constants and the leading quark mass corrections to the kinetic terms
cancel for these relations. All other ``predictions'' of the model
involve the quark mass corrections to the kinetic term. In order to be
able to study separately the quark mass expansion and the derivative
expansion we have split our systematic exploration of the linear meson
model to order $\Delta$ into several parts. Sections
\ref{LinearSigmaModel} and \ref{ParametersOfTheLinearSigmaModel} deal
with the $\Delta$--expansion of the effective potential. In section
\ref{QuarkMassCorrectionsToKineticTerms} we supplement these
considerations by a general discussion of corrections $\sim\Delta$ and
higher derivative contributions to the kinetic terms. The quark mass
corrections to the kinetic terms are computed within the linear meson
model in sect.~\ref{KineticTermsInTheLinearSigmaModel} and the higher
derivative contributions are estimated in section
\ref{HigherDerivativeContributions}.

{\bf (3)} The expansion in powers of $\Delta$ is a self--contained
systematic formalism. Nevertheless, large mixing effects sometimes
prevent a fast convergence of the series. One of the examples
encountered in this work is the relatively large momentum dependent
off--diagonal element in the inverse $\eta$--$\eta^\prime$ propagator.
In turn, the sign and size of this element can be explained by the
mixing of the pseudoscalar octet $m$ and the singlet $p$ with the
states corresponding to the divergence of the axialvector fields
$\prl_\mu\rho_A^\mu$. These states have the same quantum numbers and
the mixing corresponds to the so called ``partial Higgs effect''. We
have performed in appendix \ref{VectorMesonContributions} the
corresponding analysis of the vector and axialvector system coupled to
pseudoscalars and scalars and estimated the contribution from the
exchange of $\prl_\mu\rho_A^\mu$ to the off--diagonal element in the
$\eta$--$\eta^\prime$ inverse propagator. This estimate coincides
rather well with the value that leads to realistic numbers for
$M_\eta$, $f_\eta$ and $f_{\eta^\prime}$! This is rather encouraging
since the large $\Oc(\Delta)$ effects find now a natural explanation.
Quite generally, our results lead to the picture that the
$\Delta$--expansion as well as the derivative expansion converge well
once a large enough basis of states is included. Integrating out such
states, however, can lead to large coefficients in the formal
$\Delta$--expansion for the remaining states and therefore to a slow
convergence. Keeping the additional states or integrating them out
without a further truncation of the series to given order in $\Delta$
is equivalent to a resummation of higher order terms from the point of
view of the formal $\Delta$--expansion. This improves the convergence
substantially. We therefore have given a general discussion of mixing
effects in appendix \ref{MixingWithOtherStates}, where we also
identify the most prominent mixings relevant in our context.

{\bf (4)} The effective action $\Gamma$ includes all quantum
fluctuations. If $\Gamma$ is known no further loop calculations are
necessary. Nevertheless, it is often useful to estimate the
contributions of fluctuations for certain invariants contained in
$\Gamma$. An example are the contributions to higher derivative terms
which reflect the deviation of an inverse propagator $G^{-1}(q^2)$
{}from the leading momentum dependence $Z q^2+\ol{M}^2$. Recently, a
reformulation of perturbation theory was based on effective vertices
and propagators instead of the classical ones \cite{Wet96-1}. Here,
this means that the contributions of quantum fluctuations to the
difference $G^{-1}(q^2)-(Z q^2+\ol{M}^2)$ can be computed in terms of
$Z$, $\ol{M}^2$ and appropriate effective cubic couplings
$\gamma_i$. Since the effective vertices have been determined by our
analysis we can give in sect.~\ref{HigherDerivativeContributions} a
quantitative estimate of the contributions from scalar and
pseudoscalar fluctuations to the higher derivative terms. They turn
out to be substantially smaller than those arising from mixing effects
with other states.

Besides its main purpose of a systematic discussion of meson masses
and decay constants within the framework of the linear meson model
this work also constitutes the basis for several interesting
developments: Recently, the parameters of the two--flavor meson model
were estimated from a solution of nonperturbative flow equations
\cite{JW95-1}. The resulting values for $f_\pi$ and the chiral
condensate $\VEV{\ol{\psi}\psi}$ encourage an extension of this
approach to three flavors. In this case the nonvanishing mass of the
strange quark plays an important quantitative role and needs to be
included. The results of such a future estimate of the parameters of
the linear meson model can then be compared with the values infered in
the present work from a phenomenological analysis. The present results
also give an idea which invariants are important and should be
retained in the necessary truncation of the exact nonperturbative flow
equation.

A determination of the parameters of the linear sigma model in the
zero temperature ground state is the starting point for any analysis
of the high temperature behavior of this model. This is relevant for
the QCD phase transition in the early universe and for ongoing and
future heavy ion collision experiments. Within the linear meson model
the temperature dependence can be studied along similar lines as in
\cite{TW94-1}. On a crude qualitative level one may also use a mean
field approximation \cite{PW84-1}--\cite{MOS96-1}. The temperature
dependence of meson masses -- in particular the vector mesons -- may
be experimentally observable in heavy ion collisions. We should point
out, however, that such a study remains reliable only as long as $2\pi
T$ is smaller than the compositeness scale above which a mesonic
description of strong interaction physics becomes invalid.

Finally, we have not exploited in this work the information we have
gained concerning cubic and quartic meson couplings. Especially for
the extension of our model to include vector and axialvector mesons
(appendix \ref{VectorMesonContributions}) these couplings determine
the decay rates and branching ratios. They therefore contain a whole
lot of additional ``predictions'' that can be compared with
experiment. We plan to come back to all these issues in future work.

\sect{Meson masses to linear order}
\label{ScalarMesonMassesToLinearOrder}

In the limit of vanishing quark masses the chiral symmetry
$SU_L(3)\times SU_R(3)$ is spontaneously broken to the vector--like
$SU_V(3)$ subgroup. With respect to this symmetry breaking the $18$
real fields contained in the complex $(\ol{\bf 3},{\bf 3})$
representation decompose into a pseudoscalar octet plus singlet and a
scalar octet plus singlet. The pseudoscalar octet is associated
with the eight (pseudo--)Goldstone bosons $\pi^\pm$, $\pi^0$, $K^\pm$,
$K^0$, $\ol{K}^0$, $\eta$ and the pseudoscalar singlet with the
$\eta^\prime$ particle. The scalars comprise the
isotriplet $a_0$, the strange scalars $K_0^{*\pm}$, $K_0^{*0}$,
$\ol{K}_0^{*0}$ as well as two states with $f_0$ quantum numbers, one
of them being the ``$\sigma$--particle''. Chiral symmetry breaking is
induced by a nonvanishing expectation value $\sigma_0$ of the scalar
singlet. In presence of current quark masses the $SU_V(3)$ symmetry is
explicitly broken. This is reflected by nonzero expectation values of
the flavor neutral fields within the scalar octet.

This and the next section are devoted to a general analysis of the
effects of $SU_V(3)$--breaking without explicit reference to the
linear sigma model and spontaneously broken chiral symmetry. These
concepts will only be introduced in sect.~\ref{LinearSigmaModel}. In
this way we can understand the new relations emerging from the linear
sigma model in the context of a more general setting involving only
$SU_V(3)$ symmetry.

The octet of pseudoscalar mesons is described by a hermitian
traceless $3\times3$ matrix $m$, whereas the scalar octet is similarly
denoted by $h$. In addition, we have the pseudoscalar and scalar
$SU(3)$ singlets $p$ and $s$. These fields are normalized with
standard kinetic terms
\begin{equation}\begin{array}{rcl}
 \ds{\Lc_{\rm kin}} &=& \ds{
 \frac{1}{4}\Tr\prl^\mu m\prl_\mu m+
 \frac{1}{4}\Tr\prl^\mu h\prl_\mu h+
 \hal\prl^\mu p\prl_\mu p+
 \hal\prl^\mu s\prl_\mu s}\; .
 \label{LkinForMultiplets}
\end{array}\end{equation}
With $SU(3)$ generators $\la_z$ obeying $\Tr\la_y\la_z=2\dt_{yz}$ we
may write $m=m_z\la_z$, $h=h_z\la_z$ with $m_z$, $h_z$ real. For
vanishing quark masses the pseudoscalar octet corresponds to the
massless Goldstone bosons. The most general mass term consistent with
$SU(3)$ symmetry, charge conjugation $\Cc$ and
parity\footnote{Charge conjugation corresponds to a transposition of
  both matrices $m$ and $h$ whereas parity amounts to $m\ra-m$, $h\ra
  h$, $p\ra-p$, $s\ra s$.} $\Pc$ reads (we will work in Euclidean
space time throughout this paper)
\begin{equation}
 \Lc_{m_o}=\frac{1}{4}m_m^2\Tr m^2+\frac{1}{4}m_h^2\Tr h^2+
 \hal m_p^2 p^2+\hal m_s^2 (s-s_0)^2\; .
 \label{GerneralMassLagrangian}
\end{equation}
Small nonvanishing quark masses can be described by a linear
perturbation or external source for the scalar mesons
\begin{equation}
 \Lc_{j}=-\hal\Tr j_h h-j_s s
 \label{SourceMassTerm}
\end{equation}
with $j_h$ diagonal and traceless. Both $j_h$ and $j_s$ are linear in
the quark masses with\footnote{The normalization $c_s$ can be inferred
  {}from sect.~\ref{LinearSigmaModel}.} 
\begin{equation}
  \label{LOP0}
  j_h+c_s j_s\sim\left(
  \begin{array}{ccc}
  m_u & & \\
   & m_d & \\
   & & m_s
  \end{array}\right)\; .
\end{equation}
This leads to a shift in the expectation value of
the scalar fields
\begin{equation}\begin{array}{rcl}
 \ds{\VEV{s}} &=& \ds{u}\nnn
 \ds{\VEV{h}} &=& \ds{w\la_3-\sqrt{3}v\la_8}
\end{array}\end{equation}

There are direct relations between the parameters $v$ and $w$ and the
differences of meson decay constants which are explained in
appendix \ref{MesonDecayConstants}
\begin{equation}\begin{array}{rcl}
 \ds{\left(\frac{Z_m}{Z_h}\right)^\hal v} &=& 
 \ds{\frac{1}{3}\left(2\Dt_s-\Dt_u-\Dt_d\right)=
 \frac{1}{3}\left(\ol{f}_{K^\pm}+\ol{f}_{K^0}-2\ol{f}_\pi\right)}\nnn
 \ds{\left(\frac{Z_m}{Z_h}\right)^\hal w} &=& 
 \ds{\left(\Dt_u-\Dt_d\right)=
 \ol{f}_{K^\pm}-\ol{f}_{K^0}}\; .
 \label{DifferencesOfDecayConstants}
\end{array}\end{equation}
Here $\Dt_u$ is proportional to the difference of the $\VEV{\olu u}$
condensate for nonvanishing and vanishing quark mass and similarly for
$\Dt_d$ and $\Dt_s$. The quantities $\ol{f}_i$ are related to the
meson decay constants $f_i$ by $SU(3)$ breaking wave function
renormalizations which will be discussed in section
\ref{QuarkMassCorrectionsToKineticTerms}. Until then we take into
account the effects of nonvanishing quark masses only in the lowest
order in a derivative expansion. In this approximation the kinetic
term is given by (\ref{LkinForMultiplets}) and we may identify
$\ol{f}_i$ with $f_i$.  The wave function renormalization constants
$Z_m$ and $Z_h$ are induced by different kinetic terms for the
pseudoscalar and scalar octets within the linear sigma model and will
be explained in sect.~\ref{LinearSigmaModel}. The isospin violating
expectation value $w$ is very small and we can use the experimental
values $f_\pi\equiv f_{\pi^\pm}=92.4\MeV$, $f_{K^\pm}=113\MeV$ to
estimate
\begin{equation}
 \left(\frac{Z_m}{Z_h}\right)^\hal v=13.7\MeV\; .
 \label{vEstimate}
\end{equation}
We will see later, (\ref{NewEstimateV}), that the relation
(\ref{vEstimate}) is modified by quark mass corrections to the kinetic
terms leading to $\ol{f}_i\neq f_i$. As a result the value
(\ref{vEstimate}) will be shifted to $23.3\MeV$.

At this stage $u$ is not yet fixed, since we still have to specify
$s_0$ or, equivalently, the meaning of $s=0$. A possible choice would
be that $s=0$ denotes the minimum of the potential in the absence of
quark masses, i.e. $s_0=0$. For this choice $m_m^2$ vanishes and
$\VEV{s}=u=\sqrt{2/3}\left(\Dt_u+\Dt_d+\Dt_s\right)\left(Z_s/Z_m\right)^\hal$.
It will, however, prove to be more convenient to choose the origin of
$s$ corresponding to the minimum of the singlet field in the presence
of quark masses. In this case one has $s_0=-j_s/m_s^2$ and
\begin{equation}
 \VEV{s}=u=0\; ,\;\;\; m_m^2>0
\end{equation}
such that $m_m^2\sim\Oc(\Dt)$.  We note that more generally the choice
of $s_0$ fixes the point around which the potential $U(m,h,p,s)$ is
expanded and therefore $m_m^2$ as well as the values of all other
parameters in a polynomial expansion of $U$ depend on $s_0$. For the
choice $u=0$ we will use $m_m^2$, $v$ and $w$ instead of the current
quark masses $m_u$, $m_d$ and $m_s$ as the parameters characterizing
the explicit chiral symmetry breaking. Their relation to the quark
masses involves the solution of the field equations for $s$ and $h$ in
presence of the sources (\ref{SourceMassTerm}). With the choice $u=0$
our expansion parameter is given by
$(Z_m/Z_h)^{1/2}v\simeq\frac{2}{3}(\ol{f}_K-\ol{f}_\pi)$. The quark
mass expansion turns into a Taylor expansion in the small parameter
$(\ol{f}_K-\ol{f}_\pi)/(\ol{f}_K+\ol{f}_\pi)\simeq0.1(0.2)$. Here the
numbers in brackets include the effects of quark mass corrections to
the kinetic terms. Correspondingly, the small parameter for isospin
violating effects is $(\ol{f}_{K^\pm}-\ol{f}_{K^0})/(\ol{f}_{K^\pm}+
\ol{f}_{K^0})\simeq-2\cdot10^{-3}(-3\cdot10^{-3})$ (see below). This
implies that linear isospin breaking effects give corrections of the
same order of magnitude as quadratic quark mass effects. As a rough
estimate one expects that the squared meson masses can be computed to
linear order in the quark masses with an accuracy of $10$--$20$
percent.  This corresponds to a typical uncertainty for the masses of
the pseudoscalar octet of around $(30$--$60)\MeV$. To quadratic order
in $\Dt$ this error is expected to decrease to a few $\MeV$.

We want to investigate the dependence of the pseudoscalar meson
masses on the parameters $m_m^2$, $v$ and $w$. We will work in this
and the next three sections with a kinetic term of the form
(\ref{LkinForMultiplets}) and discuss effects from modifications of
the kinetic term in sect.~\ref{QuarkMassCorrectionsToKineticTerms}.
To linear order in the quark masses we need the contributions linear
in $m_m^2$, $v$ and $w$.  This involves the cubic couplings of two
pseudoscalars and $h$ or $s$.  Their most general form consistent
with $SU(3)$ symmetry, charge conjugation and parity reads
\begin{equation}
 \Lc_3=\frac{1}{4}\gm_1 s\Tr m^2+\frac{1}{4}\gm_2\Tr m^2 h+
 \hal\gm_3 p\Tr m h+\hal\gm_4 s p^2\; .
 \label{Lthree}
\end{equation}
We note that because of $\VEV{s}=0$ only $m_m^2$ and the couplings
$\gm_2$, $\gm_3$ contribute to the pseudoscalar mass matrix. We find
for the masses of the off--diagonal or flavored mesons $\pi^\pm$,
$K^\pm$ and $K^0$, $\olK^0$
\begin{equation}\begin{array}{rcl}
 \ds{\ol{M}_{\pi^\pm}^2} &=& \ds{m_m^2-\gm_2v}\nnn
 \ds{\ol{M}_{K^\pm}^2}  &=& \ds{m_m^2+\hal\gm_2\left( v+w\right)}\nnn
 \ds{\ol{M}_{K^0}^2}   &=& \ds{m_m^2+\hal\gm_2\left( v-w\right)}\; .
 \label{FlavoredMesonMasses}
\end{array}\end{equation}
Again, the bars indicate that the quantities $\ol{M}_i$ differ from
the physical meson masses $M_i$ by $SU(3)$ breaking wave function
renormalization effects to be discussed in section
\ref{QuarkMassCorrectionsToKineticTerms}. As long as we restrict the
discussion to the kinetic term (\ref{LkinForMultiplets}) (sections
\ref{ScalarMesonMassesToLinearOrder}--\ref{ParametersOfTheLinearSigmaModel})
we can identify $\ol{M}_i$ with $M_i$. Eq.~(\ref{FlavoredMesonMasses})
allows one to express couplings in terms of meson masses
\begin{equation}\begin{array}{rcl}
 \ds{m_m^2} &=& \ds{
 \frac{1}{3}\left(\ol{M}_{\pi^\pm}^2+\ol{M}_{K^\pm}^2+
 \ol{M}_{K^0}^2\right)}\nnn
 \ds{\gm_2w} &=& \ds{\ol{M}_{K^\pm}^2-\ol{M}_{K^0}^2}\nnn
 \ds{\gm_2v} &=& \ds{
 \frac{1}{3}\left(\ol{M}_{K^\pm}^2+\ol{M}_{K^0}^2-
 2\ol{M}_{\pi^\pm}^2\right)}
 \label{GammasMasses}
\end{array}\end{equation}
and to estimate the isospin violation to leading order in the
quark masses
\begin{equation}
 \ol{f}_{K^0}-\ol{f}_{K^\pm}=
 -\left(\frac{Z_m}{Z_h}\right)^\hal w=
 3\left(\frac{Z_m}{Z_h}\right)^\hal v
 \frac{\ol{M}_{K^0}^2-\ol{M}_{K^\pm}^2}
 {\ol{M}_{K^\pm}^2+
 \ol{M}_{K^0}^2-
 2\ol{M}_{\pi^\pm}^2}
 \simeq0.47\MeV
 \label{EstimateW}
\end{equation}
which is in good agreement with the result
$f_{K^0}-f_{K^\pm}\simeq0.45\MeV$ from chiral perturbation theory
\cite{GL82-1}. Here we have used electromagnetically corrected masses
$M_{\pi^\pm}=135.1\MeV$, $M_{K^\pm}=492.4\MeV$, $M_{K^0}=497.7\MeV$.
Including quark mass corrections to the kinetic terms these values are
shifted to $(Z_m/Z_h)^{1/2}w\simeq-0.67\MeV$, (\ref{NNN28}), and
$f_{K^0}\simeq113.28\MeV$, (\ref{NNN29}). We will occasionally also
use the isospin means $\ol{f}_K=\hal(\ol{f}_{K^\pm}+\ol{f}_{K^0})$ and
$\ol{M}_K^2=\hal(\ol{M}_{K^\pm}^2+\ol{M}_{K^0}^2)$.

To obtain the masses of the neutral pseudoscalars $\pi^0$, $\eta$
and $\etap$ we have to diagonalize the general mass term
\begin{equation}
 \hal\left( \ol{M}_3^2m_3^2+\ol{M}_8^2m_8^2+M_p^2p^2\right)+
 \ol{M}_{38}^2 m_3m_8+M_{3p}^2 m_3p+M_{8p}^2 m_8p
\end{equation}
which also involves $m_p^2$ and $\gm_3$
according to
\begin{equation}\begin{array}{rcl}
 \ds{M_p^2} &=& \ds{m_p^2}\nnn
 \ds{M_{8p}^2} &=& \ds{-\sqrt{3}\gm_3v}\nnn
 \ds{M_{3p}^2} &=& \ds{\gm_3w}\nnn
 \ds{\ol{M}_8^2} &=& \ds{m_m^2+\gm_2v=
 \frac{1}{3}\left(2\ol{M}_{K^\pm}^2+
 2\ol{M}_{K^0}^2-\ol{M}_{\pi^\pm}^2\right)}\nnn
 \ds{\ol{M}_3^2} &=& \ds{m_m^2-\gm_2v=\ol{M}_{\pi^\pm}^2}\nnn
 \ds{\ol{M}_{38}^2} &=& \ds{\frac{1}{\sqrt{3}}\gm_2w=
 \frac{1}{\sqrt{3}}\left(\ol{M}_{K^\pm}^2-\ol{M}_{K^0}^2\right)}\; .
 \label{PseudoScalarMassesToLinearOrder}
\end{array}\end{equation}
We recover the usual lowest order relations of chiral perturbation
theory or the ``eightfold way'' for $\ol{M}_8^2$, $\ol{M}_3^2$ and
$\ol{M}_{38}^2$ \cite{GMO61-1,GL82-1}. We will neglect the isospin
violating mixings of $m_3$ which only contribute to order $w^2$ to the
mass eigenvalues.  Diagonalization of the $(m_8,p)$ sector yields the
mass eigenstates
\begin{equation}\begin{array}{rcl}
 \ds{\eta} &=& \ds{m_8\cos\th_p-p\sin\th_p}\nnn
 \ds{\etap} &=& \ds{p\cos\th_p+m_8\sin\th_p}\; .
\end{array}\end{equation}
The masses of $\eta$ and $\etap$ and the octet--singlet mixing angle
$\th_p$ are given by
\begin{equation}\begin{array}{rcl}
 \ds{M_\etap^2+M_\eta^2} &=& \ds{M_p^2+\ol{M}_8^2}\nnn
 \ds{M_\etap^2-M_\eta^2} &=& \ds{
 \left[\left(M_p^2-\ol{M}_8^2\right)^2+4M_{8p}^4\right]^\hal
 }\nnn
 \ds{\tan\th_p} &=& \ds{
 \frac{\ol{M}_8^2-M_p^2+\left[\left(\ol{M}_8^2-
 M_p^2\right)^2+4M_{8p}^4\right]^\hal}
 {2M_{8p}^2}}\; .
 \label{MixingAngle}
\end{array}\end{equation}
Contrary to the meson masses the mixing angle $\th_p$ will receive
additional contributions $\sim\Oc(\Dt)$ from modifications of the
kinetic terms discussed in section
\ref{QuarkMassCorrectionsToKineticTerms}.

It will later be our aim to find relations among the parameters
$m_p^2$, $m_m^2$, $\gm_2$, $\gm_3$. For the moment we only notice
that the couplings $\gm_2$ and $\gm_3$ are directly related to the
partial decay width of the scalar octet into two mesons belonging to
the pseudoscalar octet or singlet. In particular, $\gm_2$ can be
extracted from (\ref{vEstimate}) and (\ref{GammasMasses}) and one
finds
\begin{equation}
 \left(\frac{Z_h}{Z_m}\right)^\hal \gm_2\simeq11040\MeV\; .
\end{equation}
This value changes to $(Z_h/Z_m)^{1/2}\gamma_2\simeq4942\MeV$ once
quark mass corrections to the kinetic terms are included.

We next turn to the scalar masses. To zeroth order they are given by
$m_h^2$ in (\ref{GerneralMassLagrangian}). To linear order in $\Dt$
we have to supplement (\ref{Lthree}) by
\begin{equation}
 \Dt\Lc_3=\frac{1}{4}\gm_5s\Tr h^2+
 \frac{1}{6}\gm_6\Tr h^3+\frac{1}{3}\gm_7 s^3
\end{equation}
leading to
\begin{equation}\begin{array}{rcl}
 \label{L12}
 \ds{\ol{M}_{a^\pm_o}^2} &=& \ds{
 m_h^2-2\gm_6v}\nnn
 \ds{\ol{M}_{K^{*\pm}_o}^2} &=& \ds{
 m_h^2+\gm_6(v+w)}\nnn
 \ds{\ol{M}_{K^{*0}_o}^2} &=& \ds{
 m_h^2+\gm_6(v-w)}\; .
\end{array}\end{equation}
In the following we will neglect isospin violation in the scalar
sector and use the isospin means $M_{K_o^*}$ and $M_{a_o}$. The
diagonal part of the mass matrix for the flavor neutral scalars reads
\begin{equation}\begin{array}{rcl}
 \ds{M_s^2} &=& \ds{m_s^2}\nnn
 \ds{\ol{M}_{a3}^2} &=& \ds{m_h^2-2\gm_6v}\nnn
 \ds{\ol{M}_{f8}^2} &=& \ds{m_h^2+2\gm_6v}
 \label{ScalarMasses}
\end{array}\end{equation}
whereas the off--diagonal elements are given by
\begin{equation}\begin{array}{rcl}
 \ds{M_{3s}^2} &=& \ds{\gm_5w}\nnn
 \ds{M_{8s}^2} &=& \ds{-\sqrt{3}\gm_5v}\nnn
 \ds{\ol{M}_{af}^2} &=& \ds{\frac{2}{\sqrt{3}}\gm_6w}\; .
\end{array}\end{equation}
We note that $\gm_7$ does not enter these $\Oc(\Dt)$ expressions for the
pseudoscalar meson masses.
In particular, we find the interesting
relation
\begin{equation}
 \ol{M}_{f8}^2=\frac{1}{3}
 \left(4\ol{M}_{K_o^{*}}^2-
 \ol{M}_{a_o}^2\right)
 \label{NNN2}
\end{equation}
which is the Gell-Mann--Okubo formula in the scalar
sector.

\sect{Pseudoscalar meson masses to quadratic order}
\label{PseudoScalarMesonMassesToQuadraticOrder}

Before proceeding to an estimate of the various couplings we will
analyze in this section the general structure of the pseudoscalar
meson masses to quadratic order in $v$ and $w$, assuming a kinetic
term of the form (\ref{LkinForMultiplets}). We stress, however, that
the modifications of the kinetic terms for nonvanishing quark masses
are of the same order as the effects discussed in this section. They
are postponed until sect.~\ref{QuarkMassCorrectionsToKineticTerms}
only for the sake of a simple presentation and for separating clearly
different orders in the derivative expansion. Mass corrections
quadratic in $v$ and $w$ involve the quartic couplings for two
pseudoscalars and two scalars. Their general form is given by
\begin{equation}\begin{array}{rcl}
 \ds{\Lc_4} &=& \ds{
 \frac{1}{4}\dt_1 s^2\Tr m^2+
 \frac{1}{4}\dt_2 s\Tr m^2 h+\hal\dt_3 s p\Tr m h+
 \hal\dt_4 s^2 p^2+
 \frac{1}{4}\dt_5p^2\Tr h^2}\nnn
 &+& \ds{
 \hal\dt_6p\Tr m h^2+\frac{1}{8}\dt_7\Tr h^2\Tr m^2+
 \hal\dt_8\Tr m^2h^2+\hal\dt_9\Tr(m h)^2+
 \frac{1}{8}\delta_{10}\left(\Tr mh\right)^2}\; .
 \label{Lfour}
\end{array}\end{equation}
We note that the couplings $\dt_1,\ldots,\dt_4$ do not contribute to
the pseudoscalar masses. The term $\sim\dt_5$
adds effectively to $m_p^2$ an additional piece $\dt_5(3v^2+w^2)$
whereas $m_m^2$ is supplemented by $\dt_7(3v^2+w^2)$. The quartic
terms contribute to the flavored meson masses
\begin{equation}\begin{array}{rcl}
 \ds{\Dt \ol{M}_{\pi^\pm}^2} &=& \ds{
 \dt_7(3v^2+w^2)+
 2\dt_8(v^2+w^2)+2\dt_9(v^2-w^2)}\nnn
 \ds{\Dt \ol{M}_{K^\pm}^2} &=& \ds{
 \dt_7(3v^2+w^2)+
 \dt_8(5v^2-2vw+w^2)-4\dt_9(v^2-v w)}\nnn
 \ds{\Dt \ol{M}_{K^0}^2} &=& \ds{
 \dt_7(3v^2+w^2)+
 \dt_8(5v^2+2vw+w^2)-4\dt_9(v^2+v w)}
 \label{CorrectionsToFlavoredMesonMasses}
\end{array}\end{equation}
whereas the mass matrix of the neutral pseudoscalars receives
corrections 
\begin{equation}\begin{array}{rcl}
 \ds{\Dt \ol{M}_3^2} &=& \ds{\dt_7(3v^2+w^2)+
 2(\dt_8+\dt_9)(v^2+w^2)+
 \delta_{10}w^2}\nnn
 \ds{\Dt \ol{M}_8^2} &=& \ds{\dt_7(3v^2+w^2)+
 \frac{2}{3}(\dt_8+\dt_9)(9v^2+w^2)+
 3\delta_{10}v^2}\nnn
 \ds{\Dt M_p^2} &=& \ds{
 \dt_5(3v^2+w^2)}\nnn
 \ds{\Dt \ol{M}_{38}^2} &=& \ds{
 -\frac{1}{\sqrt{3}}
 (4\dt_8+4\dt_9+3\delta_{10})v w}\nnn
 \ds{\Dt M_{3p}^2} &=& \ds{-2\dt_6vw}\nnn
 \ds{\Dt M_{8p}^2} &=& \ds{
 -\frac{1}{\sqrt{3}}\dt_6(3v^2-w^2)}\; .
 \label{ZZZ5}
\end{array}\end{equation}
The lowest order relations between the off--diagonal and diagonal
mesons are only modified by the $\dt_9$ and $\delta_{10}$ terms
\begin{equation}\begin{array}{rcl}
 \ds{\ol{M}_3^2} &=& \ds{\ol{M}_{\pi^\pm}^2+
 4\dt_9w^2+\delta_{10}w^2}\nnn
 \ds{\ol{M}_8^2} &=& \ds{
 \frac{1}{3}\left(2\ol{M}_{K^\pm}^2+
 2\ol{M}_{K^0}^2-\ol{M}_{\pi^\pm}^2\right)+
 12\dt_9v^2+3\delta_{10}v^2}\; .
 \label{YYY1}
\end{array}\end{equation}
The combination $\dt_9+\delta_{10}/4$ is therefore particularly
interesting for the mass relations.

If we neglect isospin violating contributions of order $w^2$ to the
mass eigenvalues and treat the flavored meson masses
$M_{\pi^\pm}$, $M_{K^\pm}$, $M_{K^0}$ as three input
parameters we can, in principle, predict $M_\eta$ and
$M_{\etap}$ as well as the octet--singlet mixing angle $\th_p$.
This requires information about the coupling $\dt_9$, the
off--diagonal mass term
\begin{equation}
 M_{8p}^2=-\sqrt{3}v(\gm_3+\dt_6v)
\end{equation}
as well as the singlet mass term
\begin{equation}
 M_p^2=m_p^2+3\dt_5v^2\; .
\end{equation}
It will be the aim of the next section to discuss these quantities in
the context of a linear sigma model. In addition, we can use the
relations (\ref{FlavoredMesonMasses}) and
(\ref{CorrectionsToFlavoredMesonMasses}) to obtain information about
$v$, $w$ and therefore about the differences of decay constants
(\ref{DifferencesOfDecayConstants}). Without isospin violating effects
we have at this stage four unknown observables
$(M_\eta,M_{\etap},\th_p,f_{K^\pm}-f_\pi)$ for which we want
to find relations.

Finally we note that the number of parameters can be reduced by
absorbing $\dt_5$ and $\dt_7$ into the definitions of $m_p^2$ and
$m_m^2$, respectively. This will be done in section
\ref{ParametersOfTheLinearSigmaModel} in the framework of the linear
$\si$--model. For $w=0$ also $\dt_6$ can be absorbed into $\gm_3$. The
four observables depend on the ``couplings'' $m_p^2+3\dt_5 v^2$,
$\gm_3+\dt_6 v$, $\dt_8$ and $\dt_9+\delta_10/4$.

\sect{Linear meson model}
\label{LinearSigmaModel}

In this section we will compute the couplings of sections
\ref{ScalarMesonMassesToLinearOrder},
\ref{PseudoScalarMesonMassesToQuadraticOrder} in the context of the
linear meson model (often also called linear sigma model). The fields
$m$, $p$, $h$ and $s$ are all contained in a complex $3\times3$ matrix
$\Phi$ which transforms as $(\ol{\bf 3},\bf 3)$ with respect to the
chiral flavor group $\cst$
\begin{equation}
  \label{IOU1}
  \Phi\ra\Uc_R\Phi\Uc_L^\dagger\; ;\;\;\;
  \Uc_R\in SU_R(3)\; ,\;\;\;
  \Uc_L\in SU_L(3)
\end{equation}
and carries nonvanishing axial charge.  Including up to two
derivatives the effective action of the linear $\si$--model can be
written as a sum of a potential and a kinetic term plus a source term
\begin{equation}
 \Gm[\Phi]=\int d^4x\left(U+\Lc_{\rm kin}+
 \Lc_{j}\right)\; .
\end{equation}
As a consequence of the invariance under $\cst$ symmetry and the
discrete transformations\footnote{More precisely, the transformation
  $\Phi\ra\Phi^\dagger$ corresponds to left--right symmetry which is
  closely related to the parity reflection $\Pc$.} $\Pc$
($\Phi\ra\Phi^\dagger$) and $\Cc$ ($\Phi\ra\Phi^T$) the potential is a
function of the four independent invariants \cite{JW95-1}
\begin{equation}\begin{array}{rcl}
 \ds{\rho} &=& \ds{\Tr\Phid\Phi}\nnn
 \ds{\tau_2} &=& \ds{
 \frac{3}{2}\Tr\left(
 \Phi^\dagger\Phi-\frac{1}{3}\rho\right)^2=
 \frac{3}{2}\Tr\left(\Phid\Phi\right)^2-
 \hal\rho^2}\nnn
 \ds{\tau_3} &=& \ds{
 \Tr\left(
 \Phi^\dagger\Phi-\frac{1}{3}\rho\right)^3=
 \Tr\left(\Phid\Phi\right)^3-
 \frac{2}{3}\tau_2\rho-\frac{1}{9}\rho^3}\nnn
 \ds{\xi} &=& \ds{\det\Phi+\det\Phid} \; .
 \label{Invariants}
\end{array}\end{equation}
With respect to the vector--like $SU_V(3)$ symmetry we
may decompose
\begin{equation}
 \Phi=\olsi_0+\frac{1}{\sqrt{2}}\left(
 i\Phi_p+\frac{i}{\sqrt{3}}\chi_p+\Phi_s+
 \frac{1}{\sqrt{3}}\chi_s\right)
 \label{Phi}
\end{equation}
with traceless hermitian $3\times3$ matrices $\Phi_p$, $\Phi_s$
and real singlets ($\olsi_0$ is a real positive constant)
\begin{equation}\begin{array}{rcl}
 \ds{\chi_s} &=& \ds{\frac{1}{\sqrt{6}}\left[
 \Tr\left(\Phi+\Phid\right)-6\olsi_0\right]}\nnn
 \ds{\chi_p} &=& \ds{-\frac{i}{\sqrt{6}}
 \Tr\left(\Phi-\Phid\right)}\; .
\end{array}\end{equation}

The kinetic term involving two derivatives consistent
with $\cst$ symmetry, $\Cc$ and $\Pc$ reads\footnote{By
  partial integration we bring 
  all contributions to the kinetic term into the form where the two
  derivatives act on different fields. The invariance of the last term
  follows from $V_{aa^\prime}V_{bb^\prime}V_{cc^\prime}\eps_{a^\prime
    b^\prime c^\prime}=\eps_{abc}$ for arbitrary $V\in SU(3)$.}
\begin{equation}\begin{array}{rcl}
 \ds{\Lc_{\rm kin}} &=& \ds{
 Z_\vp\Tr\prl^\mu\Phid\prl_\mu\Phi+
 \frac{1}{4}Y_\vp\prl^\mu\rho\prl_\mu\rho+
 \hal V_\vp\prl^\mu\xi\prl_\mu\xi+
 \hal\tilde{V}_\vp\prl^\mu\om\prl_\mu\om}\nnn
 &-& \ds{\frac{1}{8}X_\vp^-\Big\{\Tr\left(
 \Phid\prl_\mu\Phi-\prl_\mu\Phid\Phi\right)
 \left(\Phid\prl^\mu\Phi-\prl^\mu\Phid\Phi\right)}\nnn
 && \ds{\hspace{8.5mm}+
 \Tr\left(\Phi\prl_\mu\Phid-\prl_\mu\Phi\Phid\right)
 \left(\Phi\prl^\mu\Phid-\prl^\mu\Phi\Phid\right)
 \Big\} }\nnn
 &-& \ds{\frac{1}{8}X_\vp^+\Big\{\Tr\left(
 \Phid\prl_\mu\Phi+\prl_\mu\Phid\Phi\right)
 \left(\Phid\prl^\mu\Phi+\prl^\mu\Phid\Phi\right)}\nnn
 && \ds{\hspace{8.5mm}+
 \Tr\left(\Phi\prl_\mu\Phid+\prl_\mu\Phi\Phid\right)
 \left(\Phi\prl^\mu\Phid+\prl^\mu\Phi\Phid\right)
 \Big\} }\nnn
 &-& \ds{
 \frac{1}{4}W_\vph\Tr\left\{\left(
 \prl_\mu\Phi^\dagger\Phi\prl^\mu\Phi^\dagger\Phi+
 \Phi^\dagger\prl_\mu\Phi\Phi^\dagger\prl^\mu\Phi\right)
 \left(\Phi^\dagger\Phi-\frac{1}{3}\Tr\left(
 \Phi^\dagger\Phi\right)\right)\right\} }\nnn
 &+& \ds{\frac{1}{2}U_\vp
 \eps^{a_1a_2a_3}\eps^{b_1b_2b_3}
 \left(\Phi_{a_1b_1}\prl^\mu\Phi_{a_2b_2}\prl_\mu\Phi_{a_3b_3}+
 \Phid_{a_1b_1}\prl^\mu\Phid_{a_2b_2}\prl_\mu\Phid_{a_3b_3}
 \right) +\ldots} \; .
 \label{Lkin}
\end{array}\end{equation}
Here $Z_\vp$, $V_\vp$, etc. are functions of the four independent
scalar 
$\cst$ invariants (\ref{Invariants})
and the dots stand for other independent terms which are not relevant
for our purposes, as for example $\prl^\mu\tau_2\prl_\mu\tau_2$. (See
sect.~\ref{KineticTermsInTheLinearSigmaModel} for the precise meaning
of this statement.) We
note that the additional pseudoscalar invariant
\begin{equation}
 \om=i\left(\det\Phi-\det\Phid\right)
\end{equation}
will always appear with even powers, since there is no other
parity--odd invariant. Since its square
can be expressed in terms of (\ref{Invariants})
\begin{equation}
 \om^2+\xi^2=4\det(\Phid\Phi)=
 4\exp\left\{\Tr\ln(\Phid\Phi)\right\}\; .
 \label{OmegaSquared}
\end{equation}
it can only appear as an independent quantity in combination with
derivatives as in (\ref{Lkin}).
Evaluating (\ref{Lkin}) for a configuration
$\Phi_0={\rm diag}(\olsi_0)$ the structure of the kinetic term for
fluctuations of $\Phi$ around $\Phi_0$ simplifies to the form
\begin{equation}
 \Lc_{\rm kin}=\hal Z_m\Tr\prl^\mu\Phi_p\prl_\mu\Phi_p+
 \hal Z_h\Tr\prl^\mu\Phi_s\prl_\mu\Phi_s+
 \hal Z_p\prl^\mu\chi_p\prl_\mu\chi_p+
 \hal Z_s\prl^\mu\chi_s\prl_\mu\chi_s
 \label{Lkin1}
\end{equation}
where the normalization is adapted such that to lowest order in
$\olsi_0$ one has $Z_m=Z_h=Z_s=Z_p=Z_\vp$. The fields $\Phi_p$,
$\chi_p$, $\Phi_s$, $\chi_s$ have the same transformation properties
with respect to $SU(3)$ and parity as $m$, $p$, $h$, $s$,
respectively. 

So far $\olsi_0$ in (\ref{Phi}) has not been specified.
If we want to use an expansion around $\VEV{s}=0$ we should identify
it with the value of $\frac{1}{6}\left(\Tr\Phi+\Tr\Phid\right)$ at the
potential minimum in presence of the quark masses, i.e.
\begin{equation}
 \olsi_0=\frac{1}{3}\left(\si_u+\si_d+\si_s\right) Z_m^{-\hal}=
 \frac{1}{6}\left(\ol{f}_\pi+\ol{f}_{K^\pm}+\ol{f}_{K^0}\right) Z_m^{-\hal}\; .
 \label{CoiceForSigam0}
\end{equation}
This corresponds to our choice $u=0$ in section
\ref{ScalarMesonMassesToLinearOrder}
and leads to the
identification (cf. (\ref{LkinForMultiplets}))
\begin{equation}\begin{array}{rcl}
 \ds{\Phi_p=\left(2Z_m\right)^{-\hal}m} &,&
 \ds{\Phi_s=\left(2Z_h\right)^{-\hal}h}\nnn
 \ds{\chi_p=Z_p^{-\hal}p} &,&
 \ds{\chi_s=Z_s^{-\hal}s}\; .
 \label{Identification}
\end{array}\end{equation}
Corrections to the kinetic terms which are linear in the quark masses
involve cubic terms like $\Tr\Phi_s\prl^\mu\Phi_p\prl_\mu\Phi_p$ etc.
and will be discussed in section
\ref{KineticTermsInTheLinearSigmaModel}.

For the configuration $\Phi_0={\rm diag}(\olsi_0)$ the invariants
(\ref{Invariants}) take on the values
\begin{equation}\begin{array}{rcl}
 \ds{\rho_0} &=& \ds{3\olsi_0^2}\nnn
 \ds{\xi_0} &=& \ds{2\olsi_0^3}\nnn
 \ds{\tau_2} &=& \ds{\tau_3=0}\; .
\end{array}\end{equation}
We next decompose the invariants
$\rho-\rho_0$, $\tau_2$, $\tau_3$ and $\xi-\xi_0$ into the irreducible
representations $\Phi_p$, $\Phi_s$, $\chi_p$ and $\chi_s$.
Using occasionally the shorthand
notation
\begin{equation}
 \olchi_s=\sqrt{6}\olsi_0+\chi_s
\end{equation}
one finds
\begin{eqnarray}
 \ds{\rho-\rho_0} &=& \ds{
 \hal\Tr\Phi_p^2+\hal\Tr\Phi_s^2+
 \hal\chi_p^2+\hal\chi_s^2+\sqrt{6}\olsi_0\chi_s}\\[2mm]
 \ds{\tau_2} &=& \ds{
 \hal\olchi_s^2\Tr\Phi_s^2+\hal\chi_p^2\Tr\Phi_p^2+
 \frac{3}{8}\Tr\Phi_s^4-\frac{1}{8}\left(\Tr\Phi_s^2\right)^2+
 \frac{3}{8}\Tr\Phi_p^4-\frac{1}{8}\left(\Tr\Phi_p^2\right)^2}\nnn
 &+& \ds{
 \frac{3}{2}\Tr\Phi_s^2\Phi_p^2-
 \frac{3}{4}\Tr\left(\Phi_s\Phi_p\right)^2-
 \frac{1}{4}\Tr\Phi_s^2\Tr\Phi_p^2+
 \olchi_s\chi_p\Tr\Phi_s\Phi_p}\\[2mm]
 &+& \ds{
 \frac{\sqrt{3}}{2}\olchi_s\Tr\Phi_s^3+
 \frac{\sqrt{3}}{2}\olchi_s\Tr\Phi_s\Phi_p^2+
 \frac{\sqrt{3}}{2}\chi_p\Tr\Phi_p^3+
 \frac{\sqrt{3}}{2}\chi_p\Tr\Phi_p\Phi_s^2}\nnn
 \ds{\tau_3} &=& \ds{
 -\frac{1}{3}\left(\Tr\Phi_s^2+\Tr\Phi_p^2\right)\tau_2-
 \frac{1}{72}\left(\Tr\Phi_s^2+\Tr\Phi_p^2\right)^3}\nnn
 &+& \ds{
 \frac{1}{8}\Tr\left[\Phi_s^2+\Phi_p^2+
 \frac{2}{\sqrt{3}}\left(\olchi_s\Phi_s+\chi_p\Phi_p\right)\right]^3}\nnn
 &+& \ds{
 \frac{1}{4\sqrt{3}}\left[\olchi_s\Tr\left(
 \Phi_s^2\Phi_p\Phi_s\Phi_p-\Phi_s^3\Phi_p^2\right)+
 \chi_p\Tr\left(\Phi_p^2\Phi_s\Phi_p\Phi_s-
 \Phi_p^3\Phi_s^2\right)\right]}\\[2mm]
 &+& \ds{
 \frac{3}{8}\Tr\left[
 \Phi_s^4\Phi_p^2+\Phi_s^2\Phi_p\Phi_s^2\Phi_p-
 2\Phi_s^3\Phi_p\Phi_s\Phi_p\right.}\nnn
 && \ds{\left. \hspace{7mm}+
 \Phi_p^4\Phi_s^2+\Phi_p^2\Phi_s\Phi_p^2\Phi_s-
 2\Phi_p^3\Phi_s\Phi_p\Phi_s\right]}\; .
\nonumber\end{eqnarray}
Including terms quartic in $m$, $h$, $s$, $p$ the
invariant $\tau_3$ only contributes
\begin{equation}\begin{array}{rcl}
 \ds{\tau_3^{(4)}} &=& \ds{
 2\sqrt{2}\olsi_0^3\Tr\Phi_s^3+
 \olsi_0^2\left[
 3\Tr\left(\Phi_s^4+\Phi_s^2\Phi_p^2\right)-
 \Tr\Phi_s^2\left(\Tr\Phi_s^2+\Tr\Phi_p^2\right)
 \right. }\nnn
 && \ds{\left.\hspace{34mm}+
 2\sqrt{3}\left(\chi_s\Tr\Phi_s^3+\chi_p\Tr\Phi_s^2\Phi_p\right)
 \right]}\; .
\end{array}\end{equation}
Using the fact that for an arbitrary traceless $3\times3$ matrix T
\begin{equation}
 \det\left(\al+T\right)=
 \frac{1}{3}\Tr T^3-\hal\al\Tr T^2+\al^3
\end{equation}
one also obtains
\begin{equation}\begin{array}{rcl}
 \ds{\xi-\xi_0} &=& \ds{
 \frac{1}{3\sqrt{2}}\Tr\Phi_s^3-
 \frac{1}{\sqrt{2}}\Tr\Phi_s\Phi_p^2-
 \hal\olsi_0\left(\Tr\Phi_s^2-\Tr\Phi_p^2\right)}\nnn
 &-& \ds{
 \frac{1}{2\sqrt{6}}\chi_s\left(\Tr\Phi_s^2-\Tr\Phi_p^2\right)+
 \frac{1}{\sqrt{6}}\chi_p\Tr\Phi_s\Phi_p-\olsi_0\chi_p^2-
 \frac{1}{\sqrt{6}}\chi_s\chi_p^2}\nnn
 &+& \ds{
 \sqrt{6}\olsi_0^2\chi_s+\olsi_0\chi_s^2+
 \frac{1}{3\sqrt{6}}\chi_s^3}\; .
\end{array}\end{equation}
For the choice (\ref{CoiceForSigam0}) one has $u=0$ (c.f. section
\ref{ScalarMesonMassesToLinearOrder}) and we denote $v$ and $w$
collectively by $\Dt$. One observes that the expectation values of the
invariants in the presence of quark masses obey
\begin{equation}\begin{array}{rcl}
 \ds{\VEV{\rho-\rho_0}\sim\Oc(\Dt^2)} &,&
 \ds{\VEV{\xi-\xi_0}\sim\Oc(\Dt^2)}\nnn
 \ds{\VEV{\tau_2}\sim\Oc(\Dt^2)} &,&
 \ds{\VEV{\tau_3}\sim\Oc(\Dt^3)}
 \; .
 \label{VevsOfInvariants}
\end{array}\end{equation}
The mass squared matrix for the pseudoscalar mesons is obtained from
the potential of the linear $\si$--model by taking the second
derivatives with respect to the parity--odd representations $\Phi_p$
and $\chi_p$, which we will collectively denote by $\vp^-$. Since the
potential is parity--even, all VEVs of single derivatives of the
invariants (\ref{Invariants}) vanish. We note that
\begin{equation}\begin{array}{rcl}
 \ds{\VEV{\frac{\prl^2\rho}{\prl\vp^-\prl\vp^-}}
 \sim\Oc(1)} &,&
 \ds{\VEV{\frac{\prl^2\xi}{\prl\vp^-\prl\vp^-}}
 \sim\Oc(1)}\nnn
 \ds{\VEV{\frac{\prl^2\tau_2}{\prl\vp^-\prl\vp^-}}
 \sim\Oc(\Dt)} &,&
 \ds{\VEV{\frac{\prl^2\tau_3}{\prl\vp^-\prl\vp^-}}
 \sim\Oc(\Dt^2)}
 \; .
\end{array}\end{equation}

The most general $\cst$ symmetric potential can be expanded in a
Taylor series around $\rho=\rho_0$, $\xi=\xi_0$, $\tau_2=0$ and
$\tau_3=0$. We now see that a determination of the pseudoscalar
masses up to quadratic order in the quark masses requires
\begin{eqnarray}
 \ds{U} &=&
 \ds{\olm_g^2\left(\rho-\rho_0\right)
 -\hal\olnu\left[\xi-\xi_0-\olsi_0(\rho-\rho_0)\right]}\nnn
 &+& \ds{
 \hal\olla_1\left(\rho-\rho_0\right)^2+\hal\olla_2\tau_2+
 \hal\olla_3\tau_3
 \label{ExpansionPotential}}\\[2mm]
 &+& \ds{
 \hal\olbt_1\left(\rho-\rho_0\right)\left(\xi-\xi_0\right)+
 \hal\olbt_2\left(\rho-\rho_0\right)\tau_2+
 \hal\olbt_3\left(\xi-\xi_0\right)\tau_2+
 \hal\olbt_4\left(\xi-\xi_0\right)^2+\ldots
 \nonumber}\; .
\end{eqnarray}
Here the potential has been normalized such that it vanishes for the
configuration $\Phi={\rm diag}(\olsi_0)$. We see that to zeroth order
in the quark masses the only contributions to the pseudoscalar mass
matrix arise from $\olnu$ and $\olm_g^2$. To linear order we obtain
corrections from $\olla_2$ and $\olnu$ whereas to quadratic order the
other $\olla_i$ and the $\olbt_i$ enter. We add to the effective
action a source term
\begin{equation}
  \label{SourceTerm}
  \Lc_j=-\hal\Tr
  \left(\Phi^\dagger j+j^\dagger\Phi\right)
\end{equation}
which is linear in the real quark mass matrix $M_q={\rm
  diag}(m_u,m_d,m_s)$
\begin{equation}
  \label{Current}
  j=j^\dagger=a_q M_q\; .
\end{equation}
We denote the singlet part of the source by
\begin{equation}
  \label{SingletSource}
  j_s=\frac{1}{\sqrt{6}}Z_s^{-\hal}
 \Tr j
\end{equation}
and require
\begin{equation}
  \label{L01}
  \frac{\prl}{\prl\chi_s}\left.
  \left( U+\Lc_j\right)
  \right|_{\Phi=\olsi_0}=0\; .
\end{equation}
Our choice
(\ref{CoiceForSigam0}) for $\olsi_0$ therefore implies
\begin{equation}
 \sqrt{6}\olsi_0\olm_g^2=j_s Z_s^{\hal}
\end{equation}
and the mass term $\olm_g^2$ is linear in the quark masses.

Comparing (\ref{ExpansionPotential}) with
(\ref{GerneralMassLagrangian}), (\ref{Lthree}) and (\ref{Lfour}) we
can now determine the various couplings of the last two sections in
terms of those of the linear sigma model.  For the pseudoscalar mass
terms of (\ref{GerneralMassLagrangian}) we find
\begin{equation}\begin{array}{rcl}
 \label{L02}
 \ds{m_m^2} &=& \ds{\olm_g^2Z_m^{-1}}\nnn
 \ds{m_p^2} &=& \ds{
 \left(\olm_g^2+\frac{3}{2}\olnu\,\olsi_0\right)Z_p^{-1}}\nnn
\end{array}\end{equation}
The cubic couplings of (\ref{Lthree}) contributing to $\Oc(\Dt)$ read
\begin{equation}\begin{array}{rcl}
 \label{BBB3}
 \ds{\gm_1} &=& \ds{\sqrt{6}\left(
 \olsi_0\olla_1-\frac{1}{12}\olnu+
 \olsi_0^2\olbt_1+\olsi_0^3\olbt_4\right)
 Z_s^{-\hal}Z_m^{-1} }\nnn
 \ds{\gm_2} &=& \ds{\hal\left(3\olsi_0\olla_2+\olnu\right)
 Z_h^{-\hal}Z_m^{-1}}\nnn
 \ds{\gm_3} &=& \ds{\frac{\sqrt{6}}{2}\left(
 \olsi_0\olla_2-\frac{1}{6}\olnu\right)
 Z_p^{-\hal}Z_h^{-\hal}Z_m^{-\hal}}\nnn
 \ds{\gm_4} &=& \ds{\sqrt{6}\left(
 \olsi_0\olla_1+\frac{1}{6}\olnu-
 \hal\olsi_0^2\olbt_1-2\olsi_0^3\olbt_4\right)
 Z_s^{-\hal}Z_p^{-1}}
\end{array}\end{equation}
whereas for the quartic couplings of (\ref{Lfour}) we obtain
\begin{equation}\begin{array}{rcl}
 \label{BBB4}
 \ds{\dt_1} &=& \ds{\left(
 \hal\olla_1+\frac{5}{4}\olsi_0\olbt_1+
 2\olsi_0^2\olbt_4\right)
 Z_s^{-1}Z_m^{-1}}\nnn
 \ds{\dt_2} &=& \ds{\sqrt{\frac{3}{2}}\left(
 \hal\olla_2-\olsi_0\olbt_1+
 3\olsi_0^2\olbt_2+3\olsi_0^3\olbt_3-
 2\olsi_0^2\olbt_4\right)
 Z_s^{-\hal}Z_h^{-\hal}Z_m^{-1}}\nnn
 \ds{\dt_3} &=& \ds{\left(
 \hal\olla_2+\hal\olsi_0\olbt_1+
 3\olsi_0^2\olbt_2+3\olsi_0^3\olbt_3+
 \olsi_0^2\olbt_4\right)
 Z_s^{-\hal}Z_p^{-\hal}Z_h^{-\hal}Z_m^{-\hal}}\nnn
 \ds{\dt_4} &=& \ds{\left(
 \hal\olla_1-\olsi_0\olbt_1-4\olsi_0^2\olbt_4\right)
 Z_s^{-1}Z_p^{-1}}\nnn
 \ds{\dt_5} &=& \ds{\left(
 \hal\olla_1-\frac{3}{4}\olsi_0\olbt_1+\olsi_0^2\olbt_4+
 \frac{3}{2}\olsi_0^2\olbt_2-3\olsi_0^3\olbt_3
 \right) Z_p^{-1}Z_h^{-1}}\nnn
 \ds{\dt_6} &=& \ds{\frac{\sqrt{6}}{8}\left(
 \olla_2+4\olsi_0^2\olla_3
 \right) Z_p^{-\hal}Z_h^{-1}Z_m^{-\hal}}\nnn
 \ds{\dt_7} &=& \ds{\left(
 \hal\olla_1-\frac{1}{4}\olla_2-\hal\olsi_0^2\olbt_4-
 \olsi_0^2\olla_3+\frac{3}{2}\olsi_0^2\olbt_2+
 \frac{3}{2}\olsi_0^3\olbt_3
 \right) Z_h^{-1}Z_m^{-1}}\nnn
 \ds{\dt_8} &=& \ds{\frac{3}{8}\left(
 \olla_2+2\olsi_0^2\olla_3
 \right) Z_h^{-1}Z_m^{-1}}\nnn
 \ds{\dt_9} &=& \ds{-\frac{3}{16}\olla_2
 Z_h^{-1}Z_m^{-1}}\nnn
 \ds{\dt_{10}} &=& \ds{0}\; .
\end{array}\end{equation}

We next turn to the scalars for which we wish to relate the couplings
$m_s^2$, $m_h^2$, $\gm_5$, $\gm_6$ to those of the linear sigma model.
For this purpose we note that, contrary to the case of the
pseudoscalar mesons, the parity--even scalar meson fields,
collectively denoted by $\vp^+$, may also appear to odd powers in the
invariants $\rho$, $\xi$, $\tau_2$, $\tau_3$. We therefore need in
addition to (\ref{VevsOfInvariants})
\begin{equation}\begin{array}{rcl}
 \ds{\VEV{\frac{\prl\rho}{\prl\vp^+}}\sim\Oc(1)} &,&
 \ds{\VEV{\frac{\prl\tau_2}{\prl\vp^+}}\sim\Oc(\Dt)}\nnn
 \ds{\VEV{\frac{\prl\xi}{\prl\vp^+}}\sim\Oc(1)} &,&
 \ds{\VEV{\frac{\prl\tau_3}{\prl\vp^+}}\sim\Oc(\Dt^2)}
\end{array}\end{equation}
and
\begin{equation}\begin{array}{rcl}
 \ds{\VEV{\frac{\prl^2\rho}{\prl\vp^+\prl\vp^+}}
 \sim\Oc(1)} &,&
 \ds{\VEV{\frac{\prl^2\tau_2}{\prl\vp^+\prl\vp^+}}
 \sim\Oc(1)}\nnn
 \ds{\VEV{\frac{\prl^2\xi}{\prl\vp^+\prl\vp^+}}
 \sim\Oc(1)} &,&
 \ds{\VEV{\frac{\prl^2\tau_3}{\prl\vp^+\prl\vp^+}}
 \sim\Oc(\Dt)}
 \; .
\end{array}\end{equation}
Hence, the expansion (\ref{ExpansionPotential}) of the potential
contains exactly those terms required to obtain the scalar masses to
linear order in $\Dt$. Furthermore, we see that to zeroth order in
$\Dt$ only $\olm_g^2$, $\olnu,\olla_1,\ol{\lambda}_2,\ol{\beta}_1$ and
$\olbt_4$ contribute. We find in particular
\begin{equation}\begin{array}{rcl}
 \label{Gamma5}
 \ds{m_s^2} &=& \ds{\olm_g^2Z_s^{-1}+6\olsi_0\left(
 \olsi_0\olla_1+\olsi_0^2\olbt_1+\olsi_0^3\olbt_4-\frac{1}{12}\olnu
 \right) Z_s^{-1}}\nnn
 \ds{m_h^2} &=& \ds{\olm_g^2Z_h^{-1}+\olsi_0\left(
 3\olsi_0\olla_2+\olnu
 \right) Z_h^{-1}}
\end{array}\end{equation}
and
\begin{equation}\begin{array}{rcl}
 \label{Gamma6}
 \ds{\gm_5} &=& \ds{\sqrt{6}\left(
 \olsi_0\olla_1+\olsi_0\olla_2-\olsi_0^3\olbt_4+\frac{1}{12}\olnu+
 3\olsi_0^3\olbt_2+3\olsi_0^4\olbt_3
 \right) Z_s^{-\hal}Z_h^{-1}}\nnn
 \ds{\gm_6} &=& \ds{\frac{1}{4}\left(
 9\olsi_0\olla_2-\olnu+12\olsi_0^3\olla_3
 \right) Z_h^{-\frac{3}{2}} }\nnn
 \ds{\gm_7} &=& \ds{\sqrt{6}\left(
 \frac{3}{2}\olsi_0\olla_1+\frac{9}{4}\olsi_0^2\olbt_1+
 3\olsi_0^3\olbt_4-\frac{1}{12}\olnu\right) Z_s^{-\frac{3}{2}} }\; .
\end{array}\end{equation}

\sect{Parameters of the linear meson model}
\label{ParametersOfTheLinearSigmaModel}

In this section we will give a first estimate of the values of the
parameters of the linear $\si$--model. It is based on an expansion in
powers of $\Delta$ to lowest order in the derivative
expansion\footnote{The systematic ordering of the derivative expansion
  is ambiguous to lowest order since a minimal kinetic term must
  always be included. For our purpose we consider to lowest order the
  kinetic term (\ref{Lkin1}). The first order comprises the most
  general terms with up to two derivatives, the second order includes
  four derivatives and so on.}.  Comparison of the estimates of this
section with those of the following ones will allow us to evaluate the
quantitative influence of quark mass corrections to the kinetic terms
(which are neglected here). The results of this section can also be
compared with earlier work \cite{PW84-1}--\cite{Pis95-1} by setting
$Z_h=Z_p=Z_m$. Later we will see, however, that $Z_h/Z_m$ deviates
substantially from one.  We observe that the couplings $\olla_1$,
$\olbt_1$, $\olbt_2$, $\olbt_3$ and $\olbt_4$ influence the meson
masses only through $m_s^2$, $\dt_5$ and $\dt_7$ whereas $\olla_3$
appears in addition in $\delta_6$ and $\dt_8$. The couplings $\dt_5$
and $\dt_7$ modify the relation between the neutral and flavored
pseudoscalar masses, (\ref{ZZZ5}) and (\ref{YYY1}), through the term
$3\dt_5 v^2$ in $M_p^2$ and the term $3\dt_7 v^2$ in $\ol{M}_3^2$,
$\ol{M}_8^2$ and the flavored pseudoscalar masses. (We neglect here
corrections $\sim w^2$.) A redefinition of couplings
\begin{equation}\begin{array}{rcl}
 \ds{\olm_g^{\prime2}} &=& \ds{\olm_g^2+
 \left(3\dt_7+4\dt_8-2\dt_9\right) v^2 Z_m}\nnn
 \ds{\olnu^\prime} &=& \ds{\olnu+
 \left[2\dt_5 Z_p-\left(2\dt_7+\frac{8}{3}\dt_8-\frac{4}{3}\dt_9
 \right) Z_m\right]\frac{v^2}{\olsi_0}}\nnn
 \ds{\olla_2^\prime} &=& \ds{\olla_2-\frac{1}{3}
 \left[2\dt_5 Z_p+\left(\dt_7+\frac{4}{3}\dt_8-\frac{2}{3}\dt_9
 \right) Z_m\right]\frac{v^2}{\olsi_0^2}}
 \label{ZZZ6}
\end{array}\end{equation}
absorbs this correction in the lowest order masses $m_m^2$, $m_p^2$,
$m_h^2$, whereas the corresponding shifts in the $\gm$'s only
contribute to cubic order in $\Dt$. In terms of these couplings one
has
\begin{equation}\begin{array}{rcl}
 \label{BBB2}
 \ds{\ol{M}_{\pi^\pm}^2} &=& \ds{\olm_g^{\prime2} Z_m^{-1}-
 \frac{1}{6}\left(3\olla_2^\prime\olsi_0+\olnu^\prime\right)
 Z_m^{-\frac{3}{2}}\left( \ol{f}_{K^\pm}+\ol{f}_{K^0}-2\ol{f}_\pi\right)}\nnn
 &-& \ds{
 \frac{1}{6}\left(\olla_2^\prime+\olla_3\olsi_0^2\right)
 Z_m^{-2}\left( \ol{f}_{K^\pm}+\ol{f}_{K^0}-2\ol{f}_\pi\right)^2 }\nnn
 \ds{\ol{M}_{K^\pm}^2} &=& \ds{\olm_g^{\prime2} Z_m^{-1}+
 \frac{1}{12}\left(3\olla_2^\prime\olsi_0+\olnu^\prime\right)
 Z_m^{-\frac{3}{2}}\left(4\ol{f}_{K^\pm}-2\ol{f}_{K^0}-2\ol{f}_\pi\right)}\nnn
 &+& \ds{\frac{1}{12}\left(\olla_2^\prime+\olla_3\olsi_0^2\right)
 Z_m^{-2}\left( \ol{f}_{K^\pm}+\ol{f}_{K^0}-2\ol{f}_\pi\right)
 \left(7\ol{f}_{K^0}-5\ol{f}_{K^\pm}-2\ol{f}_\pi\right)}\nnn
 \ds{\ol{M}_{K^0}^2} &=& \ds{\olm_g^{\prime2} Z_m^{-1}+
 \frac{1}{12}\left(3\olla_2^\prime\olsi_0+\olnu^\prime\right)
 Z_m^{-\frac{3}{2}}\left(4\ol{f}_{K^0}-2\ol{f}_{K^\pm}-2\ol{f}_\pi\right)}\nnn
 &+& \ds{\frac{1}{12}\left(\olla_2^\prime+\olla_3\olsi_0^2\right)
 Z_m^{-2}\left( \ol{f}_{K^\pm}+\ol{f}_{K^0}-2\ol{f}_\pi\right)
 \left(7\ol{f}_{K^\pm}-5\ol{f}_{K^0}-2\ol{f}_\pi\right)}
\end{array}\end{equation}
or
\begin{equation}
 \label{NNN13}
 \olm_g^{\prime2}=\frac{1}{3}Z_m\left(
 \ol{M}_{K^\pm}^2+\ol{M}_{K^0}^2+\ol{M}_{\pi^\pm}^2\right)
\end{equation}
and
\begin{eqnarray}
 \ds{\ol{M}_8^2} &=& \ds{
 \frac{1}{3}\left(2\ol{M}_{K^\pm}^2+
 2\ol{M}_{K^0}^2-\ol{M}_{\pi^\pm}^2\right)-
 \frac{1}{4}\ol{\la}_2^\prime Z_m^{-2}\left(
 \ol{f}_{K^\pm}+\ol{f}_{K^0}-2\ol{f}_\pi\right)^2}\nnn
 \ds{M_p^2} &=& \ds{\left(\olm_g^{\prime2}+
 \frac{3}{2}\olnu^\prime\olsi_0\right) Z_p^{-1}}\nnn
 \ds{M_{8p}^2} &=& \ds{-\sqrt{2}\left(\frac{Z_m}{Z_p}\right)^\hal
 \Bigg[\frac{1}{3}\left(\ol{M}_{K^\pm}^2+
 \ol{M}_{K^0}^2-2\ol{M}_{\pi^\pm}^2\right)
 \label{NNN1}}\\[2mm]
 &-& \ds{
 \frac{1}{4}\olnu^\prime Z_m^{-\frac{3}{2}}
 \left( \ol{f}_{K^\pm}+\ol{f}_{K^0}-2\ol{f}_\pi\right)-
 \frac{1}{8}\olla_2^\prime
 Z_m^{-2}\left( \ol{f}_{K^\pm}+
 \ol{f}_{K^0}-2\ol{f}_\pi\right)^2 \Bigg] 
 \nonumber}\; .
\end{eqnarray}
We will use (\ref{BBB2}), (\ref{NNN13}) to determine
$\olm_g^{\prime2}$, $\olla_2^\prime$ and $\olnu^\prime$ for given
decay constants, $M_{\pi^\pm}$, $M_{K^\pm}$ and $M_{K^0}$ (once the
wave function renormalizations are known). The parameters
$\olm_g^{\prime2}$ and $\olnu^\prime$ are independent of
$\ol{\lambda}_3$ whereas the dependence of $\ol{\lambda}_2^\prime$ on
$\ol{\lambda}_3$ is linear in $\Delta$. Hence, the unflavored
pseudoscalar mass matrix given by (\ref{NNN1}) depends only to cubic
or higher order on $\ol{\lambda}_3$ and we will therefore neglect
$\ol{\lambda}_3$ in the pseudoscalar sector altogether.

In the following we will make a first attempt to estimate the
parameters $\ol{m}_g^{\prime 2}$, $\ol{\nu}^\prime$ and
$\ol{\la}_2^\prime$. We present the values to different orders in the
quark masses in order to gain some intuition for the convergence of
the quark mass expansion for these parameters. We use in this section
the simplified kinetic term (\ref{Lkin1}). To zeroth order in the
quark masses the octet--singlet mixing vanishes and $\olm_g^2=0$. This
yields the zeroth order relations for the mass of the $\etap$ and the
scalar octet
\begin{equation}\begin{array}{rcl}
 \ds{M_\etap^2} &=& \ds{\frac{3}{2}\olnu^\prime
 \olsi_0 Z_p^{-1}=
 \frac{3}{2}\nu\si_0\frac{Z_m}{Z_p}}\nnn
 \ds{m_h^2} &=& \ds{3\olla_2^\prime
 \olsi_0^2Z_h^{-1}+
 \frac{2}{3}M_\etap^2\frac{Z_p}{Z_h}=
 3\la_2\si_0^2\frac{Z_m}{Z_h}+
 \frac{2}{3}M_\etap^2\frac{Z_p}{Z_h}}\; .
\end{array}\end{equation}
Here we have used for the second identities renormalized couplings
according to
\begin{equation}\begin{array}{rcl}
 \ds{\si_0} &=& \ds{Z_m^{\hal}\olsi_0 =
 \frac{1}{6}\left(\ol{f}_\pi+
 \ol{f}_{K^\pm}+\ol{f}_{K^0}\right)
 \simeq53.1\MeV}\nnn
 \ds{\nu} &=& \ds{Z_m^{-\frac{3}{2}}
 \olnu^\prime\; ,\;\;\;
 m_g^2=Z_m^{-1}\olm_g^{\prime 2}\; ,}\\[4mm]\nonumber
 \ds{\la_1} &=& \ds{Z_m^{-2}\olla_1\; ,\;\;\;
 \la_2=Z_m^{-2}\olla_2^\prime\; ,\;\;\;
 \la_3=Z_m^{-3}\olla_3}
 \label{RenormalizedCouplings}
\end{array}\end{equation}
etc. This yields the zeroth order estimate
\begin{equation}
 \ds{\nu^{(0)}\frac{Z_m}{Z_p}} \simeq \ds{11500\MeV}\; .
 \label{NNN9}
\end{equation}
A first possibility for an estimate of $\la_2$ uses the relation
between the masses of the scalar and the pseudoscalar octets. For
this purpose we need
\begin{equation}
 m_h^2=\frac{1}{3}\left(2\ol{M}_{K^{*}_o}^2+
 \ol{M}_{a_o}^2\right)\; ,
 \label{Mh2}
\end{equation}
where $M_{K^{*}_o}\simeq1430\MeV$. The $a_0$--mesons, however, are not
unambiguously identified. Usually it is associated with the well
established $a_0(980)$ resonance. On the other hand, the neighboring
$f_0(980)$ is often assumed to be an $I=0$, $K\ol{K}$ bound state or a
four quark state.  One may therefore as well identify the $a_0(980)$
with the corresponding $I=1$, $K\ol{K}$ ``molecules''. The only
remaining candidate for the $a_0$--mesons is then the possible
$a_0(1320)$ resonance \cite{PDG94-2}.  In this section we will use
both possibilities with the notation $M_{a^\pm_o}=1320(983)\MeV$.  We
obtain
\begin{equation}\begin{array}{rcl}
 \ds{m_h} &\simeq& \ds{1394(1298)\MeV}\nnn
 \ds{\la_2^{(a)}\frac{Z_m}{Z_h}} &\simeq&
 \ds{156.9(126.5)-72.0\left(\frac{Z_p}{Z_h}-1\right)}\; .
\end{array}\end{equation}
In addition to the uncertainty in the identification of the
$a_0$--meson the value of $\la_2$ depends rather sensitively on ratios
of wave function renormalization constants. For this reason we will
compute below the value of $\la_2$ from an expression involving only
properties of the pseudoscalar mesons.

To linear order in the quark masses the mass eigenvalues of the
pseudoscalars are not affected by quark mass corrections to the
kinetic term (see sect.~\ref{QuarkMassCorrectionsToKineticTerms}).
We can therefore identify $\ol{M}_i=M_i$ and obtain
\begin{eqnarray}
 \ds{m_m^2} &=& \ds{m_g^2=\frac{1}{3}\left(
 \ol{M}_{K^\pm}^2+\ol{M}_{K^0}^2+
 \ol{M}_{\pi^\pm}^2\right)\simeq(411.7\MeV)^2}\\[2mm]
 \ds{\gm_2 v} &=& \ds{\left(
 \frac{3}{2}\la_2\si_0+\hal\nu\right)
 \left(\frac{Z_m}{Z_h}\right)^\hal v=
 \frac{1}{3}\left(
 \ol{M}_{K^\pm}^2+\ol{M}_{K^0}^2-2\ol{M}_{\pi^\pm}^2\right)}\nnn
 &\simeq& \ds{(388.9\MeV)^2}\; .
 \label{YYY3}
\end{eqnarray}
We may fix the parameter $\frac{Z_m}{Z_p}\nu$ by the relation
\begin{equation}
 M_\etap^2=m_p^2=\frac{3}{2}\frac{Z_m}{Z_p}\nu\si_0+
 \frac{1}{3}\frac{Z_m}{Z_p}\left(
 \ol{M}_{K^\pm}^2+\ol{M}_{K^0}^2+\ol{M}_{\pi^\pm}^2\right)
 \label{YYY2}
\end{equation}
and find
\begin{equation}
 \nu^{(1)}\frac{Z_m}{Z_p}\simeq
 \left(9372-2124\left[\frac{Z_m}{Z_p}-1\right]\right)\MeV\; .
\end{equation}
The coupling $\la_2$ can now be infered from (\ref{YYY3})
\begin{equation}\begin{array}{rcl}
 \ds{\la_2^{(b)}} &=& \ds{\frac{2}{3}
 \frac{(\ol{M}_{K^\pm}^2+\ol{M}_{K^0}^2-2\ol{M}_{\pi^\pm}^2)}
 {\left(\ol{f}_{K^\pm}+\ol{f}_{K^0}-2\ol{f}_\pi\right)\si_0}-
 \frac{2}{9}\frac{Z_p}{Z_m}\frac{M_\etap^2}{\si_0^2}+
 \frac{2}{27}\frac{(\ol{M}_{K^\pm}^2+
 \ol{M}_{K^0}^2+\ol{M}_{\pi^\pm}^2)}{\si_0^2} }\nnn
 &=& \ds{
 77.93-72.03\left(\frac{Z_p}{Z_m}-1\right)}\; .
 \label{Lambda2Estimate}
\end{array}\end{equation}
We note that this determination of $\la_2$ uses the ratio of two
quantities which are linear in $\Dt$. The relation between $\ol{f}_i$
and the meson decay constants $f_i$ is strongly influenced by quark
mass corrections to the kinetic terms already to leading order in the
quark mass expansion. From there we expect sizeable corrections to
$\la_2$. In addition, an estimate of $\la_2$ requires information on
$Z_p/Z_m$ and involves differences in larger quantities. As a result,
$\lambda_2$ will be poorly determined even once the quark mass
corrections to the kinetic terms are included. We will find in section
\ref{Results} typical values $\lambda_2\simeq15-30$ even for $Z_p$
equal or somewhat smaller than one.

Concerning the scalar meson masses to linear order in $\Dt$ we observe
the relation
\begin{equation}
 m_h^2-\frac{Z_m}{Z_h}m_g^2=2\si_0
 \left(\frac{Z_m}{Z_h}\right)^{1/2}\gm_2
\end{equation}
which translates with (\ref{YYY3}) into
\begin{equation}
 \label{NNN11}
 \frac{Z_h}{Z_m}=
 \frac{2\si_0}{m_h^2}
 \frac{\left(\ol{M}_{K^\pm}^2+
 \ol{M}_{K^0}^2-
 2\ol{M}_{\pi^\pm}^2\right)}
 {\left(\ol{f}_{K^\pm}+\ol{f}_{K^0}-2\ol{f}_\pi\right)}+
 \frac{m_g^2}{m_h^2}\; .
\end{equation}
Inserting experimental values this leads to the ratio
\begin{equation}
 \frac{Z_h}{Z_m}\simeq0.69(0.79)\; .
 \label{Zmh}
\end{equation}
These values are changed to $Z_h/Z_m\simeq0.40-0.65$ once quark mass
corrections to the kinetic terms are included (see (\ref{NNN12})).
{}From the mass splitting within the scalar octet we infer
\begin{equation}
 \label{NNN14}
 \frac{\gm_6}{\gm_2}=\frac{3}{2}\frac{Z_m}{Z_h}
 \left(1+\frac{12\si_0^3\la_3-4\nu}{9\si_0\la_2+3\nu}\right)=
 \frac{\ol{M}_{K^{*}_o}^2-\ol{M}_{a_o}^2}
 {\ol{M}_{K^\pm}^2+\ol{M}_{K^0}^2-2\ol{M}_{\pi^\pm}^2}\simeq
 0.67(2.38)\; .
\end{equation}
This relation can be used to estimate the size of $\la_3$.

Having computed the parameters $\nu$, $m_g^2$, $\si_0$, $\la_2$,
$\la_3$, $Z_m/Z_h$, $v$ and $w$ from $M_{\eta^\prime}^2$,
$M_{K^\pm}^2$, $M_{K^0}^2$, $M_{\pi^\pm}^2$, $M_{K_o^*}^2$,
$M_{a_o^\pm}^2$, $f_{K^\pm}$ and $f_\pi$ we can now derive other meson
properties. Within the approximation $\ol{M}_i=M_i$, $\ol{f}_i=f_i$
used in this section we discuss here briefly the non--flavored
pseudoscalar mesons. For a determination of the pseudoscalar
mixing angle $\th_p$ to $\Oc(\Dt)$ and the $\eta$ and $\etap$ masses
to $\Oc(\Dt^2)$ we need the off--diagonal element in the mass matrix
for the neutral pseudoscalars to $\Oc(\Dt)$
\begin{equation}\begin{array}{rcl}
 \ds{M_{8p}^2} &=& \ds{-\frac{1}{2\sqrt{2}}\left(
 \ol{f}_{K^\pm}+\ol{f}_{K^0}-2\ol{f}_\pi\right)\left(
 2\la_2\si_0-\frac{1}{3}\nu\right)
 \left(\frac{Z_m}{Z_p}\right)^\hal}\nnn
 &=& \ds{-\sqrt{2}\left(\frac{Z_m}{Z_p}\right)^\hal
 \Bigg\{\frac{1}{3}\left( \ol{M}_{K^\pm}^2+
 \ol{M}_{K^0}^2-2\ol{M}_{\pi^\pm}^2\right)}\nnn
 &-& \ds{
 \frac{\ol{f}_{K^\pm}+\ol{f}_{K^0}-2\ol{f}_\pi}{\ol{f}_{K^\pm}+\ol{f}_{K^0}+\ol{f}_\pi}\left[
 \frac{Z_p}{Z_m}M_\etap^2-
 \frac{1}{3}\left( \ol{M}_{K^\pm}^2+
 \ol{M}_{K^0}^2+\ol{M}_{\pi^\pm}^2\right)
 \right]\Bigg\} }\; .
 \label{YYY6}
\end{array}\end{equation}
We note that for $Z_p\simeq Z_m$ the second term almost cancels the
first one. Using (\ref{MixingAngle}) with $\ol{M}_8^2$ given by
(\ref{PseudoScalarMassesToLinearOrder}) and $M_p\simeq
M_{\etap}\simeq957.8\MeV$ as an experimental input, one finds
$\th_p=-18.7,-7.2,0.5$ for $Z_p/Z_m=0.5,1.0,1.5$.  We will see in
sect.~\ref{QuarkMassCorrectionsToKineticTerms} how quark mass
corrections to the kinetic terms modify these relations substantially
already to linear order in $\Dt$.

For the $\eta$ mass we get to linear order the Gell-Mann--Okubo
relation
\begin{equation}
 \left( M_\eta^{(1)}\right)^2=\ol{M}_8^2=\frac{1}{3}\left(
 2\ol{M}_{K^\pm}^2+2\ol{M}_{K^0}^2-\ol{M}_{\pi^\pm}^2\right)\simeq
 (566.3\MeV)^2\; .
 \label{YYY7}
\end{equation}
We can use (\ref{YYY6}) together with the estimate of $\ol{M}_8^2$ to
quadratic order in $\Dt$, (\ref{NNN1}), to compute a mass relation
between $M_\eta^2$ and $M_\etap^2$ to quadratic order in $\Dt$.
Inserting $\la_2$ from (\ref{Lambda2Estimate}) and using $M_\etap$ as
an input we find
\begin{equation}
 M_\eta^{(2)}=521.2,535.7,549.9\MeV
 \label{YYY8}
\end{equation}
for $Z_p/Z_m=0.5,1.0,1.5$.
This is already very close to the
experimental value $M_\eta=547.5\MeV$ as long as $Z_p/Z_m$ is not too
small. 

To summarize this section we find that the approximation of a quark
mass independent kinetic term (\ref{LkinForMultiplets}) or
(\ref{Lkin1}) gives already a reasonable overall picture of the scalar
and pseudoscalar mesons. The most important modifications from the
quark mass corrections of the kinetic terms are expected for the value
of $\la_2$ and the mixing angle $\theta_p$. This will, in turn,
influence the estimate of $M_\eta^2$ to second order in $\Dt$ and
similarly the relation between $M_{\eta^\prime}^2$ and $M_p^2$.

\sect{Quark mass corrections to kinetic terms}
\label{QuarkMassCorrectionsToKineticTerms}

The kinetic term (\ref{LkinForMultiplets}) obtains corrections for 
nonvanishing quark masses. Expanding around $\Phi_0={\rm
  diag}(\olsi_0)$ with 
$\VEV{\chi_s}=0$ these corrections involve only the expectation value of
$h$. We are only interested here in the kinetic terms for $m$ and $p$.
The most general corrections involving two derivatives and being linear
in $v$ and $w$ can then be written in the form
\begin{equation}
 \Lc_{\rm kin}^{(1)} =\frac{1}{4}\om_m\Tr h\prl^\mu m\prl_\mu m+
 \hal\om_{pm}\prl_\mu p\Tr h\prl^\mu m\; .
 \label{LkinCorrections}
\end{equation}
The term $\sim\om_m$ leads to different wave function renormalization
constants for pions, kaons and $m_8$ according to
\begin{equation}\begin{array}{rcl}
 \ds{Z_\pi} &=& \ds{1-\om_m v}\nnn
 \ds{Z_{K^\pm}} &=& \ds{1+\hal\om_m (v+w)}\nnn
 \ds{Z_{K^0}} &=& \ds{1+\hal\om_m (v-w)}\nnn
 \ds{Z_8} &=& \ds{1+\om_m v}\; .
 \label{WaveFunctionCorrections}
\end{array}\end{equation}
This implies that the renormalized pion mass $M_{\pi^\pm}$ obeys
\begin{equation}
 \label{BBB0}
 M_{\pi^\pm}^2=\ol{M}_{\pi^\pm}^2 Z_\pi^{-1}
\end{equation}
where $\ol{M}_{\pi^\pm}^2$ is the mass computed in the previous sections,
e.g.
\begin{equation}
 \ol{M}_{\pi^\pm}^2=m_m^2-\gm_2 v+\dt_7(3v^2+w^2)+
 2\dt_8(v^2+w^2)+2\dt_9(v^2-w^2)\; .
\end{equation}
Similar relations hold for $M_{K^\pm}^2$, $M_{K^0}^2$, $M_3^2$ and $M_8^2$
whereas $M_{38}^2=\ol{M}_{38}^2Z_\pi^{-\hal}Z_8^{-\hal}$. 
There is also a mixed term $\sim\om_m w\prl^\mu m_3\prl_\mu m_8$. It
gives corrections $\sim w^2$ to the $\pi^0$ and $\eta$ masses and will
be neglected here.

One should note that the choice of $Z_\pi$, $Z_{K^\pm}$, $Z_{K^0}$ and
$Z_8$ is somewhat arbitrary. It depends on the convention for $Z_m$,
since a rescaling of $Z_m$ would result in a rescaling of $Z_\pi$,
$Z_{K^\pm}$, $Z_{K^0}$ and $Z_8$. We employ here a convention for $Z_m$
where (cf. (\ref{WaveFunctionCorrections}))
\begin{equation}
 Z_\pi+Z_{K^\pm}+Z_{K^0}=1\; .
 \label{WFC}
\end{equation}
Neglecting corrections $\sim w^2$ we may then use
\begin{equation}
 \om_m v=\frac{1}{3}\left(
 \frac{Z_{K^\pm} Z_{K^0}}{Z_\pi^2}-1\right)\; .
\end{equation}

The difference between $\ol{M}^2$ and $M^2$ influences the symmetry
relations once expressed in terms of physical masses $M^2$.
In particular, we observe a modification of
the relation (\ref{YYY1}) between $M_8^2$, $M_{\pi^\pm}^2$, $M_{K^\pm}^2$ and
$M_{K^0}^2$. Including the terms (\ref{LkinCorrections}) this relation
now reads (see eq.~(\ref{A69}) for the definition of $\ol{f}$)
\begin{equation}
 M_8^2=\frac{1}{3}\left(2M_{K^\pm}^2\frac{Z_{K^\pm}}{Z_8}+
 2M_{K^0}^2\frac{Z_{K^0}}{Z_8}-M_{\pi^\pm}^2\frac{Z_\pi}{Z_8}\right)-
 \frac{1}{4}\la_2\left(\ol{f}_{K^\pm}+\ol{f}_{K^0}-2\ol{f}_\pi\right)^2\frac{1}{Z_8}\; .
 \label{NewPseudoScalarMassRelation}
\end{equation}
In consequence, the corrections due to the modification of the kinetic
term influence 
the pseudoscalar mass eigenvalues to second order in the quark
masses. Expanding (\ref{NewPseudoScalarMassRelation}) to this order one
finds neglecting terms $\sim w^2$
\begin{equation}\begin{array}{rcl}
 \ds{M_8^2} &=& \ds{
 \frac{1}{3}\left(2M_{K^\pm}^2+2M_{K^0}^2-M_{\pi^\pm}^2\right)-
 \frac{1}{3}\om_m v\left(
 M_{K^\pm}^2+M_{K^0}^2-2M_{\pi^\pm}^2\right)}\nnn
 &-& \ds{
 \frac{1}{4}\la_2\left[
 \left(f_{K^\pm}+f_{K^0}-2f_\pi\right)-
 \frac{1}{4}\om_m v
 \left(f_{K^\pm}+f_{K^0}+4f_\pi\right)\right]^2}\; .
 \label{NewPseudoScalarMassRelation2}
\end{array}\end{equation}

Here we have used that the deviation of $Z_\pi$, $Z_{K^\pm}$,
etc.~from unity also influences the relation between $v$, $w$ and the
decay constants $f_\pi$, $f_{K^\pm}$, $f_{K^0}$. The effect of the
wave function renormalization on the meson decay constants is
discussed in appendix \ref{MesonDecayConstants} and leads to
\begin{equation}\begin{array}{rcl}
 \ds{\left(\frac{Z_m}{Z_h}\right)^\hal w} &=& \ds{
 Z_{K^\pm}^{-\hal}f_{K^\pm} -Z_{K^0}^{-\hal}f_{K^0}=
 \ol{f}_{K^\pm}-\ol{f}_{K^0}}\nnn
 \ds{\left(\frac{Z_m}{Z_h}\right)^\hal v} &=& \ds{\frac{1}{3}
 \left(Z_{K^\pm}^{-\hal}f_{K^\pm}+
 Z_{K^0}^{-\hal}f_{K^0}-2Z_\pi^{-\hal}f_\pi\right)=
 \frac{1}{3}\left(\ol{f}_{K^\pm}+\ol{f}_{K^0}-2\ol{f}_\pi\right)}
 \label{A69}
\end{array}\end{equation}
with
\begin{equation}
 \label{BBB1}
  \olf_\pi=Z_\pi^{-\hal}f_\pi 
\end{equation} 
and similarly for $f_{K^\pm}$, $f_{K^0}$.

In summary, we will denote the physical meson masses and decay
constants by $M_i$ and $f_i$. They correspond to a normalization of
the fields with inverse propagator $q^2+M_i^2$ in the vicinity of
$q^2=-M_i^2$. This is also the relevant field normalization for the
decay constants --- see appendix \ref{MesonDecayConstants}. On the
other hand, the quantities $\ol{M}_i$ and $\olf_i$ correspond to a
common $SU(3)$ symmetric wave function renormalization for the whole
octet.  Symmetry relations are therefore most easily expressed in
terms of $\ol{M}_i$ and $\olf_i$. For an approximation of the kinetic
term to lowest order in the quark masses there is no difference
between $M_i$ and $\ol{M}_i$ or $f_i$ and $\olf_i$. This approximation
is sufficient to compute the meson masses to linear order in the quark
masses, but not for decay constants and the mixing angle $\th_p$. In
sections
\ref{ScalarMesonMassesToLinearOrder}--\ref{ParametersOfTheLinearSigmaModel}
we employed a lowest order approximation to the kinetic term and
therefore omitted the distinction between $M$ and $\ol{M}$ or $f$ and
$\olf$ for the quantitative estimates. On the other hand, the
algebraic relations in the preceding sections are all expressed in
terms of $\ol{M}$ and $\ol{f}$ and are therefore not altered by
modifications of the kinetic terms. In consequence, the only necessary
change for the quantitative estimates involves the relations between
$\ol{f}$ and $f$ or $\ol{M}$ and $M$.

We also note that the relations (\ref{WaveFunctionCorrections}) use a
wave function renormalization which is defined for a common momentum
$q_0^2$ for the whole octet. In fact, one may view the derivative
expansion as a Taylor expansion of the inverse propagators $G^{-1}(q)$
around some fixed nonvanishing momentum $q_0^2$ rather than an
expansion around zero momentum, i.e.~an expansion of the type
$G_i^{-1}(q)=\ol{M}_i^2+q^2\{Z_i+\Oc(q^2-q_0^2)\}$. Here the
$Z_i$ depend on the choice of $q_0^2$ through the normalization
condition 
\begin{equation}\begin{array}{rcl}
 \ds{Z_i} &=& \ds{\frac{1}{q_0^2}\left(
 G_i^{-1}(q_0^2)-G_i^{-1}(0)\right)}\nnn
 \ds{\ol{M}_i^2} &=& \ds{G_i^{-1}(0)}\; .
 \label{DefinitionOfZi}
\end{array}\end{equation}
This condition, together with (\ref{WFC}), also specifies the precise
meaning of\footnote{For an expression of the propagators in terms of
  the fields $\Phi$ or $\Phi_p$ the $Z_i$ stand for $Z_{\pi^\pm}Z_m$
  etc.~and $\ol{M}_i^2$ should be replaced by $\ol{M}_{\pi^\pm}^2 Z_m$
  etc.} $Z_m$. Similar definitions also apply for $Z_p$ and $Z_h$, but
the momentum used can now be different from the pseudoscalar octet
momentum $q_0^2$. For the definition of $Z_p$ it seems convenient to
replace\footnote{For convenience we will later also use
  $q_p^2=-M_\eta^2$.} $q_0^2$ by $q_{p}^2=-m_p^2$, whereas for $Z_h$
one may use $q_h^2=-m_h^2$. We note that our definition
(\ref{DefinitionOfZi}) of $Z_i$ also specifies the precise meaning of
the couplings $\om_m$ and $\om_{pm}$ in
(\ref{WaveFunctionCorrections}). They multiply three--point functions
at zero momentum for\footnote{The normalization of $h$, however, is
  specified by $Z_h$ and therefore adapted to the behavior of the
  scalar propagator in the vicinity of its pole.} $h$ and momenta
given by $q_0^2$ for $p$ and $m$. All these specifications are
irrelevant as long as only terms with two derivatives are included.
The conceptual setting becomes crucial, however, once we go beyond
this approximation and include higher derivative terms.

In fact, the definition $M_i^2=\ol{M}_i^2/Z_i$ yields the physical
pole masses $M_i^2$ only if we replace in (\ref{DefinitionOfZi}) the
common momentum $q_0^2$ by the individual locations of the poles at
$q_i^2=-M_i^2$. This involves corrections of the inverse propagators
$\sim(q_i^2-q_0^2)$. We discuss these additional contributions from
higher derivative terms more explicitly in section
\ref{HigherDerivativeContributions}. Here we note only the following
general properties: Except for the mixing angle the higher derivative
corrections can be absorbed completely in the wave function
renormalization constants $Z_i$. They lead to additional terms on the
right hand side of (\ref{WaveFunctionCorrections}) which are
proportional to $(q_i^2-q_0^2)$ according to
\begin{equation}\begin{array}{rcl}
 \label{NNN24}
 \ds{G_i^{-1}(q)} &=& \ds{\ol{M}_i^2+\ol{Z}_i q^2+
 \ol{H}_i(q^2-q_0^2)q^2+\hal\ol{H}_i^{(2)}(q^2-q_0^2)^2 q^2+
 \ldots}\nnn
 &=& \ds{\ol{M}_i^2+Z_i q^2+H_i(q^2-q_i^2)q^2+\ldots}\; .
\end{array}\end{equation}
Here the quantities $\ol{Z}_i$ correspond to a normalization at
$q_0^2$ and are given by (\ref{WaveFunctionCorrections}) whereas the
true wave function renormalizations $Z_i$ are defined at
$q_i^2=-M_i^2$, with
\begin{equation}\begin{array}{rcl}
 \label{NNN16}
 \ds{Z_i} &=& \ds{\ol{Z}_i+\Dt Z_i}\nnn
 \ds{\Dt Z_i} &=& \ds{\ol{H}_i(q_i^2-q_0^2)+
 \hal\ol{H}_i^{(2)}(q_i^2-q_0^2)^2+\ldots\; .}
\end{array}\end{equation}
The difference $q_i^2-q_0^2$ is given by pseudoscalar mass
differences and is therefore linear in the quark masses. More precisely,
it is again linear in $v$ and $w$. The Taylor expansion of
$G^{-1}(q)-G^{-1}(0)$ around $q_0^2$ seems most reliable for
$\sqrt{-q_0^2}$ somewhere inbetween the kaon and pion masses. We will
choose the renormalization scale for $Z_m$ as
\begin{equation}
  \label{TTT8}
  q_0^2=-\frac{1}{3}
  \left(M_{K^\pm}^2+M_{K^0}^2+M_{\pi^\pm}^2\right)
\end{equation}
such that (\ref{WFC}) remains valid. This implies that $\Delta
Z_\pi=-(\Delta Z_{K^\pm}+\Delta Z_{K^0})$ and we can absorb the higher
derivative effects partially in a redefinition of an effective
\begin{equation}
  \ol{\om}_m=\om_m-\frac{\Delta Z_\pi}{v}
\end{equation}
leading to
\begin{equation}\begin{array}{rcl}
 \ds{Z_\pi} &=& \ds{1-\ol{\om}_m v}\nnn
 \ds{Z_{K^\pm}} &=& \ds{1+\hal\ol{\om}_m(v+w)}\nnn
 \ds{Z_{K^0}} &=& \ds{1+\hal\ol{\om}_m(v-w)}\nnn
 \ds{Z_8} &=& \ds{1+\ol{\om}_m v+K_8}\; .
 \label{NWFRC}
\end{array}\end{equation}
In terms of $\ol{\om}_m$ only $Z_8$ receives an additional
correction
\begin{equation}
 K_8=\Dt Z_8-2\Dt Z_K\; .
\end{equation}
This yields the complete expression for $M_8^2$ to quadratic order in
the quark masses
\begin{equation}\begin{array}{rcl}
 \ds{M_8^2} &=& \ds{
 \frac{1}{3}\left(1-K_8\right)
 \left(2M_{K^\pm}^2+2M_{K^0}^2-M_{\pi^\pm}^2\right)-
 \frac{1}{3}\ol{\om}_m v\left(
 M_{K^\pm}^2+M_{K^0}^2-2M_{\pi^\pm}^2\right)}\nnn
 &-& \ds{
 \frac{1}{4}\la_2\left[
 \left(f_{K^\pm}+f_{K^0}-2f_\pi\right)-
 \frac{1}{4}\ol{\om}_m v
 \left(f_{K^\pm}+f_{K^0}+4f_\pi\right)\right]^2}\; .
 \label{AP2}
\end{array}\end{equation}
For a computation of $M_\eta$ to quadratic order in the quark masses
one needs, in addition, the complete expression for the mixing angle
$\th_p$ to linear order in the quark masses.

In order to get a rough estimate of the size of the higher derivative
corrections one may assume that the true inverse propagator does not
deviate by more than $10\%$ from the lowest order form $M^2+q^2$ over
a momentum range between $-M_\pi^2$ and $-M_\eta^2$. This results in a
typical bound $\abs{\ol{H}_m}\lta0.1\abs{q_0^2}^{-1}$ and we expect the
higher derivative corrections to be unimportant. Furthermore, within a
systematic derivative expansion we can take to leading order a common
$\ol{H}_m$ for the whole octet. This yields
\begin{equation}\begin{array}{rcl}
 \ds{K_8} &=& \ds{\ol{H}_m
 \left(\frac{4}{3}M_K^2-M_\eta^2-\frac{1}{3}M_\pi^2\right)}\; .
 \label{AP1}
\end{array}\end{equation}
Inserting the leading order for $M_\eta^2$ on the right hand
side of (\ref{AP1}) we find
\begin{equation}
 K_8\sim\Oc(\Delta^2)\;.
 \label{K8}
\end{equation}
We conclude that $K_8$ gives corrections to the pseudoscalar masses
which are formally cubic in the $m_q$.

The deviation of $Z_\pi$, $Z_{K^\pm}$, etc.~from one can lead to a
sizeable change of the infered value for $v$ and therefore to
important modifications of the values of the couplings of the linear
sigma model. Inserting (\ref{NWFRC}) into (\ref{A69}) and
(\ref{RenormalizedCouplings}) yields to leading order (neglecting
isospin violation)
\begin{equation}\begin{array}{rcl}
 \ds{\left(\frac{Z_m}{Z_h}\right)^\hal v} 
 &=& \ds{
 \frac{2}{3}\left[f_K-f_\pi-
 \frac{1}{4}\ol{\om}_m v\left(f_K+
 2f_\pi\right)\right] }\nnn
 \ds{\si_0} &=& \ds{
 \frac{1}{6}
 \left[2f_K+f_\pi-
 \frac{1}{2}\ol{\om}_m v
 \left( f_K-f_\pi\right)\right]
 }\; .
 \label{NewEstimateV}
\end{array}\end{equation}
We observe that the correction to the relation
(\ref{DifferencesOfDecayConstants}) is of the order
$(Z_h/Z_m)^{1/2}\ol{\om}_m\si_0$. It is not suppressed by quark mass
terms. On the other hand, the correction to $\si_0$ is of second order
in the quark masses and therefore small. Taking $\ol{\om}_m v=-0.20$
(see sect.~\ref{Results}) the relations
(\ref{NewEstimateV}) result in
\begin{equation}
 \label{NNN27}
 \left(\frac{Z_m}{Z_h}\right)^\hal v\simeq23.3\MeV\; ,\;\;\;\;
 \si_0\simeq53.8\MeV
\end{equation}
to be compared with (\ref{vEstimate}) and
(\ref{RenormalizedCouplings}). We conclude that the quark mass
corrections to the kinetic terms are important for a quantitative
understanding of the linear sigma model! These corrections also matter
for a determination of the decay constant $f_{K^0}$. We first notice
the change in $w$ induced by $\ol{\om}_m v=-0.20$ in
(\ref{EstimateW}). Using (\ref{NWFRC}) we find
\begin{equation}
 \label{NNN28}
 \left(\frac{Z_m}{Z_h}\right)^\hal w=
 -\frac{3}{2}
 \frac{\left(\frac{Z_m}{Z_h}\right)^\hal v}
 {(1-\ol{\omega}_m v)}
 \frac{(M_{K^0}^2-M_{K^\pm}^2)}
 {(M_K^2-M_\pi^2)}\simeq
 -0.67\MeV\; .
\end{equation}
This yields
\begin{equation}
 \label{NNN29}
 f_{K^0}-f_{K^\pm}=
 -\left(\frac{Z_m}{Z_h}\right)^\hal w
 \left[\left(1+
 \frac{1}{2}
 \ol{\omega}_m v\right)^{\frac{1}{2}}+
 \frac{f_{K^\pm}}{2\left(\frac{Z_m}{Z_h}\right)^\hal v}
 \frac{\ol{\om}_m v}
 {\left(1+
 \frac{1}{2}\ol{\omega}_m v\right)}
 \right]\simeq
 0.28\MeV
\end{equation}
to be compared with (\ref{EstimateW}), $f_{K^0}-f_{K^\pm}=0.47\MeV$,
for $\ol{\om}_m v=0$. The quark mass corrections to the kinetic terms
reduce the isospin violation of the decay constants significantly!

Let us next turn to the term $\sim\om_{pm}\prl_\mu p\Tr h\prl^\mu m$ in
(\ref{LkinCorrections}) which influences the octet--singlet mixing
angle $\th_p$. This term leads to an off--diagonal kinetic term
\begin{equation}
 \hat{\om} \prl^\mu m_8\prl_\mu p=
 -\sqrt{3}\om_{pm}v\prl^\mu m_8\prl_\mu p\; .
\end{equation}
Using $m_{R8}=Z_8^\hal m_8$ the inverse propagator for the fields
$p$, $m_{R8}$ is given in momentum space by the matrix
\begin{equation}
 G^{-1}=\left(\ba{ccc}
 z_p(q^2)q^2+M_p^2 &,& Z_8^{-\hal}
 \left(\hat{\om}(q^2) q^2+M_{8p}^2\right) \\
 Z_8^{-\hal}\left(\hat{\om}(q^2)q^2+
 M_{8p}^2\right) &,&
 z_8(q^2)q^2+M_8^2
 \ea\right)\; .
 \label{ZZZ7}
\end{equation}
Here $M_{8p}^2$ is given to quadratic order in $m_q$ by
(\ref{NNN1}), whereas $M_p^2=m_p^2$, (\ref{YYY2}), and $M_8^2$ obeys
(\ref{NewPseudoScalarMassRelation}).  We parameterize corrections from
higher derivative terms for the diagonal elements of (\ref{ZZZ7})
by the functions $z_p(q^2)$, $z_8(q^2)$. Our normalization of $Z_p$
and $Z_8$ is chosen such that $z_p(-M_\eta^2)=1$, $z_8(-M_\eta^2)=1$.
Higher derivative corrections to the off--diagonal elements of
(\ref{ZZZ7}) result in an effective $q^2$--dependence of $\hat{\om}$.
For $q^2=-M_\eta^2$ this correction is formally of third order in the
quark masses, whereas for $q^2=-M_{\etap}^2$ it is of first order
(since $M_{\etap}^2-M_\eta^2$ is counted as being of zeroth order). In
real life, however, $M_{\etap}^2$ and $M_\eta^2$ are separated by a
factor less than four.  With the reasonable assumption that there is
no dramatic $q^2$--dependence of $\hat{\om}$ we will treat
$\hat{\om}q^2$ as a term linear in the quark masses for both
$-q^2=M_\eta^2$ and $M_{\etap}^2$, and correspondingly count the
higher derivative corrections as terms quadratic in the quark masses.
For a lowest order estimate they can therefore be neglected,
$z_p(q^2)=z_8(q^2)=1$, $\hat{\omega}(q^2)=\hat{\omega}$.

The diagonalization of (\ref{ZZZ7}) can not be performed
independently of $q^2$ anymore. As a consequence, the effective
octet--singlet mixing angle $\th_p$ will depend on $q^2$:
\begin{equation}
 \tan\th_p(q^2)=\frac{M_8^2-M_p^2+\sqrt{(M_8^2-M_p^2)^2+
 4(M_{8p}^2+\hat{\om} q^2)^2 Z_8^{-1}}}
 {2(M_{8p}^2+\hat{\om} q^2)Z_8^{-\hal}}\; .
 \label{YYY5}
\end{equation}
The propagator in a diagonal basis corresponds to the eigenvalues
of $G$, i.e.
\begin{equation}
 G_{\etap,\eta}^{-1}(q^2)=q^2+\hal\left( M_p^2+M_8^2\right)\pm
 \left[\frac{1}{4}\left( M_p^2-M_8^2\right)^2+
 \frac{M_{8p}^4}{Z_8}+\frac{\hat{\om} ^2}{Z_8}q^4+
 2\frac{\hat{\om} M_{8p}^2}{Z_8}q^2\right]^\hal\; .
\end{equation}
The masses $M_\etap^2$ and $M_\eta^2$ are given by the location of
the poles of $G$ for negative $q^2$:
\begin{equation}\begin{array}{rcl}
 \label{BBB7}
 \ds{M_{\etap,\eta}^2} &=& \ds{\left(
 1-\frac{\hat{\om} ^2}{Z_8}\right)^{-1}\Bigg[
 \hal\left( M_p^2+M_8^2\right)-
 \frac{\hat{\om} }{Z_8}M_{8p}^2 }\nnn
 &\pm& \ds{\left\{
 \frac{1}{4}\left( M_p^2-M_8^2\right)^2+
 \frac{M_{8p}^4}{Z_8}-
 \frac{\hat{\om} }{Z_8}M_{8p}^2\left( M_p^2+M_8^2\right)+
 \frac{\hat{\om} ^2}{Z_8}M_p^2M_8^2\right\}^\hal\Bigg]}\; .
\end{array}\end{equation}
Expanding to quadratic order in the quark masses (with $\hat{\om} $ and
$M_{8p}^2$ linear in $m_q$) one finds the relations
\begin{eqnarray}
 \ds{M_\etap^2+M_\eta^2} &=& \ds{M_p^2+M_8^2-
 2\hat{\om} M_{8p}^2+\hat{\om} ^2\left( M_p^2+M_8^2\right)}\\[2mm]
 \ds{M_\etap^2-M_\eta^2} &=& \ds{M_p^2-M_8^2+
 \hat{\om} ^2\left( M_p^2-M_8^2\right)}\nnn
 &+& \ds{
 \frac{2}{M_p^2-M_8^2}\left[
 M_{8p}^4-\hat{\om} M_{8p}^2\left(M_p^2+M_8^2\right)+
 \hat{\om} ^2M_p^2M_8^2\right]\label{YYY4} }\\[2mm]
 &=& \ds{M_\etap^2+M_\eta^2-2M_8^2+
 \frac{2M_{8p}^4}{M_\etap^2+M_\eta^2-2M_8^2}
 \nonumber}\; .
\end{eqnarray}
It is remarkable that the relation between $M_\etap^2-M_\eta^2$ and
$M_\etap^2+M_\eta^2$ becomes independent of $\hat{\om} $ in this
approximation. 

For an experimental determination of the octet--singlet mixing through
the decay of $\etap$ or $\eta$ into two photons the relevant
quantities will be $\th_p(q^2=-M_\etap^2)$ or $\th_p(q^2=-M_\eta^2)$,
respectively. The mixing angle depends strongly on $\hat{\om} $ if
$\hat{\om} M_\etap^2$ is of the same order of magnitude as $M_{8p}^2$.
This leads to a sizeable dependence on $q^2$. If one intends to
compute the octet--singlet mixing angle to quadratic order in the
quark masses one needs in addition contributions to $\hat{\om} $
quadratic in $m_q$. They arise from a modification of the kinetic term
through
\begin{equation}
 \Lc_{\rm kin}^{(2)}=\hal\om_{pm}^\prime\prl_\mu p
 \Tr h^2\prl^\mu m+\ldots\; .
 \label{LkinModification}
\end{equation}
This leads to a second order correction for $\hat{\om} $
\begin{equation}
 \hat{\om}(q^2) =-\sqrt{3}\left[\om_{pm}v+
 \om^\prime_{pm}v^2\right]
 \frac{f_\omega(q^2)}{f_\omega(-m_m^2)}
 \label{ZZZ4}
\end{equation}
where $f_\omega(q^2)$ contains the higher derivative corrections where
$f_\omega(-M_\eta^2)=1$ and $f_\omega(-m_m^2)$ reflects the
normalization of $\omega_{pm}$, $\omega_{pm}^\prime$ at $q^2=-m_m^2$.
Also the deviation of $Z_8$ from one has to be included to this order.
We note that for $z_p(q^2)=z_8(q^2)$ the formula (\ref{YYY5}) remains
valid if $\hat{\omega}$ is replaced by $\hat{\omega}(q^2)$,
(\ref{ZZZ4}).

\sect{$M_\eta$ and $M_{\eta^\prime}$ to quadratic order}
\label{MassRelaionsAndCouplingConstantsToQuadraticOrder}

We are now ready to address the question of mass relations for the
pseudoscalar mesons to quadratic order in the quark masses. First, we
replace the coupling $\nu$, or equivalently $M_p^2$, by $M_\etap^2$
as a phenomenological input parameter. Our aim is to compute
$M_\eta^2$ and the mixing angle $\th_p$ as functions of $M_\etap^2$,
$M_{\pi^\pm}^2$, $M_{K^\pm}^2$, $M_{K^0}^2$, $f_K$ and $f_\pi$.  For
this purpose we use the relations (\ref{YYY4}) and (\ref{YYY5}) which
involve the quantities $M_8^2$, $M_{8p}^2 Z_8^{-1/2}$ and $\hat{\om}
Z_8^{-1/2}$. The difference $M_\etap^2-M_\eta^2$ is to quadratic order
independent of $\hat{\om}$.  The dependence on $Z_8$ arises only
indirectly through the correction $\sim K_8$ in $M_8^2$ (\ref{AP2})
and gives no correction to quadratic order in the quark masses,
(\ref{K8}). The mixing angle depends already to linear order on
$\hat{\om}$, but the difference between $Z_8$ and one is only needed
to quadratic order. For the octet mass term $M_8^2$ we use (\ref{AP2})
with
\begin{equation}
 12\dt_9 v^2=-\frac{1}{4}\la_2\left(
 \olf_{K^\pm}+\olf_{K^0}-2\olf_\pi\right)^2
\end{equation}
and determine $\la_2$ by (\ref{Lambda2Estimate}), with $M_\etap^2$
replaced by the lowest order expression
$M_\etap^2+M_\eta^2-\frac{1}{3}\left(2M_{K^\pm}^2+2M_{K^0}^2-M_{\pi^\pm}^2\right)$.
To the order relevant for our estimate $M_{8p}^2$ is given by
(\ref{YYY6}). Inserting this into the second expression (\ref{YYY4})
one finally obtains the relation
\begin{eqnarray}
 \ds{M_\eta^2} &=& \ds{\frac{1}{3}\left(
 2M_{K^\pm}^2+2M_{K^0}^2-M_{\pi^\pm}^2\right)+
 \frac{1}{3}\left(
 M_{K^\pm}^2+M_{K^0}^2-2M_{\pi^\pm}^2\right)
 \frac{\olf_{K^\pm}+\olf_{K^0}-2\olf_\pi}
 {\olf_{K^\pm}+\olf_{K^0}+\olf_\pi} }\nnn
 &\times& \ds{
 \left(1-\frac{1}{3}\left(\frac{Z_h}{Z_m}\right)^\hal
 \ol{\om}_m\left(\olf_{K^\pm}+\olf_{K^0}+\olf_\pi\right)-
 2 \frac{\olf_{K^\pm}+\olf_{K^0}-2\olf_\pi}
 {\olf_{K^\pm}+\olf_{K^0}+\olf_\pi}
 \right)}\nnn
 &-& \ds{\frac{2}{9}
 \frac{\olf_\pi^2}{\left(\olf_{K^\pm}+\olf_{K^0}+\olf_\pi\right)^2}
 \frac{\left( 3M_{K^\pm}^2+3M_{K^0}^2-2M_{\pi^\pm}^2\right)^2}
 {\left[M_\etap^2-\frac{1}{3}\left( 
 2M_{K^\pm}^2+2M_{K^0}^2-M_{\pi^\pm}^2\right)\right]} 
 \label{YYY9}}\\[2mm]
 &-& \ds{\frac{2}{3}
 \left(\frac{Z_p}{Z_m}-1\right)
 \frac{\left(\olf_{K^\pm}+\olf_{K^0}-2\olf_\pi\right)^2}
 {\left(\olf_{K^\pm}+\olf_{K^0}+\olf_\pi\right)^2}
 \left(2M_{K^\pm}^2+2M_{K^0}^2-M_{\pi^\pm}^2\right)}\nnn
 &\times& \ds{\left[
 1+\frac{2M_{K^\pm}^2+2M_{K^0}^2-M_{\pi^\pm}^2}
 {3M_\etap^2-2M_{K^\pm}^2-2M_{K^0}^2+M_{\pi^\pm}^2}\right]}\nnn
 &-& \ds{\frac{2}{3}
 \left(\frac{Z_m}{Z_p}-1\right)
 \frac{\left[\left(M_{K^\pm}^2+M_{K^0}^2\right)
 \left(2\olf_{K^\pm}+2\olf_{K^0}-\olf_\pi\right)-
 M_{\pi^\pm}^2\left(\olf_{K^\pm}+\olf_{K^0}\right)\right]^2}
 {\left(3M_\etap^2-2M_{K^\pm}^2-2M_{K^0}^2+M_{\pi^\pm}^2\right)
 \left(\olf_{K^\pm}+\olf_{K^0}+\olf_\pi\right)^2}}\nnn
 &-& \ds{\frac{1}{3}K_8\left(2M_{K^\pm}^2+2M_{K^0}^2-M_{\pi^\pm}^2\right)
 \nonumber}\; .
\end{eqnarray}
For $\ol{\om}_m=0$ and $K_8=0$ we recover the mass relation
(\ref{YYY8}) of sect.~\ref{ParametersOfTheLinearSigmaModel}. Based
on (\ref{YYY9}) one can now evaluate the corrections $\sim\ol{\om}_m$
and $\sim K_8$ quantitatively. Neglecting higher orders in
$\left(\olf_{K^\pm}+\olf_{K^0}-2\olf_\pi\right)$ one has the
approximate relations
\begin{equation}\begin{array}{rcl}
 \ds{\frac{\left(\olf_{K^\pm}+\olf_{K^0}-2\olf_\pi\right)}
 {\left(\olf_{K^\pm}+\olf_{K^0}+\olf_\pi\right)}} &\simeq& \ds{
 \frac{\left( f_{K^\pm}+f_{K^0}-2f_\pi\right)}
 {\left( f_{K^\pm}+f_{K^0}+f_\pi\right)}-
 \hal\ol{\om}_m v }\nnn
 \ds{\frac{\olf_\pi}
 {\left(\olf_{K^\pm}+\olf_{K^0}+\olf_\pi\right)}} &\simeq& \ds{
 \frac{f_\pi}
 {\left( f_{K^\pm}+f_{K^0}+f_\pi\right)}+
 \frac{1}{6}\ol{\om}_m v }\; .
\end{array}\end{equation}
In the following we also drop $K_8$ since it is formally of third
order in the quark masses.  If we linearize (\ref{YYY9}) in
$\ol{\om}_m
v=\frac{1}{3}\left(\frac{Z_h}{Z_m}\right)^{1/2}\ol{\om}_m\left(
\ol{f}_{K^\pm}+\ol{f}_{K^0}-2\ol{f}_\pi\right)$ and
$\left(\frac{Z_m}{Z_p}-1\right)$ we find
\begin{equation}
 M_\eta=(535.7\MeV)\left[1-0.04\left(\frac{Z_m}{Z_p}-1\right)-
 0.511\,\ol{\om}_m v\right]\; .
 \label{K9}
\end{equation}
Within the range $Z_p/Z_m=1.0\pm0.2$ we see that the
corrections $\sim(Z_p/Z_m-1)$ only amount to at most $7\MeV$. For a
rough estimate of the size of $\ol{\om}_m v$ we can neglect them.
Comparison of (\ref{K9}) with the measured $\eta$ mass yields for
$Z_p=Z_m$
\begin{equation}
 \ol{\om}_m v\simeq-0.043\; .
 \label{OmegamV}
\end{equation}

\begin{figure}
\unitlength1.0cm
\begin{picture}(13.,9.)
\put(.7,5.){\bf $\ds{\frac{M_\eta}{\MeV}}$}
\put(7.5,0.8){\bf $\om_m v$}

\put(7.9,6.4){$\frac{Z_p}{Z_m}=1.3$}
\put(5.7,3.0){$\frac{Z_p}{Z_m}=0.7$}

\put(-0.8,-11.5){
\epsfysize=22.cm
\epsffile{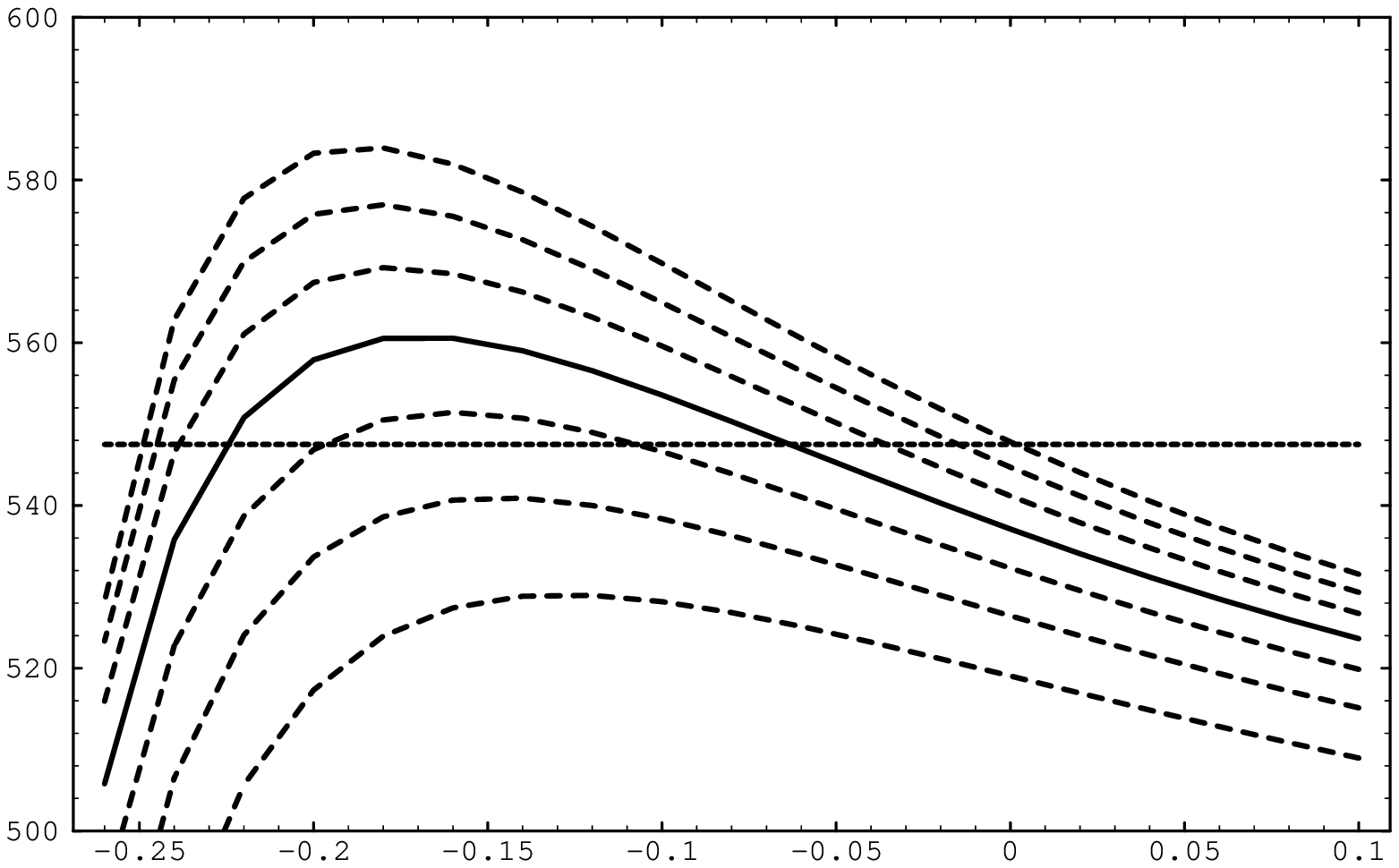}
}
\end{picture}
\caption{\footnotesize The plot shows $M_\eta$ as a function of
  $\om_m v$ for various values of $Z_p/Z_m$ and $\om_m=\ol{\om}_m$.
  The solid line corresponds to $Z_p/Z_m=1$ and the difference in
  $Z_p/Z_m$ between two adjacent lines is $0.1$. The horizontal dotted
  line indicates the experimental value $M_\eta\simeq547.5\MeV$.}
\label{Plot1} 
\end{figure}

\begin{figure}
\unitlength1.0cm
\begin{picture}(13.,9.)
\put(.7,5.){\bf $\ds{\frac{M_p}{\MeV}}$}
\put(7.5,0.8){\bf $\om_m v$}
\put(-0.8,-11.5){
\epsfysize=22.cm
\epsffile{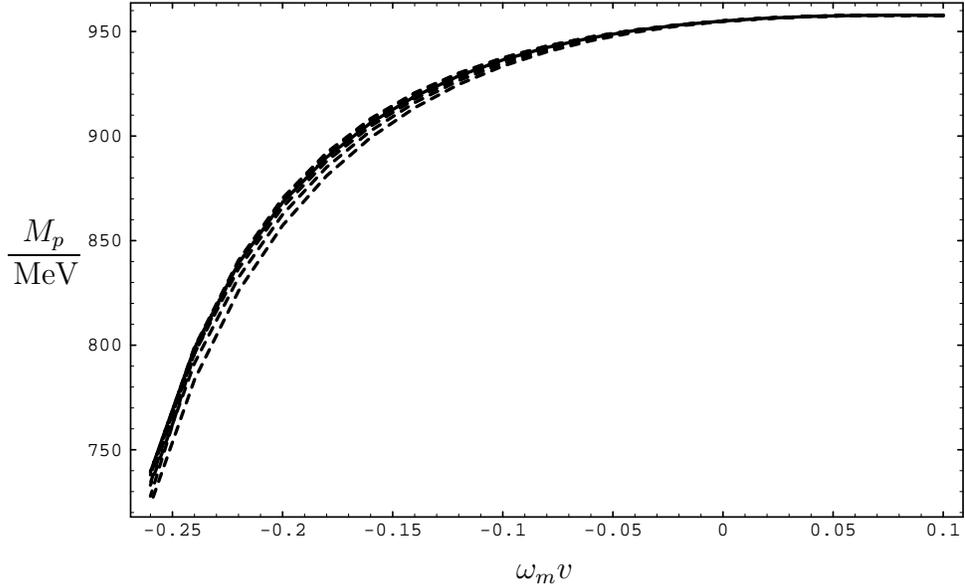}
}
\end{picture}
\caption{\footnotesize The plot shows $M_p$ as a function of
  $\omega_m v$ for given $M_{\eta^\prime}=957.8\MeV$,
  $\ol{\omega}_m=\omega_m$ and $Z_p/Z_m$ varying between $0.7$
  and $1.3$ in steps of $0.1$. The solid line corresponds to
  $Z_p/Z_m=1.0$.}
\label{Plot1b} 
\end{figure}

We should note, however, that the linearization in $\ol{\om}_m v$ is
not reliable anymore for $\ol{\om}_m v\lta-0.15$. This is
demonstrated in fig.~\ref{Plot1} where we plot $M_\eta$ as a
function of $\om_m v$ for various values of $Z_p/Z_m$.
For this plot we have evaluated all matrix elements in (\ref{ZZZ7}) to
second order in $\Delta$ (keeping the full $\ol{\omega}_m
v$--dependence, e.g., in $Z_8^{-1/2}$, though) and diagonalized the
matrix without further expansion in $\Delta$. We have neglected higher
derivative contributions which can not be absorbed into the wave
function renormalizations, i.e.~we used $z_p=z_8=1$, $f_\omega=1$. For
$\hat{\omega}$ we have used (\ref{TTT9}) and $\omega_m=\ol{\omega}_m$.
The nonlinear effects due to terms which are formally
$\sim\Oc(\Delta^3)$ and higher are reflected by the deviation of the
curves in fig.~\ref{Plot1} from the tangents at $\omega_m v=0$.
Because of the important nonlinearities for $\ol{\omega}_m v\lta-0.15$
one finds that there is a second solution with $Z_p\simeq Z_m$, namely
for $\om_m v\simeq-0.22$. For large values of $|\om_m v|$ the
$\eta$--$\eta^\prime$ mixing starts playing an important role.  We
will see below that realistic values for the decay rates
$\eta\ra2\gamma$ and $\eta^\prime\ra2\gamma$ are only consistent with
this second solution for $\omega_m v$. In this region the formal quark
mass expansion does not converge well anymore!  This breakdown of the
quark mass expansion is even more apparent in the relation between
$M_{\eta^\prime}^2$ and $M_p^2$ since linear and higher order mixing
effects occur with the same sign for this quantity. We plot in figure
\ref{Plot1b} the value of $M_p$ as a function of $\om_m v$ for a fixed
value of $M_{\eta^\prime}=957.8\MeV$ and various values of
$Z_p/Z_m$ neglecting all higher derivative corrections to the kinetic
terms.
As a first observation one sees that the dependence of $M_p$ on
$Z_p/Z_m$ is rather week. Furthermore, for $\om_m v\simeq-0.22$ the
ratio $M_p^2/M_{\eta^\prime}^2$ has decreased to about $0.77$ despite
the fact that this effect is formally of second order in $\Delta$ if
$M_p^2$ is counted as $\Oc(1)$. A partial explanation of this strong
mixing effect is related to the observation that for $\om_m
v\simeq-0.22$ the value $M_p\simeq839\MeV$ is actually almost
comparable to $\ol{M}_8=579\MeV$. (The values are for $Z_p/Z_m=1.0$.)
A counting where $\ol{M}_8^2/M_p^2=\Oc(\Delta)$ becomes therefore
doubtful. As an alternative, we may count the anomaly contribution to
$M_p^2$, i.e.~$M_p^2Z_p/Z_m-m_m^2\simeq(742\MeV)^2$ also as an effect
of order $\Delta$ since its size is not too different from the $SU(3)$
breaking induced by the mass of the strange quark. In this counting
scheme all elements of the $\eta$--$\eta^\prime$ mass matrix have the
same order of magnitude. Eigenvalues to a given order in $\Delta$
involve then all matrix elements to this order and the mixing effects
are naturally large. Nevertheless, since $\ol{M}_8^2/M_p^2=0.69$ is
still substantially smaller than one the real situation is somewhere
in the transition region between the two counting rules. Our
approximations to order $\Delta^2$ are consistent with both ways of
counting.

In the remaining sections of this paper we will gradually collect
information on the quantities $Z_m/Z_p$, $\ol{\om}_m v$, $\hat{\om}$
and $K_8$. We should mention that the quantities $\ol{\om}_m v$ and
$K_8$ only involve properties of the effective action for the
pseudoscalar octet. There is, in principle, no information which goes
beyond the one contained in chiral perturbation theory. The quantities
$Z_m/Z_p$ and $\hat{\om}$ also involve the $\etap$ and go beyond
standard chiral perturbation theory. Of course, if one were able to
predict directly quantities like $\ol{\om}_m v$ one would gain in
addition information on some parameters appearing in chiral
perturbation theory.

\sect{Kinetic terms in the linear meson model}
\label{KineticTermsInTheLinearSigmaModel}

In this section we discuss in more detail the derivative terms in the
context of the linear $\si$--model. Our first aim is to gain
information about the size of $Z_m/Z_p$, $\om_m$ and $\hat{\om}$.
Expanding (\ref{Lkin}) in powers of $\Phi_p$, $\Phi_s$ and $\chi_p$ we
observe that only those terms contribute which have both derivatives
acting on pseudoscalar fields. Also all fields without derivatives
must be scalars. We can therefore replace in (\ref{Phi})
\begin{equation}\begin{array}{rcl}
 \ds{\prl_\mu\Phi} &\lra& \ds{
 \frac{i}{\sqrt{2}}\prl_\mu\Phi_p+
 \frac{i}{\sqrt{6}}\prl_\mu\chi_p}\nnn
 \ds{\prl_\mu\Phid} &\lra& \ds{
 -\frac{i}{\sqrt{2}}\prl_\mu\Phi_p-
 \frac{i}{\sqrt{6}}\prl_\mu\chi_p}
\end{array}\end{equation}
and for the fields without derivatives
\begin{equation}\begin{array}{rcl}
 \ds{\Phi} &\lra& \ds{\olsi_0+\frac{1}{\sqrt{2}}\Phi_s}\nnn
 \ds{\Phid} &\lra& \ds{\olsi_0+\frac{1}{\sqrt{2}}\Phi_s}\; .
\end{array}\end{equation}
The terms appearing in $\Lc_{\rm kin}$ (\ref{Lkin}) have been selected
such that the leading terms contributing to $Z_p/Z_m$, $\om_m$ and
$\om_{pm}$ have been included.  For this purpose we first classify the
invariants contributing to terms with two derivatives acting on
pseudoscalars which are at most linear in $\Phi_s$.  The
reasoning is somewhat lengthy but straightforward: By partial
integration all invariants involve at most one derivative acting on a
given $\Phi$. We can then distribute the two derivatives either within
the same index contraction (with $\delta_{ab}$ or $\epsilon_{abc}$)
with respect to $SU_L(3)\times SU_R(3)$ or among two different such
structures.  Traces involving six and more powers of $\Phi$ have at
least one combination $\Phi^\dagger\Phi$ (or $\Phi\Phi^\dagger$)
without derivatives in the chain. By a suitable definition of the
invariants this may be replaced by
$\left(\Phid\Phi-\frac{1}{3}\Tr(\Phid\Phi)\right)$ and such invariants
contribute therefore only to higher order in the quark masses.  For
instance, combinations involving two factors of $\Phi^\dagger\Phi$ or
$\Phi\Phi^\dagger$ (without derivatives) contribute at most to order
$\Phi_s^2$. The only invariant involving a trace of six factors
$\Phi^\dagger$ or $\Phi$ that may contribute linearly in $\Phi_s$ is
therefore the term $\sim W_\vph$. There are also two terms quartic in
$\Phi$ with couplings $X_\vph^+$, $X_\vph^-$. To linear order in
$\Phi_s$ the structures involving $\epsilon$--tensors are the one
$\sim U_\vph$ and terms not listed in (\ref{Lkin}), namely
\begin{eqnarray}
  \label{KKL1}
  \ds{\Lc_{\rm kin}(T)} &=& \ds{
  \frac{1}{2}\epsilon^{a_1 a_2 a_3}
  \epsilon^{b_1 b_2 b_3}\Bigg\{
  T_\vph^{(1)}\prl^\mu\Phi_{a_1 b_1}
  \prl_\mu\Phi_{a_2 b_2}
  \left(\Phi_{a_3 b_4}\Phi^\dagger_{b_4 a_4}\Phi_{a_4 b_3}-
  \frac{1}{3}\Phi_{a_3 b_3}\Tr\Phi^\dagger\Phi\right)}\nnn
  &+& \ds{
  T_\vph^{(2)}\prl^\mu\Phi_{a_1 b_1}\Phi_{a_2 b_2}
  \left(\prl_\mu\Phi_{a_3 b_4}
  \Phi^\dagger_{b_4 a_4}\Phi_{a_4 b_3}+
  \Phi_{a_3 b_4}\Phi^\dagger_{b_4 a_4}\prl_\mu\Phi_{a_4 b_3}-
  \frac{2}{3}\prl_\mu\Phi_{a_3 b_3}\Tr\Phi^\dagger\Phi\right) }\nnn
  &+& \ds{
  T_\vph^{(3)} \prl^\mu\Phi_{a_1 b_1}\Phi_{a_2 b_2}
  \left(\Phi_{a_3 b_4}\prl_\mu\Phi^\dagger_{b_4 a_4}
  \Phi_{a_4 b_3}+
  \frac{1}{3}\prl_\mu\Phi_{a_3 b_3}\Tr\Phi^\dagger\Phi \right)}\\[2mm]
  &+& \ds{
  T_\vph^{(4)}\left(
  \Phi_{a_1 b_1}\Phi_{a_2 b_2}\prl^\mu\Phi_{a_3 b_4}
  \Phi^\dagger_{b_4 a_4}\prl_\mu\Phi_{a_4 b_3}+
  \frac{1}{3}\Phi_{a_1 b_1}\Phi_{a_2 b_2}
  \Phi_{a_3 b_3}\Tr\prl^\mu\Phi^\dagger\prl_\mu\Phi\right)}\nnn
  &+& \ds{
  T_\vph^{(5)}\Phi_{a_1 b_1}\Phi_{a_2 b_2}
  \left(\prl^\mu\Phi_{a_3 b_4}\prl_\mu\Phi^\dagger_{b_4 a_4}
  \Phi_{a_4 b_3}+
  \Phi_{a_3 b_4}\prl^\mu\Phi^\dagger_{b_4 a_4}
  \prl_\mu\Phi_{a_4 b_3}\right.}\nnn
  &+& \ds{\left.
  2\prl^\mu\Phi_{a_3 b_4}\Phi^\dagger_{b_4 a_4}
  \prl_\mu\Phi_{a_4 b_3}\right)
  +\left(\Phi\leftrightarrow\Phi^\dagger\right)
  \Bigg\}
  \nonumber}
\end{eqnarray}
These invariants do not contribute to $Z_m$ or $Z_p$ but they may
contribute to $\omega_m$ or $\omega_{pm}$. We treat $\Lc_{\rm kin}(T)$
as a higher order correction to the term $\sim U_\vph$ and neglect
this piece in the following. Next we turn to the case where the two
derivatives act within two different $SU_L(3)\times SU_R(3)$ invariant
index structures.  Since the invariants $\rho$, $\tau_2$, $\tau_3$ and
$\xi$ are at least quadratic in the pseudoscalar fields the terms
$\sim\prl_\mu\rho\prl^\mu\rho$, $\prl_\mu\tau_2\prl^\mu\tau_2$,
$\prl_\mu\tau_3\prl^\mu\tau_3$, or $\prl_\mu\xi\prl^\mu\xi$ do not
contribute to $Z_m$, $Z_p$, $\om_m$ or $\hat{\om}$. The same holds for
mixed terms like $\prl_\mu\rho\prl^\mu\xi$ and for index structures of
the type $\Tr\Phid\prl^\mu\Phi\Tr\Phid\Phi\prl_\mu\Phid\Phi$ etc. On
the other hand, the pseudoscalar invariant $\om=i(\det\Phi-\det\Phid)$
contains a term linear in the pseudoscalar fields
\begin{equation}
 \prl_\mu\om=-\sqrt{6}\olsi_0^2\prl_\mu\chi_p+
 \frac{1}{2\sqrt{6}}\Tr\Phi_s^2\prl_\mu\chi_p+
 \olsi_0\Tr\Phi_s\prl_\mu\Phi_p-
 \frac{1}{\sqrt{2}}\Tr\Phi_s^2\prl_\mu\Phi_p+\ldots\; .
\end{equation}
More generally, we can construct invariants with a possible
contribution to $Z_p/Z_m$, $\omega_m$ or $\omega_{pm}$ by Lorentz
contraction of $\Cc\Pc$--odd factors like $\prl_\mu\omega$,
\begin{displaymath}\begin{array}{lll}
  &\ds{i\epsilon^{a_1 a_2 a_3}\epsilon^{b_1 b_2 b_3}
  \left\{\prl_\mu\Phi_{a_1 b_1}\Phi_{a_2 b_2}
  \Phi_{a_3 b_4}\Phi^\dagger_{b_4 a_4}\Phi_{a_4 b_3}-
  \prl_\mu\Phi^\dagger_{a_1 b_1}\Phi^\dagger_{a_2 b_2}
  \Phi^\dagger_{a_3 b_4}\Phi_{b_4 a_4}\Phi^\dagger_{a_4 b_3}
  \right\}}&, \nnn
  &\ds{
  i\epsilon^{a_1 a_2 a_3}\epsilon^{b_1 b_2 b_3}\prl_\mu
  \left\{\Phi_{a_1 b_1}\Phi_{a_2 b_2}\Phi_{a_3 b_4}
  \Phi^\dagger_{b_4 a_4}\Phi_{a_4 b_3}-
  \Phi^\dagger_{a_1 b_1}\Phi^\dagger_{a_2 b_2}
  \Phi^\dagger_{a_3 b_4}\Phi_{b_4 a_4}
  \Phi^\dagger_{a_4 b_3}
  \right\}}&,\nnn
  &\ds{
  i\epsilon^{a_1 a_2 a_3}\epsilon^{b_1 b_2 b_3}
  \left\{\Phi_{a_1 b_1}\Phi_{a_2 b_2}
  \Phi_{a_3 b_4}\prl_\mu\Phi^\dagger_{b_4 a_4}\Phi_{a_4 b_3}-
  \Phi^\dagger_{a_1 b_1}\Phi^\dagger_{a_2 b_2}
  \Phi^\dagger_{a_3 b_4}\prl_\mu
  \Phi_{b_4 a_4}\Phi^\dagger_{a_4 b_3}
  \right\}}&\; .
\end{array}\end{displaymath}
Within these structures we may replace subsets $\Phi\Phi^\dagger\Phi$
by the combination
$\Phi\Phi^\dagger\Phi-\frac{1}{3}\Phi\Tr\Phi^\dagger\Phi$ which is
$\sim\Phi_s$. We will keep here only the term which is
$\sim\prl_\mu\omega\prl^\mu\omega$ and consider the other contractions
as higher order corrections to the term $\sim\tilde{V}_\vph$. The
phenomenological analysis below will indicate that the ``anomaly''
terms $\sim U_\vph$ and $\sim\tilde{V}_\vph$ are not very
important. This justifies to neglect corrections to them. We have also
included in (\ref{Lkin}) some invariants which do not contribute to
$Z_p/Z_m$, $\omega_m$ and $\omega_{pm}$. This is partly for the
purpose to demonstrate that not much can be learned about the ratio
$Z_h/Z_m$ from exploiting the symmetries of the linear sigma model.

The contribution of the term $\sim\tV_\vp$ to $Z_m$, $Z_p$ and $\hat{\om}$
can be read off from
\begin{eqnarray}
 \ds{\Lc_{\rm kin}(\tilde{V}_\vph)=
 \hal\tV_\vp\prl^\mu\om\prl_\mu\om} &=& \ds{
 3\tV_\vp\olsi_0^4\prl^\mu\chi_p\prl_\mu\chi_p-
 \sqrt{6}\tV_\vp\olsi_0^3\prl^\mu\chi_p\Tr\Phi_s\prl_\mu\Phi_p}\nnn
 &+& \ds{
 \sqrt{3}\tV_\vp\olsi_0^2\prl^\mu\chi_p\Tr\Phi_s^2\prl_\mu\Phi_p+
 \ldots}\; .
 \label{ZZZ3}
\end{eqnarray}
Next we turn to the terms $\sim X_\vp^-$ and $X_\vp^+$ in
(\ref{Lkin}). Whereas $X_\vp^+$ gives no contribution linear in the
quark masses the term $\sim X_\vp^-$ yields
\begin{equation}\begin{array}{rcl}
 \ds{\Lc_{\rm kin}(X_\vp^-)} &=& \ds{
 \hal X_\vp^-\olsi_0^2
 \left(\Tr\prl^\mu\Phi_p\prl_\mu\Phi_p+\prl^\mu\chi_p\prl_\mu\chi_p\right)+
 \frac{1}{\sqrt{2}}X_\vp^-
 \olsi_0\Tr\Phi_s\prl^\mu\Phi_p\prl_\mu\Phi_p}\nnn
 &+& \ds{
 \frac{2}{\sqrt{6}}X_\vp^-\olsi_0\prl^\mu\chi_p\Tr\Phi_s\prl_\mu\Phi_p+
 \frac{1}{2\sqrt{3}}X_\vp^-\prl^\mu\chi_p\Tr\Phi_s^2\prl_\mu\Phi_p}\; .
 \label{ZZZ2}
\end{array}\end{equation}
Similarly, one obtains
\begin{eqnarray}
  \label{KKL2}
  \ds{\Lc_{\rm kin}(W_\vph)} &=& \ds{
  \frac{1}{2\sqrt{2}}W_\vph\ol{\sigma}_0^3
  \Tr\Phi_s\prl^\mu\Phi_p\prl_\mu\Phi_p+
  \frac{1}{\sqrt{6}}W_\vph\ol{\sigma}_0^3
  \prl^\mu\chi_p\Tr\Phi_s\prl_\mu\Phi_p}\nnn
  &+& \ds{
  W_\vph\ol{\sigma}_0^2\Bigg\{
  \frac{3}{8}\Tr\Phi_s^2\prl^\mu\Phi_p\prl_\mu\Phi_p-
  \frac{1}{24}\Tr\Phi_s^2\Tr\prl^\mu\Phi_p\prl_\mu\Phi_p}\\[2mm]
  &+& \ds{
  \frac{1}{4}\Tr\Phi_s\prl^\mu\Phi_p\Phi_s\prl_\mu\Phi_p+
  \frac{5}{4\sqrt{3}}\prl^\mu\chi_p
  \Tr\Phi_s^2\prl_\mu\Phi_p+
  \frac{1}{6}\Tr\Phi_s^2\prl^\mu\chi_p\prl_\mu\chi_p
  \Bigg\}+\ldots\nonumber}\; .
\end{eqnarray}
For the piece $\sim U_\vp$ we use
\begin{equation}\begin{array}{rcl}
 \ds{\eps^{a_1 a_2 a_3}\eps^{b_1 b_2 b_3}} &=& \ds{
 \dt_{a_1 b_1}\dt_{a_2 b_2}\dt_{a_3 b_3}+
 \dt_{a_1 b_2}\dt_{a_2 b_3}\dt_{a_3 b_1}+
 \dt_{a_1 b_3}\dt_{a_2 b_1}\dt_{a_3 b_2}}\nnn
 &-& \ds{
 \dt_{a_1 b_1}\dt_{a_2 b_3}\dt_{a_3 b_2}-
 \dt_{a_1 b_3}\dt_{a_2 b_2}\dt_{a_3 b_1}-
 \dt_{a_1 b_2}\dt_{a_2 b_1}\dt_{a_3 b_3}}
\end{array}\end{equation}
and find
\begin{equation}\begin{array}{rcl}
 \ds{\Lc_{\rm kin}(U_\vp)} &=& \ds{U_\vp\olsi_0\left(\hal
 \Tr\prl^\mu\Phi_p\prl_\mu\Phi_p-\prl^\mu\chi_p\prl_\mu\chi_p\right)}\nnn
 &-& \ds{
 \frac{1}{\sqrt{2}}U_\vp\Tr\Phi_s\prl^\mu\Phi_p\prl_\mu\Phi_p+
 \frac{1}{\sqrt{6}}U_\vp\prl^\mu\chi_p\Tr\Phi_s\prl_\mu\Phi_p}\; .
 \label{ZZZ1}
\end{array}\end{equation}

\begin{figure}
\unitlength1.0cm
\begin{picture}(15.,5.)
\put(0.,4.){\bf $\th_p(\eta)$}
\put(4.,0.2){\bf $\om_m v$}

\put(14.5,4.){\bf $\th_p(\eta^\prime)$}
\put(11.2,0.2){\bf $\om_m v$}
\put(-2.5,-19.5){
\epsfxsize=19.cm
\epsffile{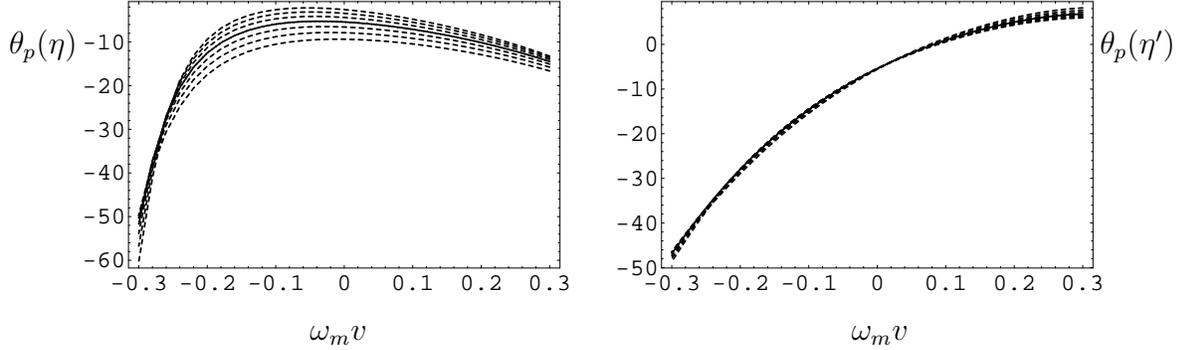}
}
\end{picture}
\caption{\footnotesize The plots show the pseudoscalar mixing angles
  $\th_p(\eta)$ and $\th_p(\eta^\prime)$ as functions of $\om_m v$ for
  various values of $Z_p/Z_m$ and $\om_m=\ol{\om}_m$. The solid line
  corresponds to $Z_p/Z_m=1$ and the difference in $Z_p/Z_m$ between
  two adjacent lines is $0.1$.}
\label{Plot2} 
\end{figure}

In our truncation no other second derivative invariant (except the one
$\sim Z_\vp$) contributes to $Z_m$, $Z_p$ $\om_m$, $\om_{pm}$ or
$\om_{pm}^\prime$. Combining (\ref{ZZZ3}), (\ref{ZZZ2}) and
(\ref{ZZZ1}) with the term $\sim Z_\vp$ and comparing with
(\ref{Lkin1}), (\ref{LkinCorrections}), (\ref{LkinModification}) and
(\ref{ZZZ4}) yields
\begin{equation}\begin{array}{rcl}
 \ds{Z_m} &=& \ds{Z_\vp+X_\vp^-\olsi_0^2+U_\vp\olsi_0}\nnn
 \ds{Z_p} &=& \ds{Z_\vp+X_\vp^-\olsi_0^2+
 6\tV_\vp\olsi_0^4-2U_\vp\olsi_0}\nnn
 \ds{\om_m} &=& \ds{\left(X_\vp^-\ol{\sigma}_0
 +\frac{1}{2}W_\vph\ol{\sigma}_0^3
 -U_\vp\right)
 Z_h^{-\hal}Z_m^{-1}}\nnn
 \ds{\om_{pm}} &=& \ds{\left(
 \frac{2}{\sqrt{6}}X_\vp^-\ol{\sigma}_0
 +\frac{1}{\sqrt{6}}W_\vph\ol{\sigma}_0^3
 -\sqrt{6}\tV_\vp\olsi_0^3+
 \frac{1}{\sqrt{6}}U_\vp\right)
 Z_p^{-\hal}Z_h^{-\hal}Z_m^{-\hal}}\nnn
 \ds{\om_{pm}^\prime} &=& \ds{\left(
 \frac{1}{2\sqrt{6}}X_\vp^-+
 \frac{5}{4\sqrt{6}}W_\vph\ol{\sigma}_0^2+
 \frac{\sqrt{6}}{2}\tV_\vp\olsi_0^2\right)
 Z_p^{-\hal}Z_h^{-1}Z_m^{-\hal}}\nnn
 \ds{\hat{\om}} &=& \ds{-
 \left(\olf_{K^\pm}+\olf_{K^0}-2\olf_\pi\right)
 \Bigg[\frac{1}{3\sqrt{2}}U_\vp+\frac{2}{3\sqrt{2}}X_\vp^-\olsi_0
 +\frac{1}{3\sqrt{2}}W_\vph\ol{\sigma}_0^3-
 \sqrt{2}\tV_\vp\olsi_0^3}\nnn
 &+& \ds{
 \frac{\olf_{K^\pm}+\olf_{K^0}-2\olf_\pi}
 {\olf_{K^\pm}+\olf_{K^0}+\olf_\pi}
 \left(\frac{1}{3\sqrt{2}}X_\vp^-\olsi_0+
 \frac{5}{6\sqrt{2}}W_\vph\ol{\sigma}_0^3+
 \sqrt{2}\tV_\vp\olsi_0^3\right)\Bigg]
 Z_p^{-\hal}Z_m^{-1}}\; .
 \label{A3}
\end{array}\end{equation}
We observe the relation
\begin{equation}
 \om_{pm}=\frac{1}{\sqrt{6}}\left[
 2\left(\frac{Z_m}{Z_p}\right)^\hal\om_m+
 \frac{1}{\si_0}\left(\frac{Z_m}{Z_p}-1\right)
 \left(\frac{Z_p}{Z_h}\right)^\hal\right]
\end{equation}
which is independent of the values of $X_\vph^-$, $W_\vph$, $U_\vph$
and $\tilde{V}_\vph$. It leads to an estimate of $\hat{\om}$ to lowest
order in the quark masses:
\begin{equation}
 \hat{\om}^{(1)}=\sqrt{2}
 \left(\frac{Z_m}{Z_p}\right)^{\frac{1}{2}}
 \left[
 \frac{\left(\frac{Z_m}{Z_h}\right)^{\frac{1}{2}}v}
 {2\sigma_0}
 \left(\frac{Z_p}{Z_m}-1\right)-
 \om_m v\right]\; .
 \label{A5}
\end{equation}
If we furthermore assume $|6\tilde{V}_\vph\ol{\sigma}_0^2|\,
,\,|\frac{5}{2}W_\vph\ol{\sigma}_0^2|\ll X_\vph^-$ (see section
\ref{ExpansionInTheChiralCondensate}) we find\footnote{The coupling
  $\om_{pm}^\prime$ cannot be expressed in terms of $\om_m$ and
  $Z_m-Z_p$. This reflects that the kinetic invariants are linearly
  independent.} (for $\tilde{V}_\vph=0$, $W_\vph=0$)
\begin{equation}
  \label{TTT10}
  \omega_{pm}^\prime=
  \frac{1}{2\sqrt{6}\sigma_0}
  \left(\frac{Z_m}{Z_p}\right)^{\frac{1}{2}}
  \frac{Z_m}{Z_h}
  \left[\omega_m
  \left(\frac{Z_h}{Z_m}\right)^{\frac{1}{2}}+
  \frac{1}{3\sigma_0}
  \left(1-\frac{Z_p}{Z_m}\right)\right]\; .
\end{equation}
This leads to an estimate to second order in $\Delta$
\begin{equation}
  \label{TTT9}
  \hat{\omega}^{(2)}=\hat{\omega}^{(1)}-
  \frac{1}{2\sqrt{2}}
  \left(\frac{Z_m}{Z_p}\right)^{\frac{1}{2}}
  \frac{\left(\frac{Z_m}{Z_h}\right)^{\frac{1}{2}}v}
  {\sigma_0}
  \left[\omega_m v+
  \frac{\left(\frac{Z_m}{Z_h}\right)^{\frac{1}{2}}v}
  {3\sigma_0}
  \left(1-\frac{Z_p}{Z_m}\right)\right]\; .
\end{equation}
These results can be used for a computation of the $\eta$--$\etap$
mixing angle to second order in the quark masses.  If we neglect the
higher derivative corrections ($\ol{\om}_m=\om_m$, $K_8=0$,
$z_p=z_8=f_\omega=1$) the mixing angles $\th_p(\eta)$ and
$\theta_p(\eta^\prime)$ relevant for the two photon decay of the
$\eta$ and $\eta^\prime$, respectively, depend on two parameters
$\om_m v$ and $Z_p/Z_m-1$. We plot in fig.~\ref{Plot2} these
quantities as functions of $\om_m v$ for various values of
$Z_p/Z_m-1$.
Assuming
$\abs{\frac{Z_p}{Z_m}-1}<\frac{1}{4}$ and comparing this value with
$\abs{\frac{1}{3}\left(\frac{Z_h}{Z_m}\right)^{1/2}\om_m(\ol{f}_{K^\pm}+
  \ol{f}_{K^0}+\ol{f}_\pi)}\simeq\abs{\ol{\om}_m
  v\frac{2f_K+f_\pi}{2(f_K-f_\pi)}}>\frac{1}{3}$, (\ref{OmegamV}), we
find that the first contribution is smaller than the second one. In
the approximation $Z_p=Z_m$ the quantity $\hat{\om}$ is simply related
to $\om_m v$ by
\begin{equation}
 \label{XXXA0}
 \ds{\hat{\omega}^{(1)}=-\sqrt{2}\omega_m v\; ,\;\;\;
 \hat{\om}^{(2)}} =-\sqrt{2}\om_m v
 \left[1+
 \frac{\left(\frac{Z_m}{Z_h}\right)^{\frac{1}{2}}v}
 {4\sigma_0}\right]\; .
\end{equation}
The existence of a relation between $\hat{\omega}$ and $\omega_m$ is
crucial for the predictive power of our model since it is necessary in
order to relate the $\eta$--$\eta^\prime$ mixing angle to other
observables. Within our truncation and to lowest order in $\Delta$ the
relation (\ref{A5}) is a pure symmetry relation without any assumption
on the values of the couplings $X_\vph^-$, $W_\vph$, $U_\vph$,
$\tilde{V}_\vph$! In contrast, the kinetic term of the scalar mesons
is independent of $\tV_\vp$, $U_\vp$ or $X_\vp^-$, whereas $Z_h$
receives contributions $\sim X_\vp^+$, etc. No new relations can be
obtained in this way.

We observe that the couplings $U_\vph$ and $\tV_\vph$ violate the
axial $U_A(1)$ symmetry and are therefore connected to effects from
the axial anomaly. Anomaly contributions to kinetic terms of the
pseudoscalar octet are often counted as higher order corrections. If
we decide to do so we obtain the leading order relations
\begin{equation}\begin{array}{rcl}
 \ds{Z_p} &=& \ds{Z_m}\nnn
 \ds{\om_m} &=& \ds{
 X_\vph^-\olsi_0 Z_h^{-\hal}Z_m^{-1}}\; .
 \label{K10}
\end{array}\end{equation}
In this approximation only the kinetic term $\sim X_\vph^-$ is
essential. The importance of $X_\vph^-$ is also manifest in the
leading mixing approximation. Only this term is generated by the
mixing with the divergence of the axialvector field
$\prl_\mu\rho_A^\mu$ (``partial Higgs effect''). In fact, the
effective action for the (pseudo)scalars contained in $\Phi$ receives
contributions from the exchange of other particles.  Prominent
candidates are, of course, the vector and axialvector fields. We have
computed in appendix \ref{VectorMesonContributions} the contributions
{}from the exchange of the vector and axialvector meson octets as well
as the associated $0^{-+}$ and $0^{+-}$ states corresponding to the
divergence of the (axial)vector fields. Up to order $\Phi^4$ this
exchange only contributes to derivative terms. In terms of the
couplings which contribute to the kinetic terms to linear order in
$\Delta$ only $X_\vp^-$ gets a negative contribution from the exchange
of the $0^{-+}$ state whereas all other couplings remain unaffected.
We find a large effect
\begin{equation}
  \label{TTT0}
  \om_m^{(\rho)}v\simeq-0.15\; .
\end{equation}
Comparison of this value with figs.~\ref{Plot1}--\ref{Plot1b}
indicates a large $\eta$--$\eta^\prime$ mixing in a range where
nonlinear effects in $\omega_m v$ are already important!

\sect{Decay constants of $\eta$ and $\eta^\prime$}
\label{DecayConstantsOfEtaAndEtap}

The decay constants $f_{\pi^0}$, $f_\eta$ and $f_{\eta^\prime}$
are experimentally determined from the partial decay width of the
$\pi^0$, $\eta$ and $\eta^\prime$ into two photons
\begin{eqnarray}
  \label{INO2}
  \ds{\Gamma(\eta\ra2\gamma)} &=& \ds{
  \frac{\alpha^2}{64\pi^3}
  \frac{M_{\eta}^3}{f_{\eta}^2}}
\end{eqnarray}
and similarly for $\eta^\prime$ and $\pi^0$. The experimental values
for the decay widths are \cite{PDG94-1}
\begin{eqnarray}
  \label{UUI0}
  \ds{\Gamma(\pi^0\ra2\gamma)} &=& \ds{
  (7.78\pm0.56)\eV}\nnn
  \ds{\Gamma(\eta\ra2\gamma)} &=& \ds{
  (0.46\pm0.04)\keV}\nnn
  \ds{\Gamma(\eta^\prime\ra2\gamma)} &=& \ds{
  (4.26\pm0.19)\keV}
\end{eqnarray}
which yields ($f_\pi=(92.4\pm0.3)\MeV$)
\begin{eqnarray}
  \label{TTT3}
  \ds{\left(\frac{f_{\pi^0}}{f_\pi}\right)^{\rm exp}}
  &\simeq& \ds{1.00\pm0.04}\nnn
  \ds{\left(\frac{f_\eta}{f_\pi}\right)^{\rm exp}}
  &\simeq& \ds{1.06\pm0.05}\nnn
  \ds{\left(\frac{f_{\eta^\prime}}{f_\pi}\right)^{\rm exp}}
  &\simeq& \ds{0.81\pm0.02}\; .
\end{eqnarray}
To lowest order in the quark mass expansion the decay
constant\footnote{We use for $f_\eta$ and $f_{\eta^\prime}$ the
  conventions of \cite{PDG94-1}, with the warning that factors
  $\sqrt{3}$ and $\sqrt{3/8}$ appear here as compared to a perhaps
  more natural convention based on $SU(3)$ symmetry. In the limit of
  exact $SU(3)$ symmetry the quantity corresponding to $f_\pi$ is
  $f_{\eta8}$.} $f_\eta$ is related to $f_K=f_\pi$ by $SU(3)$
symmetry (cf.~appendix \ref{MesonDecayConstants})
\begin{equation}
  \label{TTT1}
  f_\eta^{(0)}=\sqrt{3}f_{\eta8}=\sqrt{3}f_\pi\; .
\end{equation}
Similarly, for $Z_p=Z_m$ one finds for $f_{\eta^\prime}$
\begin{equation}
  \label{TTT2}
  f_{\eta^\prime}^{(0)}=
  \sqrt{\frac{3}{8}}f_{\eta^\prime0}=
 \sqrt{\frac{3}{8}}f_\pi\; .
\end{equation}
The experimental values (\ref{TTT3})
differ substantially from this estimate. In appendix
\ref{MesonDecayConstants} we have computed
((\ref{TTT4})--(\ref{TTT5})) the corrections to $f_{\eta8}$ and
$f_{\eta^\prime0}$ as well as the corresponding constants for the
singlet current in the $\eta$ ($f_{\eta0}$) and the octet current in
the $\eta^\prime$ ($f_{\eta^\prime8}$). Expanding the $Z^{1/2}$
factors to linear order in $\ol{\om}_m v$ and neglecting additional
higher derivative corrections one obtains the
ratios
\begin{equation}\begin{array}{rcl}
 \ds{\frac{f_{\eta8}}{f_\pi}} &=& \ds{
 \frac{1}{3f_\pi}\Bigg[
 2f_{K^\pm}+2f_{K^0}-f_\pi+\hal\ol{\om}_m v
 \left(f_{K^\pm}+f_{K^0}-2f_\pi\right)}\nnn
 &+& \ds{
 \hal K_8\left(2f_{K^\pm}+2f_{K^0}-f_\pi\right)\Bigg]}\nnn
 \ds{\frac{f_{\eta^\prime0}}{f_\pi}} &=&
 \ds{\frac{1}{3f_\pi}\Bigg[
 f_{K^\pm}+f_{K^0}+f_\pi-
 \frac{1}{4}\ol{\om}_m v
 \left(f_{K^\pm}+f_{K^0}-2f_\pi\right)\Bigg]
 \left(\frac{Z_p}{Z_m}\right)^\hal}\; .
\end{array}\end{equation}
Extracting the terms linear in the quark masses and neglecting
isospin violation yields
\begin{eqnarray}
 \ds{\frac{f_{\eta8}}{f_\pi}} &=& \ds{
 \frac{4f_K-f_\pi}{3f_\pi}\simeq1.3}\\[2mm]
 \ds{\frac{f_{\eta^\prime0}}{f_\pi}} &=& \ds{
 \frac{2f_K+f_\pi}{3f_\pi}
 \left(\frac{Z_p}{Z_m}\right)^\hal\simeq
 1.15\left(\frac{Z_p}{Z_m}\right)^\hal}\; .
\end{eqnarray}
We note that this estimate for $f_{\eta8}/f_\pi$ agrees well with the
one obtained in chiral perturbation theory $(f_{\eta8}/f_\pi)_{\chi
  PT}\simeq1.25$ \cite{GL82-1,DHL85-1}. Denoting by
$\theta_p(\eta)\equiv\theta_p(q^2=-M_\eta^2)$ and
$\theta_p(\etap)\equiv\theta_p(q^2=-M_{\eta^\prime}^2)$ (see
(\ref{YYY5})) the octet--singlet mixing angles relevant for the two
photon decay of the $\eta$ and $\etap$, respectively, one can compute
\cite{DHL85-1} the effective decay constants for these decays from
\begin{eqnarray}
 \label{BBB9}
 \ds{\frac{1}{f_\eta}} &=& \ds{
 \frac{1}{\sqrt{3}}\left(\frac{\cos\theta_p(\eta)}{f_{\eta8}}-
 \frac{\sqrt{8}\sin\theta_p(\eta)}{f_{\eta0}}\right)}\\[2mm]
 \ds{\frac{1}{f_\etap}} &=& \ds{
 \frac{1}{\sqrt{3}}\left(\frac{\sin\theta_p(\etap)}{f_{\etap8}}+
 \frac{\sqrt{8}\cos\theta_p(\etap)}{f_{\etap0}}\right)}
 \label{BBB9a}
\end{eqnarray}
Here we have neglected the mixing of $\pi_0$ with $\eta$ and $\etap$
which induces corrections to the decay constants $\sim w^2$. Since
also $Z_{\pi^0}$ equals $Z_\pi$ up to corrections $\sim w^2$ we will
not distinguish between $f_{\pi^0}$ and $f_\pi$ in this paper. 

\begin{figure}
\unitlength1.0cm
\begin{picture}(13.,15.)
\put(1.2,4.6){\bf $\ds{\frac{f_{\eta^\prime}}{f_\pi}}$}
\put(1.2,11.7){\bf $\ds{\frac{f_{\eta}}{f_\pi}}$}
\put(7.5,0.5){\bf $\omega_m v$}

\put(4.0,13.6){$\frac{Z_p}{Z_m}=1.3$}
\put(7.0,11.0){$\frac{Z_p}{Z_m}=0.7$}
\put(7.0,4.6){$\frac{Z_p}{Z_m}=1.3$}
\put(4.0,2.2){$\frac{Z_p}{Z_m}=0.7$}

\put(-0.8,-4.5){
\epsfysize=22.cm
\epsffile{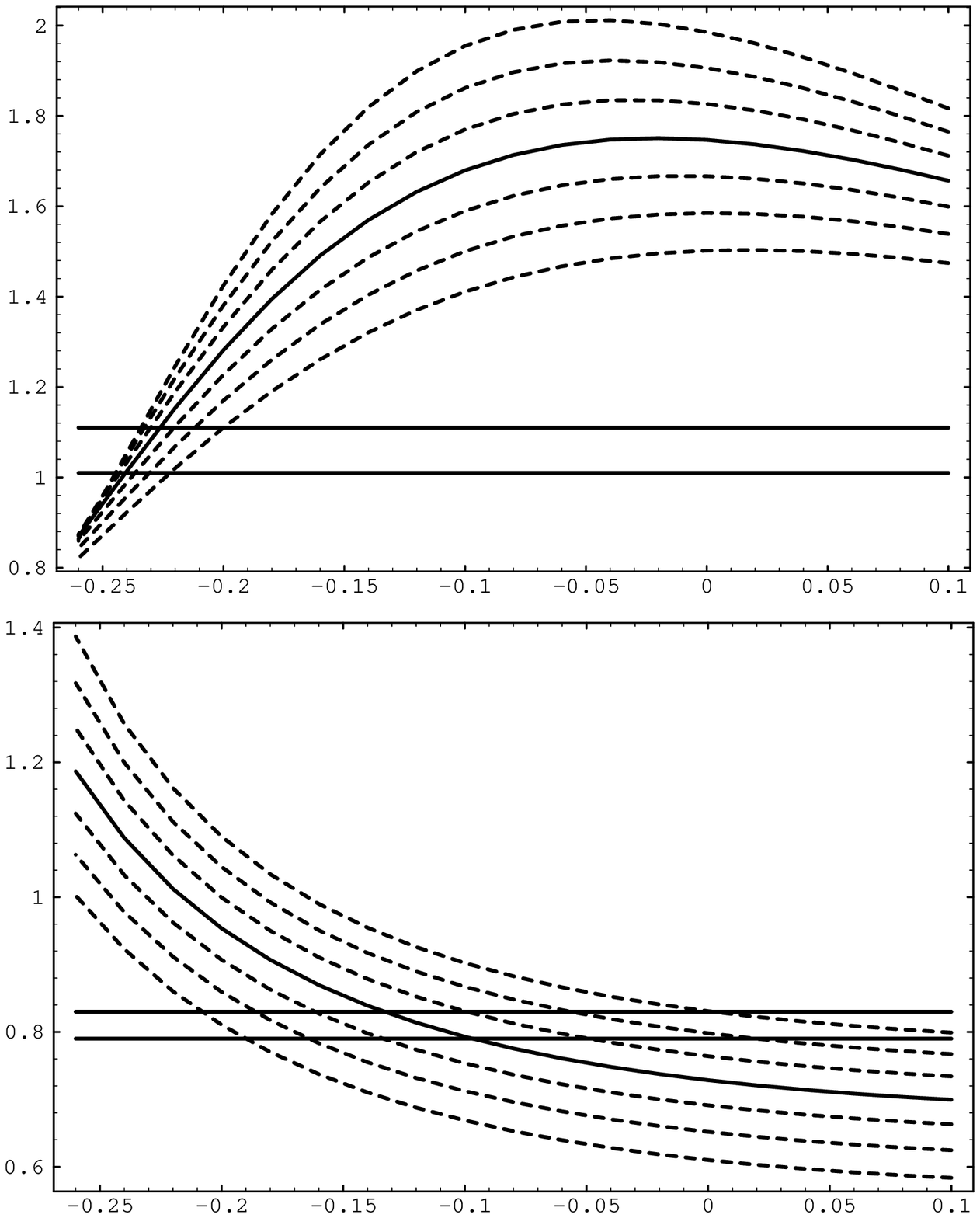}
}
\end{picture}
\caption{\footnotesize The plots show the ratios $f_\eta/f_\pi$ and
  $f_{\eta^\prime}/f_\pi$ as functions of $\omega_m v$ for various
  values of $Z_p/Z_m$ and $\om_m=\ol{\om}_m$.  The solid lines
  correspond to $Z_p/Z_m=1$ and the difference in $Z_p/Z_m$ between
  two adjacent lines is $0.1$. The experimentally allowed windows
  ($1\sigma$) for both quantities are bounded by the horizontal solid
  lines.}
\label{Plot2b} 
\end{figure}

We plot in fig.~\ref{Plot2b} the
decay constants $f_\eta$ and $f_{\eta^\prime}$ as functions of
$\omega_m v$ for various values of $Z_p/Z_m$, assuming
$\ol{\omega}_m=\omega_m$.
It is obvious from these plots that no satisfactory solution exists
for the value $\omega_m v\simeq -0.05$, corresponding to the first
solution for $M_\eta$ (cf.~fig.~\ref{Plot1}) which remains within
the range of convergence of the quark mass expansion. On the other
hand, it is encouraging that for the second solution $\omega_m
v\simeq-0.22$ ($Z_p/Z_m=1$) $f_\eta$ is already very close to the
experimentally allowed window!  The deviation of $f_{\eta^\prime}$
{}from its experimental value by around $25\%$ is not completely
unexpected for the approximations employed so far. In fact, already
the uncertainty of a first order computation of $f_\eta$ and
$f_{\eta^\prime}$ in powers of quark masses should be of the order of
$10\%$. On top of this, the neglect of higher derivative terms is less
accurate for $q^2=-M_{\eta^\prime}^2$ and one expects a less
convergent expansion for $f_{\eta^\prime}$. If $Z_p/Z_m$ is treated as
a second free parameter we see that $\omega_m v=-0.20$,
$Z_p/Z_m=0.9$ provides a solution for which $f_\eta$ and
$f_{\eta^\prime}$ are within $10\%$ of the experimentally allowed
windows.  Furthermore, these values of $\omega_m v$ are quite close to
the ones estimated in appendix \ref{VectorMesonContributions} from the
exchange of higher $0^{-+}$ states (cf.~(\ref{TTT0}))! We conclude
that all observations fit together in a picture with large mixing in
the $\eta$--$\eta^\prime$ sector. As discussed at the end of section
\ref{MassRelaionsAndCouplingConstantsToQuadraticOrder} the anomaly
term comes out relatively small in this case. The naive quark mass
expansion is then expected to converge well only for the flavored
mesons whereas its convergence is unsatisfactory in the
$\eta$--$\eta^\prime$ sector. There it can be replaced by a modified
expansion where $M_p$ is also counted as $\Oc(\Delta)$.

\begin{figure}
\unitlength1.0cm
\begin{picture}(13.,12.)
\put(7.7,0.2){\bf $Z_P$}
\put(12.3,14.3){(a)}
\put(12.3,6.8){(b)}

\put(3.7,11.4){\bf $\frac{M_\eta}{M_\eta^{\rm exp}}$}
\put(5.8,14.0){\bf $\frac{f_\eta}{f_\eta^{\rm exp}}$}
\put(9.4,14.2){\bf $\frac{f_{\eta^\prime}}{f_{\eta^\prime}^{\rm exp}}$}

\put(3.7,3.7){\bf $\frac{M_\eta}{M_\eta^{\rm exp}}$}
\put(5.8,6.3){\bf $\frac{f_\eta}{f_\eta^{\rm exp}}$}
\put(9.4,5.9){\bf $\frac{f_{\eta^\prime}}{f_{\eta^\prime}^{\rm exp}}$}

\put(-0.8,-4.5){
\epsfysize=22.cm
\epsffile{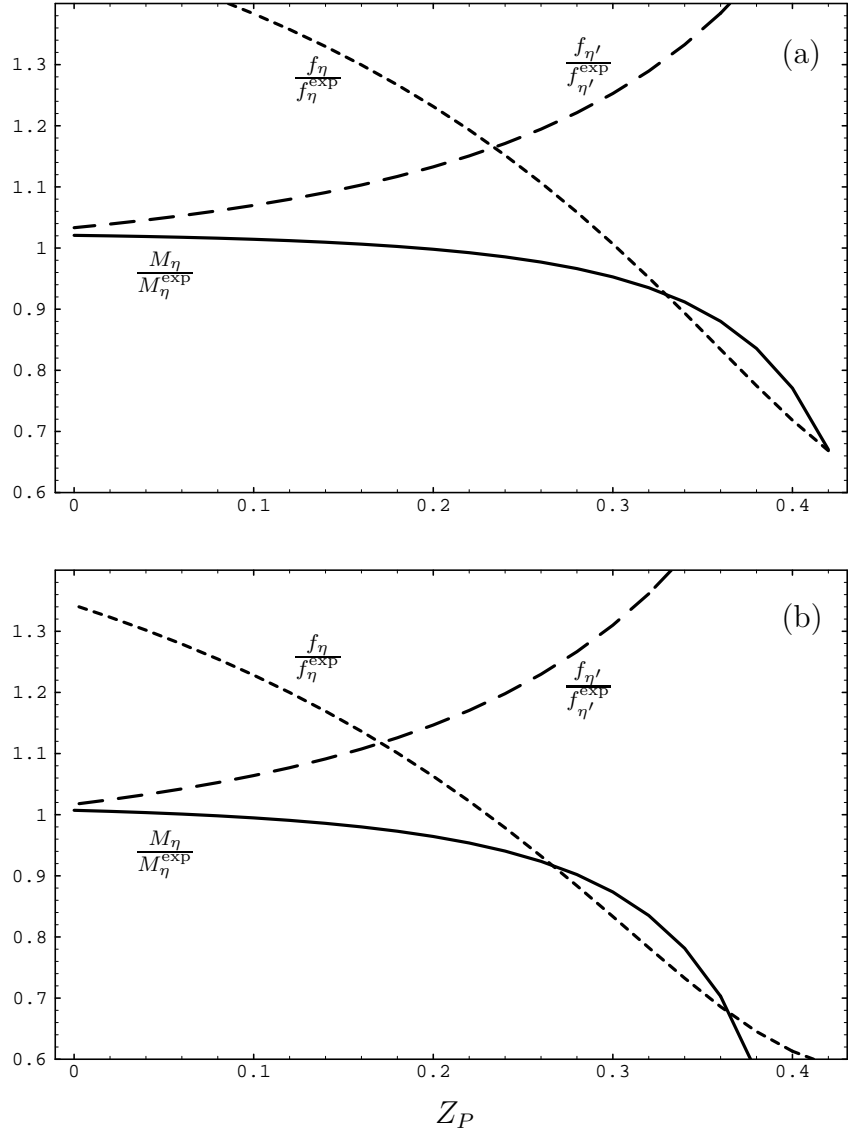}
}
\end{picture}
\caption{\footnotesize The plots show the curves for
  $M_\eta/M_\eta^{\rm exp}$ (solid line), $f_\eta/f_\eta^{\rm exp}$
  (dotted line) and $f_{\eta^\prime}/f_{\eta^\prime}^{\rm exp}$
  (dashed line) in the ``leading mixing approximation'' as functions
  of $Z_P$ for $x_\rho=1$, $Z_p/Z_m=1$ (a) and $Z_p/Z_m=0.9$ (b).}
\label{Plot5} 
\end{figure}

Apparently, the kinetic term $\sim X_\vph^-$ which is induced by the
mixing with the higher $0^{-+}$ multiplet plays a very important role
in our picture. One is tempted to assume that this term dominates the
rich structure in the kinetic terms for the pseudoscalars. The
hypothesis that all deviations from a standard kinetic term for the
pseudoscalars (i.e., eq.~(\ref{LkinForMultiplets}) with $Z_p=Z_m$) are
due to mixing leads to a highly predictive scheme. In the limit where
$X_\vph^-$ is independent of momentum this leads to $Z_p\simeq Z_m$,
$\ol{\omega}_m\simeq\omega_m$. It is far from trivial that for
$Z_p\simeq Z_m$ and $\ol{\omega}_m\simeq\omega_m$ there exists a value
$\omega_m v=-0.20$ for which the quantities $M_\eta$, $f_\eta$ and
$f_{\eta^\prime}$ are all compatible with observation! The
``robustness'' of these ``predictions'' can be estimated from table
\ref{tab6} in sect.~\ref{Results} where we also give numbers for
various values of $Z_p/Z_m$ and $\ol{\omega}_m/\omega_m$. A more
detailed study of the effect of mixing with the higher $0^{-+}$ states
can be found in appendix \ref{MixingWithOtherStates}. In particular,
the coupling $X_\vph^-$ becomes effectively momentum dependent due to
propagator effects. This leads to
\begin{description}
\item[(i)] a contribution to a $q^4$--kinetic term and therefore to
  $(\ol{\omega}_m-\omega_m)/\omega_m\simeq0.1$ (cf.~(\ref{NNN20})); 
\item[(ii)] an effective momentum dependence of $\hat{\omega}$ which
  gets multiplied by the factor
  $f_\omega=(M_P^2-M_\eta^2)/(M_P^2+q^2)$ with
  $M_P\gta2000\MeV$;
\item[(iii)] similar contributions to $z_p$ and $z_8$, (\ref{TTT17}) in
  (\ref{ZZZ7});
\item[(iv)] a correction to $f_{\eta^\prime}$ (\ref{TTT5}) given by
  (\ref{TTT20});
\item[(v)] an effective $Z_p$ (normalized here for $q_0^2=-M_\eta^2$)
  obeying $Z_p/Z_m=0.99$.
\end{description}
Due to the large uncertainty in the value of $M_P$ relevant in the
$\eta$--$\eta^\prime$ sector we have not included these higher
derivative effects in the figs.~\ref{Plot1}--\ref{Plot2b}. A
quantitative discussion can be found in sections
\ref{HigherDerivativeContributions} (fig.~\ref{Plot5}),
\ref{Results} and appendix \ref{MixingWithOtherStates}.

\sect{Expansion in the chiral condensate}
\label{ExpansionInTheChiralCondensate}

There are various scales characteristic for the amount of spontaneous
chiral symmetry breaking: the chiral condensate
$|\langle\ol{\psi}\psi\rangle |^{1/3}\simeq200\MeV$, the pion decay
constant $f_\pi\simeq90\MeV$ and the constituent quark mass
$m_q\simeq300\MeV$. All these scales are typically smaller than the
characteristic scale for the formation of the mesonic bound states,
$k_\vph\gta600\MeV$ \cite{EW94-1,JW95-1} or typical mesonic mass
scales unrelated to the Goldstone phenomenon --- the latter being
around $1\GeV$ with the lowest one given by\footnote{The $\rho$--meson
  is perhaps somewhat special because of approximate gauge symmetry,
  see appendix \ref{VectorMesonContributions}.} $M_{\rho^0}=770\MeV$.
The question emerges if the typical scale appearing in the parameters
$\ol{\nu}$, $X_\vph^-$, $\ol{\la}_3$, etc.~is larger than
characteristic scales associated with chiral symmetry breaking. Could
it be that chiral symmetry breaking is related to an additional small
parameter leading to a suppression of contributions with high powers
of $\si_0$? The existence of a small parameter associated to $\si_0$
would enhance the predictive power of the linear sigma model since the
terms with lower powers of $\si_0$ would dominate. Together with a
derivative expansion it would allow to classify invariants according
to their dimension with a suppression of higher dimensional operators.

Within the linear sigma model we observe distinct mass scales of very
different origin: Whereas the expectation value $\si_0$ measures the
strength of spontaneous chiral symmetry breaking, the scale $\nu$
indicates the size of the explicit breaking of the axial $U_A(1)$
because of the chiral anomaly. Furthermore, there are hadronic mass
scales which are not directly related to chiral symmetry breaking or
the axial anomaly as, for example, the string tension or the glueball
masses. One expects that the last type of scales dominates the
dimensionfull parameters in the effective potential and the kinetic
terms in the limit of vanishing anomaly and $\sigma_0=0$. Since for
mass terms etc.~$\si_0$ is multiplied by some dimensionless coupling
constant, a typical parameter for testing the convergence of a
$\si_0$--expansion could be $x_\si=\la_2\si_0/\nu$. Using the values
of the parameters determined in section
\ref{ParametersOfTheLinearSigmaModel} yields $x_\si\simeq0.3$ whereas
including the quark mass corrections to the kinetic terms gives
$x_\sigma\simeq0.2$. This seems indeed to allow for the possibility
that an expansion in powers of $\si_0$ does not converge too badly. We
will see that this picture is confirmed by the size of other
dimensionless ratios involving powers of $\si_0$.

In order to make a guess for the size of contributions with higher
powers of $\si_0$ we first note that the smallness of $\si_0$ is
partly due to the smallness of $Z_m$ \cite{JW95-1}. Since the physics
cannot depend on the choice of the scaling for the field $\Phi$, only
ratios which are independent of $Z_m$ can appear in measurable
quantities. This includes combinations like
$\olla_2\olsi_0/\olnu=\la_2\si_0/\nu$,
$\olla_3\olsi_0^3/\olnu=\la_3\si_0^3/\nu$, $X_\vph^-\olsi_0^2/Z_m$,
$U_\vph\olsi_0/Z_m$, $\tilde{V}_\vph\olsi_0^4/Z_m$. The physical
scales hidden in the renormalized parameters can be better appreciated
if we choose (somewhat arbitrarily) a fixed $Z_m^{(0)}$ such that the
dimensionless couplings are of order one, say $Z_m^{(0)}=0.15$ such
that $\olla_2^{(0)}\simeq\la_2/50$. With this scaling of the field one
has $\olsi_0^{(0)}=\si_0( Z_m^{(0)})^{-1/2}\simeq137\MeV$ and
$\olnu^{(0)}=\nu( Z_m^{(0)})^{3/2}\simeq540\MeV$. We will now assume
that dimensionfull parameters like $\olla_3^{(0)}$ are given by powers
of a characteristic scale which we take to be around $700\MeV$,
i.e.~$\olla_3^{(0)}\simeq(700\MeV)^{-2}$,
$X_\vph^{-(0)}\simeq(700\MeV)^{-2}$. This suggests typical ratios
$|\la_3\si_0^3/\nu|\simeq0.015$, $|\la_3\si_0^2/\la_2|\simeq0.08$,
$|X_\vph^-\olsi_0^2/Z_m|\simeq0.25$, $|U_\vph\olsi_0/Z_m|\simeq1.3$,
$|\tV_\vph\olsi_0^4/Z_m|\simeq0.01$. Obviously, these numbers can only
be used as rough guesses. There may be additional small coefficients
--- this is obviously necessary for the contribution $\sim U_\vph$ if
$Z_p$ is in the vicinity of $Z_m$ (see (\ref{A3}) --- or relatively
large group theoretical or dynamical factors. If the contribution from
$U_\vph$ does not dominate $\om_m$ we may estimate
$|X_\vph^-\olsi_0^2/Z_m|\simeq|\om_m Z_h^{1/2}\olsi_0|\simeq0.5$ (for
$\om_m v=-0.20$) which is somewhat larger but still compatible with
the above guess. We conclude that the $\si_0$--expansion converges at
best slowly.  For low powers of $\si_0$ group theoretical factors or
dynamically small quantities (i.e. $Z_m^{(0)}$, $U_\vph\olsi_0$)
remain very relevant.  Nevertheless, we find it very unlikely that
terms with high powers of $\si_0$ dominate those with low powers. Even
the conservative assumption that terms with high powers of $\si_0$ are
bounded in size by the strength of terms with lower powers has
important implications!

As an example we compare the contributions
$\sim\frac{9}{4}\olsi_0\olla_2$ and $\sim3\olsi_0^3\olla_3$ to the
cubic coupling $\gm_6$ (\ref{Gamma6}) which determines the mass
split within the scalar octet. If we assume that the second term
does not exceed in size the first one we obtain the bound
\begin{equation}
  \label{L11}
  \gm_6<\frac{1}{4}
  \left(18\si_0\la_2-\nu\right)
  \left(\frac{Z_m}{Z_h}\right)^{\frac{3}{2}}\simeq
  17.1\GeV
\end{equation}
where we have used $\la_2=21.3$, $\nu=6447\MeV$ and $Z_h/Z_m=0.35$
(see sect.~\ref{Results}).  With the help of (\ref{L12}) this can be
transformed into a bound for the mass difference between the $K_0^*$
and $a_0$ mesons in the scalar octet
\begin{equation}
  \label{L13}
  \ol{M}_{K_o^*}^2-\ol{M}_{a_o}^2=3\gm_6 v<0.7\GeV^2\; .
\end{equation}
This relatively conservative bound seems to disfavor the interpretation of the
$a_0(980)$ resonance as a member of the same octet as the
$K_0^*(1430)$, since in this case the difference in mass squared would
have to exceed $1\GeV^2$. In simpler words, it seems at first sight
unlikely that a strange quark mass of about $180\MeV$ produces a mass
difference between strange and non--strange scalar mesons of
$450\MeV$. Yet, we notice that (\ref{L13}) is subject to quark mass
corrections from kinetic terms, and we will come back to this issue in
sect.~\ref{MassRelationsForTheScalarOctet}.

Before closing this section let us comment on the question if the
limiting case $\sigma_0\ra0$ can be used as an expansion point within
a generalized class of linear sigma models. (Of course, the sigma
model corresponding to low--energy QCD has a fixed value of
$\sigma_0$.) Let us consider an effective quark--meson theory which is
supposed to be valid for momentum scales below some cutoff (or
compositeness) scale $k_\vph$.  The ``classical action'' of such a
model is parameterized by a potential and, in particular, a mass term
$\ol{m}^2(k_\vph)$. (Quantum fluctuations of modes with momenta
$q^2<k_\vph^2$ change the form of the effective action and lead to an
effective potential as parameterized by (\ref{ExpansionPotential})).
The size of $\si_0$ can be influenced by the meson mass term
$\ol{m}^2(k_\vph)$ at the scale $k_\vph$. If the phase transition
associated to a variation of $\ol{m}^2(k_\vph)$ were of second order
the order parameter $\si_0$ could be arbitrarily small. An expansion
in powers of $\si_0$ would then always be meaningful for small
enough\footnote{At nonvanishing temperature $\sigma_0(T)$ typically
  decreases as $T$ increases. For a second order high temperature
  phase transition one can always expand such a system for a very
  small even though perhaps not arbitrarily small value of $\si_0$. In
  the limit $\si_0\ra0$ one may encounter nonanalyticities associated
  to the critical three dimensional behavior at the transition.}
$\si_0$. For three flavors the anomaly induces a first order
transition and excludes arbitrarily small values of $\si_0$.
Nevertheless, for small $\si_0$ a polynomial expansion of $U(\si)$
should be meaningful and we may stop after the term quartic in $\si$.
In the limit of equal quark masses the potential
(\ref{ExpansionPotential}) gives
\begin{equation}
  \label{L10}
  U=-3m_g^2\si_0^2-
  \hal\nu\si_0^3+
  \frac{9}{2}\la_1\si_0^4+
  \left(3m_g^2+\frac{3}{2}\nu\si_0-
  9\la_1\si_0^2\right)\si^2-
  \nu\si^3+
  \frac{9}{2}\la_1\si^4\; .
\end{equation}
The requirement $U(\si_0)<U(0)$ implies a lower bound
\begin{equation}
  \label{L08}
  \si_0^2>\frac{1}{9\la_1}
  \left(\nu\si_0+6m_g^2\right)
\end{equation}
whereas the positivity of the mass term at $\si_0$ requires
\begin{equation}
  \label{L09}
  \si_0^2>\frac{1}{12\la_1}
  \left(\nu\si_0-2m_g^2\right)\; .
\end{equation}
On the other hand, the dimensionless coupling $\la_1$ is typically
bounded from above as a result of the ``triviality'' of
$\Phi^4$--theory. (More precisely, the infrared interval of allowed
renormalized quartic couplings is bounded.) Comparing (\ref{L09}) with
the definition of $x_\si$ we find that for vanishing quark masses
($m_g^2=0$) the expansion coefficient must obey
$x_\si>\frac{\la_2}{12\la_1}$ and can therefore not be arbitrarily
small. Despite this caveat there seems to be enough room for a
meaningful $\si_0$--expansion. It is interesting to note that for
given $\sigma_0$ the inequalities (\ref{L08}) and (\ref{L09}) can also
be used to establish lower bounds for $\lambda_1$ which hold for a
polynomial approximation (\ref{L10}). Taking $\nu$, $m_g^2$ and
$\sigma_0$ from section (\ref{Results}) one finds $\lambda_1>48.9$ and
$\lambda_1>1.1$, respectively.

\sect{Mass relations for the scalar octet}
\label{MassRelationsForTheScalarOctet}

In this section we want to determine the masses of the members of the
$0^{++}$ octet contained in $\Phi$. Together with the $0^{++}$ singlet
these scalars play for spontaneous chiral symmetry breaking the same
role as the Higgs scalar in the electroweak theory. Since the isospin
singlet member of the octet has the same quantum numbers as a possible
scalar glueball the determination of its mass is also important for
the identification of glueball state candidates. We restrict most of
the discussion to the linear order in an expansion in powers of quark
masses and we neglect isospin violation. The quark mass corrections to
the kinetic terms for the scalar octet to linear order in $\Dt$ arise
{}from an interaction analogous to (\ref{LkinCorrections})
\begin{equation}
  \label{NNN4}
  \Lc_{\rm kin}^{(1,s)}=
  \frac{1}{4}\om_h\Tr h\prl^\mu h\prl_\mu h\; .
\end{equation}
Since in the linear sigma model several of the generalized kinetic
terms (\ref{Lkin}) contribute to this invariant we will treat
$\omega_h v$ as a free parameter to order $\Delta$.  In addition, we
will consider a particular term contributing to second order in
$\Delta$
\begin{equation}
  \label{OPQ1}
  \Lc_{\rm kin}^{(2,s)}=
  \frac{1}{4}\zeta_h\Tr\left\{\left(
  h\prl^\mu h-\prl^\mu h h\right)\left(
  h\prl_\mu h-\prl_\mu h h\right)\right\}\; .
\end{equation}
This term is induced by the exchange of the scalar state contained in
the divergence of the vector meson field $\prl_\mu\rho_V^\mu$
(cf.~appendix \ref{VectorMesonContributions}) with a sizeable
coefficient $\zeta_h v^2\gta0.02$. Combining (\ref{NNN4}) and
(\ref{OPQ1}) and neglecting terms $\sim\omega_h\zeta_h v^3$ this leads
to different wave function renormalizations for $a_0$, $K^*$ and $f_8$
\begin{equation}\begin{array}{rcl}
 \ds{Z_{a_o}} &=& \ds{1-\om_h v}\nnn
 \ds{Z_{K_o^*}} &=& \ds{
 \left(1+\hal\om_h v\right)
 \left(1-9\zeta_h v^2\right)}\nnn
 \ds{Z_{f8}} &=& \ds{1+\om_h v}\; .
 \label{NNN5}
\end{array}\end{equation}
We next have to include the effects of higher derivative terms using
the definition of the wave function renormalization constants
(\ref{DefinitionOfZi}). The discussion is completely analogous to
sect.~\ref{QuarkMassCorrectionsToKineticTerms} and the dominant
higher derivative terms lead to a replacement of $\om_h$ by
$\ol{\om}_h=\om_h+2\Dt Z_{K_o^*}/v$. (Here we have adopted a
definition of $Z_h$ such that $2Z_{K_o^*}+Z_{a_o}=1$ for $\zeta_h=0$.)
We note that the effects from a nonvanishing $\zeta_h$ can be absorbed
in an effective mass term
\begin{equation}
  \label{OPQ2}
  \hat{M}_{K_o^*}^2=M_{K_o^*}^2
  \left(1-9\zeta_h v^2\right)\; .
\end{equation}
Up to the replacement of the physical mass $M_{K_o^*}^2$ by the
$\zeta_h$ dependent quantity $\hat{M}_{K_o^*}^2$ our discussion
systematically only includes terms linear\footnote{The exception from
  a systematic procedure of keeping only terms in the effective action
  for scalars that contribute to linear order in $\Delta$ is motivated
  by the well identified mechanism that induces a sizeable $\zeta_h
  v^2$ (cf.~appendix \ref{VectorMesonContributions}).} in $\Delta$.

{}From (\ref{Mh2}) we now obtain
\begin{equation}
  \label{NNN10}
  m_h^2=\frac{1}{3}
  \left[ 2\ol{M}_{K_o^*}^2+
  \ol{M}_{a_o}^2\right]=
  \frac{1}{3}
  \left[2\hat{M}_{K_o^*}^2+
  M_{a_o}^2+
  \ol{\om}_h v
  \left(\hat{M}_{K_o^*}^2-M_{a_o}^2\right)\right]\; .
\end{equation}
Hence, the linear quark mass corrections to the kinetic terms
$\sim\ol{\omega}_h$ modify $m_h^2$ only to quadratic order in $\Dt$.
For $M_{a_o}=1320(983)\MeV$, $\ol{\omega}_h=0$ and $\zeta_h=0$ we
find\footnote{A similar observation holds in the pseudoscalar sector
  for $m_g^2$ and we obtain $m_g\simeq393\MeV$.}
$m_h\simeq1394(1298)\MeV$.  The dominant correction to the lowest
order relation is most likely due to the term $\sim\zeta_h$. For
$\zeta_h v^2=0.02$ one obtains $m_h=1303(1200)\MeV$ whereas $\zeta_h
v^2=0.04$ yields already a large shift to $m_h=1206(1093)\MeV$.  We
are now in a position to compute $Z_h/Z_m$ from (\ref{NNN11}). For
$\ol{\om}_m v=-0.20$ we obtain
\begin{equation}
  \label{NNN12}
  \frac{Z_h}{Z_m}\simeq0.39(0.45)\;,\;\;\;
  0.45(0.53)\; ,\;\;\;
  0.53(0.64)
\end{equation}
where the three values correspond to $\zeta_h v^2=0,0.02,0.04$.
Comparing with (\ref{Zmh}) we see that the quark mass corrections to
the kinetic terms strongly influence the determination of $Z_h/Z_m$.
We conclude that $1-\frac{Z_h}{Z_m}$ can not be treated as a very
small number. The difference between $Z_h$ and $Z_m$ has to be
included for any systematic discussion of the scalars within the
linear sigma model! The neglect of this difference in earlier works
\cite{PW84-1}--\cite{Pis95-1} partially explains the quantitative
differences with our results.

\begin{figure}
\unitlength1.0cm
\begin{picture}(13.,9.)
\put(.7,5.){\bf $M_{a_o}$}
\put(8.,0.5){\bf $\ol{\om}_h v$}
\put(-0.8,-11.5){
\epsfysize=22.cm
\epsffile{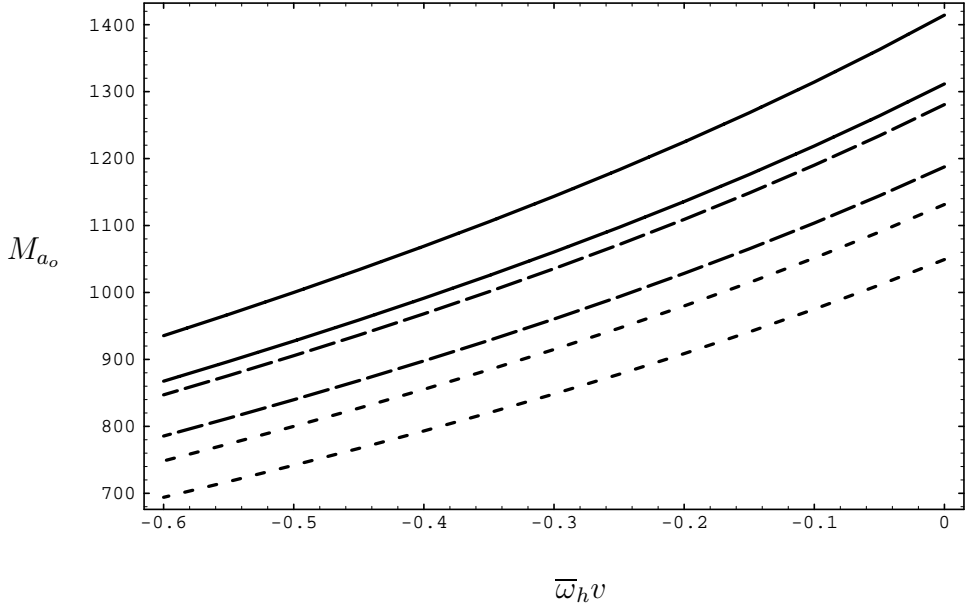}
}
\end{picture}
\caption{\footnotesize The plot shows
  $M_{a_o}$ as a function of $\ol{\om}_h v$ for $Z_p/Z_m=0.9$,
  $\omega_m v=\ol{\omega}_m v=-0.20$ and three values of $\zeta_h
  v^2=0$ (solid lines), $0.02$ (dashed lines) and $0.04$ (dotted
  lines). The upper curve in each band corresponds to
  $\sigma_0^2\lambda_3=-\lambda_2/4$ and the lower one to
  $\sigma_0^2\lambda_3=\lambda_2/4$.}
\label{Plot4} 
\end{figure}

We are now ready to reexamine the mass splitting in the scalar octet
(\ref{L13}). Including the corrections arising from (\ref{NNN5}) we
find
\begin{equation}
  \label{NNN31}
  \hat{M}_{K_o^*}^2-M_{a_o}^2=
  \frac{3\gamma_6 v-\frac{3}{2}
  m_h^2\ol{\om}_h v}
  {1-\frac{1}{2}\ol{\om}_h v-
  \frac{1}{2}(\ol{\om}_h v)^2}\; .
\end{equation}
One sees that large negative values of $\ol{\om}_h v$ can considerably
increase the mass difference between the $K_0^*$ and the $a_0$ for
given $\gamma_6 v$. The same holds for $\zeta_h v^2>0$. This weakens
the argument of the preceding section against the association of the
$a_0$ meson with the resonance $a_0(980)$. In fig.~\ref{Plot4} we
plot $M_{a_o}$ as a function of $\ol{\om}_h v$ for $Z_p=Z_m$ and three
different values of $\zeta_h v^2$, with bands corresponding to ranges
of $\lambda_3$ between $\sigma_0^2\lambda_3=-\lambda_2/4$ (lower
curves) and $\sigma_0^2\lambda_3=\lambda_2/4$ (upper curves).
Here we have used (\ref{NNN14}) together with
\begin{equation}
  \label{NNN15}
  \frac{\gamma_6}{\gamma_2}=
  \hal\frac{\hat{M}_{K_o^*}^2-M_{a_o}^2+
  \ol{\om}_h v\left(\hal
  \hat{M}_{K_o^*}^2+M_{a_o}^2\right)}
 {M_K^2-M_\pi^2+
  \ol{\om}_m v\left(
  \hal M_K^2+M_\pi^2\right)}\; .
\end{equation}
In fact, large quark mass corrections to the scalar kinetic terms seem
the only plausible possibility for the choice of the $a_0(980)$. This
would indicate a large mixing between two--kaon states and the
$a_0(980)$ (see appendix \ref{MixingWithOtherStates}). Consequently,
it could explain why the $a_0(980)$ behaves in many respects similarly
to a $\ol{q}q\ol{q}q$ state even though it may belong to an octet of
$\ol{q}q$ states. We should also point out that for large values of
$|\ol{\om}_h v|$ the quark mass expansion becomes questionable in the
scalar sector. From the linearized expressions
\begin{eqnarray}
  \label{NNN33}
  \ds{M_{a_o}^2} &=& \ds{
  m_h^2-2\left(\gamma_6 v-
  \hal m_h^2\ol{\om}_h v\right)}\nnn
  \ds{\hat{M}_{K_o^*}^2} &=& \ds{
  m_h^2+\left(\gamma_6 v-
  \hal m_h^2\ol{\om}_h v\right)}
\end{eqnarray}
we infer the ratio of the first order correction for $M_{a_o}^2$ as
compared to
$m_h^2$
\begin{equation}
  \label{NNN34}
  \frac{2\left(\gamma_6 v-
  \hal m_h^2\ol{\om}_h v\right)}
  {m_h^2}=
  \frac{2}{3}\frac{\hat{M}_{K_o^*}^2-M_{a_o}^2}
  {m_h^2}\simeq 0.10(0.43)\; ,\;\;\;
  -0.03(0.33)\; ,\;\;\;
  -0.20(0.19)
\end{equation}
for $\zeta_h v^2=0,0.02,0.04$. Apparently, for $\zeta_h v^2=0$ a good
convergence of the quark mass expansion is only realized for the
assignment $a_0(1320)$. 
For larger values of $\zeta_h v^2$ as infered from the leading mixing
approximation in appendix \ref{VectorMesonContributions} a reasonable
convergence can also be obtained for $a_0(980)$.

Summarizing the various aspects of the problem of the correct
assignment of the isotriplet belonging to the same octet as the
$K_0^*(1430)$ we may state that the association $a_0(1320)$ would make
the understanding easier only in case of a standard kinetic term for
$h$. Taking into account nonminimal kinetic terms there is no
conclusive argument to rule out the $a_0(980)$ as a member of the
scalar octet.  For the latter assignment one expects important mixing
effects with two--kaon states.  Actually, such large mixing effects
concern presumably only the $a_0$ and not the other members of the
scalar octet. It may therefore be preferable not to include these
mixing effects into the parameter $\omega_h v$ appearing in
(\ref{NNN5}) but to treat them as additional corrections to the $a_0$
propagator only.  In this case the size of $\omega_h v$ is expected to
remain small, $|\omega_h v|\lta0.1$, but the physical mass of the
$a_0$ is related to $\ol{M}_{a_0}$ by an unknown factor reflecting the
mixing. The value of $M_{a_o}$ appearing in formulae like
(\ref{NNN10}) or (\ref{NNN7}) below should then be replaced by an
effective mass $\hat{M}_{a_o}$ somewhat above $1\GeV$ (say around
$1100\MeV$). A natural mechanism of ``threshold mass shifting''
leading to a physical mass $M_{a_o}=980\MeV$ is described in appendix
\ref{MixingWithOtherStates}.

\begin{figure}
\unitlength1.0cm
\begin{picture}(13.,9.)
\put(.7,5.){\bf $\ds{\frac{M_f}{\MeV}}$}
\put(7.5,0.5){\bf $\ds{\frac{M_{a_o}}{\MeV}}$}
\put(-0.8,-11.5){
\epsfysize=22.cm
\epsffile{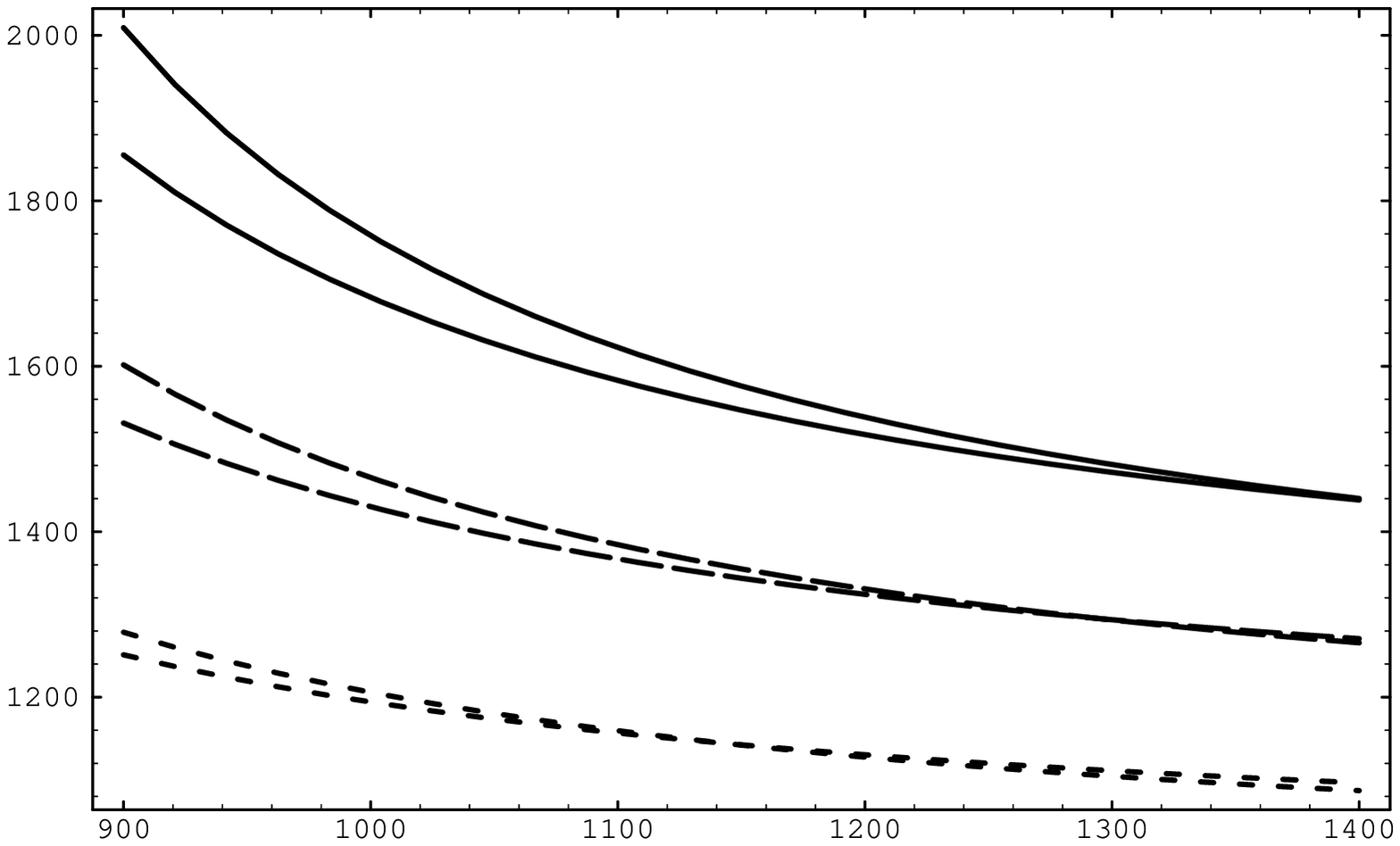}
}
\end{picture}
\caption{\footnotesize The plot shows $M_f$ as a function of $M_{a_o}$
  according to the scalar Gell-Mann--Okubo relation (\ref{NNN2}) with
  (\ref{NNN5}), $Z_p/Z_m=0.9$, $\omega_m v=\ol{\omega}_m v=-0.20$ and
  fixed $M_{K_o^*}=1430\MeV$. The bands correspond to values of
  $\lambda_3\sigma_0^2$ between $-\lambda_2/4$ (upper curves) and
  $\lambda_2/4$ (lower curves) and we give results for $\zeta_h
  v^2=0$ (solid lines), $0.02$ (dashed lines) and $0.04$ (dotted
  lines).}
\label{Plot3} 
\end{figure}

We finally want to show that the Gell-Mann--Okubo type mass relation
(\ref{NNN2}) is not affected by linear quark mass corrections to the
kinetic terms. Inserting $M_{a_o}^2=\ol{M}_{a_o}^2 Z_{a_o}^{-1}$,
$M_{K_o^*}^2=\ol{M}_{K_o^*}^2 Z_{K_o^*}^{-1}$,
$M_{f_8}^2=\ol{M}_{f8}^2 Z_{f8}^{-1}$ into (\ref{NNN2}) one obtains to
linear order in $\ol{\om}_h v$ the relation
\begin{equation}
  \label{NNN7}
  M_{f_8}^2=\frac{4}{3}\hat{M}_{K_o^*}^2-
  \frac{1}{3}M_{a_o}^2-
  \frac{2}{3}\ol{\om}_h v
  \left(\hat{M}_{K_o^*}^2-M_{a_o}^2\right)\; .
\end{equation}
The correction to the relation (\ref{NNN2}) is indeed only quadratic
in $\Dt$. For given $M_{a_o}=1320\MeV$ and $M_{K_o^*}=1430\MeV$ the
symmetry relation (\ref{NNN7}) yields for $\zeta_h v^2=0$
\begin{equation}
  \label{NNN8}
  M_{f_8}\simeq 1465\MeV\; .
\end{equation}
On the other hand, for $\zeta_h v^2=0.02$ and $\hat{M}_{a_o}=1100\MeV$
we find
\begin{equation}
  \label{OPE12}
  M_{f_8}\simeq1354\MeV\; .
\end{equation}
Taking into account the uncertainties from the mixing with the scalar
singlet $s$ both values are consistent with the observed broad
resonance\footnote{The mass of the $f_0(1300)$ is not determined very
  precisely. It could easily be around $1400\MeV$.} $f_0(1300)$.  In
fig.~\ref{Plot3} we have plotted $M_{f_8}$ as a function\footnote{In
  figs.~\ref{Plot4} and \ref{Plot3} we have not distinguished between
  $M_{a_o}$ and $\hat{M}_{a_o}$. Taking into account the additional
  mixing with two--kaon states for the $a_0$ the relevant axis
  actually shows $\hat{M}_{a_o}$.} of $\hat{M}_{a_o}$ in order to
demonstrate the relative insensitivity of $M_{f_8}$ on the
identification of the $a_0$ meson.
For this purpose we have used $\zeta_h=0,0.02$ and $0.04$ and
$\lambda_3\sigma_0^2=\pm\lambda_2/4$. For each set of parameters we
have determined $\ol{\omega}_h v$ from fig.~\ref{Plot4}. Figure
\ref{Plot3} demonstrates that for $\zeta_h v^2\gta0.02$ values of
$M_{f_8}$ below $1500\MeV$ are preferred. On the other hand, a value
of $M_{f8}$ below $1100\MeV$ would require a very substantial mixing
of $K_0^*$ with a state in $\prl_\mu\rho_V^\mu$ ($\zeta_h
v^2\gta0.04$.). The identification of either the $f_0(1590)$
or the $f_0(980)$ with the octet seems therefore disfavored.

In this context it is perhaps interesting to note that the
branching ratio \cite{PDG94-2}
\begin{equation}
  \label{OPE14}
  R=\frac{\Gamma\left(f_0(1300)\ra2\pi\right)}
  {\Gamma\left(f_0(1300)\ra K\ol{K}\right)}\simeq12.5
\end{equation}
is consistent with an octet assignment of the $f_0(1300)$. For a pure
octet this ratio should be around $6$ and a relatively small admixture
of a singlet state could easily explain a further enhancement. (Whereas
a pure singlet state would lead to a much smaller ratio $R\simeq1.5$ a
second solution for large $R$ corresponds to a large octet--singlet
mixing angle, i.e.~$\tan\vth_s\simeq-0.9$ for equal $s m^2$ and $h
m^2$ couplings.) On the other hand, the branching ratios of the higher
mass resonance $f_0(1590)$ seem compatible with a singlet with large
cubic coupling $\sim p^2 s$, but not with an octet. For the $f_0(980)$
an assignment is difficult in view of the presumably large mixing with
two--kaon states.

In summary, two natural scenarios for the scalar nonet seem to be
compatible with the parameters of the linear meson model extracted
{}from the pseudoscalar sector: For one scenario the isotriplet
corresponds to the $a_0(980)$ and the singlet (or dominant $\ol{s}s$
state in case of large $|\vth_s|$) is associated with the $f_0(980)$.
These four states are largely influenced by mixing with two--kaon
states. The other $f_0$ state of the nonet (dominantly an octet state
in case of small $|\vth_s|$) corresponds to the $f_0(1300)$. In this
case a relatively large mixing in the strange sector with
$\prl_\mu\rho_V^\mu$ (large $\zeta_h v^2$) explains why the
$K_0^*(1430)$ has the highest mass in the nonet. The other scenario
has a larger average mass $m_h$ of the octet. The triplet is
associated with the proposed $a_0(1320)$ which is not far below the
doublets $K_0^*(1430)$. The $a_0(980)$ and $f_0(980)$ are dominantly
four--quark states or $K\ol{K}$ molecules in this case. Again, the
octet state is the $f_0(1300)$. The singlet corresponds\footnote{There
  are other not so well established resonances $f_0(1510)$ and
  $f_0(1525)$ \cite{PDG94-1} which may be identical with the
  $f_0(1590)$ or also be possible candidates for the singlet.} either
to $f_0(1590)$ or its width is too large to be detected.  Reliable
information about the value of $\lambda_1$ would certainly be of great
interest for further pinning down the possible options.

\sect{Higher derivative contributions}
\label{HigherDerivativeContributions}

In this section we investigate deviations of the meson propagator from
the approximated form $G_i=(Z_i q^2+\ol{M}_i^2)^{-1}$. We will first
concentrate on the flavored pseudoscalars. In the language of
sect.~\ref{QuarkMassCorrectionsToKineticTerms} we want to make an
estimate of the corrections $\Dt Z_i$ (\ref{NNN16}).  Within a
systematic quark mass expansion we need to order $\Dt^2$ only the
$q^4$ correction to the inverse propagator in an approximation where
it is independent of the quark masses. This correction arises from a
term involving four derivatives
\begin{equation}
  \label{NNN17}
  \Lc_{\rm kin(4)}=
  \frac{\ol{H}_m}{4}
  \Tr\left(\prl^2 m\prl^2 m-
  q_0^2\prl^\mu m\prl_\mu m\right)
\end{equation}
where $q_0^2$ is chosen according to (\ref{AP1}). It involves one
additional parameter $\ol{H}_m$ which determines the ratio
\begin{equation}
  \label{NNN18}
  \frac{\ol{\om}_m}{\om_m}=
  1+\delta_\omega=
  1-\frac{2}{3}\ol{H}_m
  \frac{M_K^2-M_\pi^2}{\om_m v}\; .
\end{equation}

As a first observation we notice that $\ol{H}_m$ receives
contributions from the mixing of the pseudoscalar mesons with other
states. We infer from appendix \ref{MixingWithOtherStates} that the
mixing with other $0^{-+}$ octets indeed induces higher derivative
corrections because of the momentum dependence of the propagators of
the additional states that are integrated out. If we assume that the
dominant contribution to $\om_m$ arises through mixing with other
states we can identify in (\ref{NNN19}) $\om_m$ with $\om_m^{(\rho)}$
and find
\begin{equation}
  \label{NNN20}
  \delta_\omega^{(\rho)}=
  \frac{1}{6}\frac{(M_K^2-M_\pi^2)}
  {(M_P^2-m_m^2)}
  \frac{[2\ol{f}_K+\ol{f}_\pi]}
  {[\ol{f}_K-\ol{f}_\pi]}\simeq0.07(0.05)
\end{equation}
where we used $M_P=2280\MeV(2670\MeV)$ (cf.~appendix
\ref{MixingWithOtherStates} and table \ref{TTA1} in appendix
\ref{VectorMesonContributions}).
One may also estimate $K_8$ according to (\ref{AP1}) and finds
\begin{equation}
  \label{OOJ2}
  K_8^{(\rho)}\simeq0.002(0.001)\; .
\end{equation}
This is indeed negligible for the wave function renormalizations as
compared to $\ol{\omega}_m v$.

\begin{figure}
\unitlength1.0cm
\begin{picture}(8.,4.)
\put(5.3,2.0){$m_i(q)$}
\put(10.2,2.0){$m_i(q)$}
\put(8.0,2.8){$m_j$}
\put(8.05,0.5){$h_k$}
\put(7.2,1.7){$\gamma_2$}
\put(8.6,1.7){$\gamma_2$}
\put(3.3,1.7){$\Sigma_i(q^2)\simeq$}
\put(5.2,0.5){
\epsfysize=2.5cm
\epsfxsize=5.5cm
\epsffile{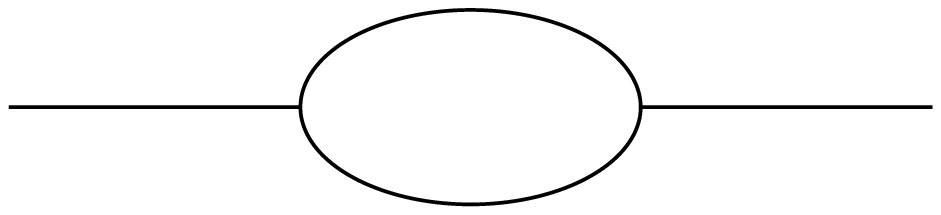}
}
\end{picture}
\caption{\footnotesize One--loop diagram for the dominant contribution
  to $\Sigma_i(q^2)$.}
\label{Feyn} 
\end{figure}

A different contribution to the higher derivative term (\ref{NNN17})
arises from loops of meson fluctuations. For an estimate of their
importance we use the modified loop expansion of \cite{Wet96-1}
(``systematically resummed perturbation theory''). This allows to
compute the deviations of the inverse propagator from the form
$q^2+M_i^2$ in terms of $M_i^2$ and effective $1PI$ cubic vertices.
It is crucial in our context that instead of ``classical vertices''
only the $1PI$ Green functions appear in the perturbative series since
only the latter are directly calculable from the present
phenomenological analysis. Also the loop expansion is only used for
``higher order couplings'' where it converges reasonably well. (We do
not expect a good convergence for quantities like $Z_h/Z_m$
etc.) Let us write the inverse propagator for a member of the
pseudoscalar octet as
\begin{eqnarray}
  \label{KKK0}
  \ds{G_i^{-1}(q^2)} &=& \ds{
  q^2+M_i^2+
  \tilde{\Sigma}_i(q^2)}\; .
\end{eqnarray}
Here we have subtracted from the usual self energy $\Sigma_i(q^2)$ those
parts which are already contained in the effective wave function
renormalizations $\ol{Z}_i$ and mass terms $\ol{M}_i^2$
\begin{equation}
  \label{XXXU0}
  \tilde{\Sigma}_i(q^2)=
  \Sigma_i(q^2)-\Sigma_i(0)-
  \frac{q^2}{q_0^2}
  \left(\Sigma_i(q_0^2)-\Sigma_i(0)\right)\; .
\end{equation}
This definition implies
\begin{equation}
  \label{XXXU1}
  \tilde{\Sigma}_i(0)=\tilde{\Sigma}_i(q_0^2)=0
\end{equation}
and we use $q_0^2=-m_m^2$. In consequence, $\tilde{\Sigma}_i(q^2)$
contains only contributions to higher derivative terms. The dominant
one--loop contribution to $\Sigma_m(q^2)$ for the pseudoscalar octet
is depicted in fig.~\ref{Feyn}. 
It involves the propagation of a
scalar and a pseudoscalar in the loop and we therefore need the cubic
coupling $\gamma_2$ (cf.~section
\ref{ScalarMesonMassesToLinearOrder}). We observe that the subtraction
(\ref{XXXU0}) makes the usual one--loop expression ultraviolet
finite. In fact, if the momentum dependence of the effective
three--point vertex is not too strong the momentum integral for the
difference $\Sigma_m(q^2)-\Sigma_m(0)$ is dominated by momenta in the
range between the masses of the two particles propagating in the
loop. We will use here the approximation of a constant cubic vertex
which we approximate by its value to lowest order in the quark mass
expansion given by $\gamma_2$. We also neglect the octet mass
splitting for the particles propagating in the loop for which we use
average masses $m_m=412\MeV$ and $m_h=1394\MeV$. To lowest order in
the quark mass expansion we are interested in the effective coupling
$\ol{H}_m$ (\ref{NNN24}) which is given by
\begin{eqnarray}
  \label{KKK1}
  \ds{\ol{H}_m} &=& \ds{
  \frac{\prl}{\prl q^2}
  \left(\frac{\tilde{\Sigma}_m}{q^2}
  \right)_{q^2=q_0^2} }\; .
\end{eqnarray}
We are interested in the momentum range $-q^2<(m_h-m_m)^2$ for which
the one--loop contribution is given by
\begin{eqnarray}
  \label{KKK2}
  \ds{ \Sigma_m^{(1)}(q^2)-
  \Sigma_m^{(1)}(0)} 
  &\simeq& \ds{
  \frac{5\gamma_2^2}{48\pi^2}
  \Bigg\{ 1-
  \left[\frac{m_m^2-m_h^2}{q^2}+
  \frac{m_m^2+m_h^2}{m_m^2-m_h^2}\right]
  \ln\frac{m_h}{m_m}}\nnn
  &-& \ds{
  \frac{1}{q^2}\sqrt{(m_m+m_h)^2+q^2}
  \sqrt{(m_m-m_h)^2+q^2} }\\[2mm]
  &\times& \ds{
  \ln\frac{\sqrt{(m_m+m_h)^2+q^2}+\sqrt{(m_m-m_h)^2+q^2}}
  {\sqrt{(m_m+m_h)^2+q^2}-\sqrt{(m_m-m_h)^2+q^2}}
  \Bigg\}\; . }\nonumber
\end{eqnarray}
Here we have neglected contributions $\sim\gamma_1^2,\gamma_3^2$ as
well as the $\eta$--$\eta^\prime$ mixing. For $q_0^2=-m_m^2$ this
yields the one--loop contribution to $\ol{H}_m$
\begin{eqnarray}
  \label{KKK3}
  \ds{\ol{H}_m^{(1)}} &\simeq& \ds{
  -\frac{5}{48\pi^2}
  \frac{\gamma_2^2}{m_m^4}\Bigg\{
  2-\frac{2m_h^4-5m_h^2 m_m^2+m_m^4}
  {m_m^2(m_h^2-m_m^2)}
  \ln\frac{m_h}{m_m}}\\[2mm]
  &+& \ds{
  m_h\sqrt{m_h^2-4m_m^2}
  \left(\frac{2}{m_m^2}+
  \frac{1}{m_h^2-4m_m^2}\right)
  \ln\frac{\sqrt{m_h+2m_m}+\sqrt{m_h-2m_m}}
  {\sqrt{m_h+2m_m}-\sqrt{m_h-2m_m}}
  \Bigg\} \nonumber}\; .
\end{eqnarray}
Using $\gamma_2\simeq8000\MeV$ and $\omega_m v\simeq-0.20$ (see
sect.~\ref{Results}) this results in
\begin{equation}
  \label{KKK5}
  \ol{H}_m^{(1)}\simeq1.64\cdot10^{-8}\MeV^{-2}
\end{equation}
or
\begin{equation}
  \label{KKK6}
  \delta_\omega^{(1)}\simeq0.012\; .
\end{equation}
We see that loop corrections to the higher derivative terms are
negligible as compared to contributions arising through the mixing
with other states. We will therefore assume that the higher derivative
terms are dominated by such mixings and estimate
\begin{equation}
  \label{OOJ1}
  \frac{\omega_m}{\ol{\omega}_m}=
  \frac{1}{1+\delta_\omega^{(\rho)}}\simeq0.95\; .
\end{equation}
Even though $(\ol{\om}_m-\om_m)/\om_m$ is formally not
suppressed by powers of $\Dt$ we see that this higher derivative
effect is actually small.

In the $\eta$--$\eta^\prime$ sector we need information about the
momentum dependence of $z_8(q^2)$, $z_p(q^2)$ and $\hat{\omega}(q^2)$
(\ref{ZZZ7}). The relevant quantities are
$d_8=z_8(-M_{\eta^\prime}^2)-1$, $d_p=z_p(-M_{\eta^\prime}^2)-1$ and
$d_\omega=(\hat{\omega}(-M_{\eta^\prime}^2)-\hat{\omega})/\hat{\omega}$
where we remind the reader of the definitions
$z_8(-M_\eta^2)=z_p(-M_\eta^2)=1$ and
$\hat{\omega}(-M_\eta^2)=\hat{\omega}$. Since the mass difference
$M_{\eta^\prime}^2-M_\eta^2$ exceeds substantially $M_K^2-M_\pi^2$ the
higher derivative corrections in the $\eta$--$\eta^\prime$ sector
could be somewhat larger than for the flavored mesons. Altogether,
the contributions beyond the lowest order in the derivative expansion
involve four additional dimensionless parameters, $\delta_\omega$,
$d_8$, $d_p$ and $d_\omega$. Their absolute size is expected to be
small if the derivative expansion converges. Since the predictions of
the lowest order in the derivative expansion come already very close
to the experimental values of $f_\eta$ and $f_{\eta^\prime}$ it seems
not difficult to achieve agreement with observation by using small but
otherwise arbitrary values for these four parameters. The expansion in
powers of $\sigma_0$ may lead to some approximate relations between
$\delta_\omega$, $d_8$, $d_p$ and $d_\omega$ but we will not pursue
this issue here further.

Instead, we conclude this section by a description of the predictions
of the ``leading mixing approximation''. For this purpose we assume
that all corrections to the kinetic terms --- both, quark mass and
higher derivative corrections --- are due to a mixing with higher
states contained in the divergence of the axialvector field
$\prl_\mu\rho_A^\mu$. The formalism is described in appendices
\ref{VectorMesonContributions} and \ref{MixingWithOtherStates}. All
parameters appearing in the kinetic terms can be computed in terms of
masses and couplings of the vector-- and axialvector fields. Most of
these couplings can be determined from observation (cf.~appendix
\ref{VectorMesonContributions}). There remains essentially only one
important free parameter $Z_P$ which appears in the term
$(1/4)\Tr\left\{\tilde{Z}_P(\prl_\mu\rho_V^\mu)^2+
Z_P(\prl_\mu\rho_A^\mu)^2\right\}$. This parameter determines the
strength of the higher derivative terms induced by the mixing with
$M_P^2\sim Z_P^{-1}$ in (\ref{NNN20}). In fig.~\ref{Plot5}a we plot
the values of $M_\eta/M_\eta^{\rm exp}$, $f_\eta/f_\eta^{\rm exp}$ and
$f_{\eta^\prime}/f_{\eta^\prime}^{\rm exp}$ as functions of $Z_P$.
(The higher derivative corrections vanish for $Z_P=0$.)
For this plot we use $x_\rho=1$ (cf.~appendix
\ref{VectorMesonContributions}) and employ the leading mixing
approximation which assumes $\omega_m=\omega_m^{(\rho)}$ and
$\ol{\omega}_m/\omega_m$ given by (\ref{NNN20}), (\ref{KKK6}).  The
higher derivative contributions in (\ref{ZZZ7}) and (\ref{TTT5}) are
now included according to (\ref{TTT12}), (\ref{TTT17}) and
(\ref{TTT20}). We see that a reasonable picture emerges for
$Z_P\simeq0.22$. It is consistent with the assumption that the
nonminimal kinetic terms for the pseudoscalars are dominated by the
mixing with $\prl_\mu\rho_A^\mu$ or the ``partial Higgs effect''. The
remaining differences of the curves from one can reasonably be
attributed to subleading effects beyond the leading mixing
approximation, as described in the more general framework of the main
text. For a demonstration we also show in fig.~\ref{Plot5}b the
situation which arises if in addition to the contributions from the
leading mixing approximation one also includes a nonvanishing $U_\vph$
in the kinetic terms (\ref{Lkin}). For this plot we have chosen
$U_\vph$ such that $Z_p/Z_m=0.9$ (cf.~(\ref{A3}). The agreement with
observation improves and the optimal value of $Z_P$ is shifted to
somewhat smaller values. We emphasize that the leading mixing
approximation is complementary to the formal expansion in powers of
quark masses. It is encouraging that a simple mechanism (the partial
Higgs effect) can apparently explain the dominant parts of the
parameters appearing in the systematic quark mass expansion.

\sect{Results}
\label{Results}

For the convenience of the reader we summarize in this section the
results of a numerical solution of our equations. We observe that the
pseudoscalar sector can be treated independently from the scalar
sector. For the flavored pseudoscalars there are two small parameters
whose influence is rather modest, namely $Z_p/Z_m-1$ and
$\delta_\omega=\ol{\omega}_m/\omega_m-1$. Three additional small
parameters $d_\omega$, $d_8$, $d_p$ characterize the most general form
of the higher derivative terms in the $\eta$--$\eta^\prime$ sector.
For the first two lines $(a)$ and $(b)$ in our tables we use the first
order in the derivative expansion,
i.e.~$\delta_\omega=d_\omega=d_8=d_p=0$.  We present two values
$Z_p/Z_m=1.0$ and $0.9$. The value of the dominant free parameter
$\omega_m v$ is chosen such that $M_\eta$ comes out close to its
experimental value (cf.~fig.~\ref{Plot1}).  Going beyond the first
order in the derivative expansion we include in line $(c)$ the higher
derivative corrections corresponding to a nonvanishing
$\delta_\omega$. It is taken according to the leading mixing estimate
(\ref{OOJ1}) such that $\omega_m/\ol{\omega}_m=0.9$.  Finally, we
present the results of the leading mixing approximation in line $(d)$.
Here all quark mass corrections to the kinetic and all higher
derivative terms are determined from the simple assumption that they
are induced by the exchange of the axialvector field
$\prl_\mu\rho_A^\mu$. Line $(d)$ corresponds to fig.~\ref{Plot5}b with
$Z_P\simeq0.16$. In the leading mixing approximation $\omega_m v$ and
$\omega_m/\ol{\omega}_m$ are not anymore free parameters and can
therefore not be adapted to fix $M_\eta$ to its experimental value.
For line $(d)$ the values of $M_\eta$, $f_\eta$ and $f_{\eta^\prime}$
differ somewhat from those obtained for the optimal value $\omega_m
v=-0.20$. The leading mixing approximation comes nevertheless quite
close to the experimental results.  In table \ref{tab1} we give our
input values for the pseudoscalar sector.
\begin{table}
\begin{center}
\begin{tabular}{|c|c|c|c||c|c|} \hline
  $M_{\pi^\pm}$ &
  $M_{K^\pm}$ & 
  $M_{K^0}$ &
  $M_{\eta^\prime}$ &
  $f_\pi$ &
  $f_{K^\pm}$
  \\[0.5mm] \hline\hline
  $135.1$ &
  $492.4$ &
  $497.7$ &
  $957.8$ &
  $92.4$ &
  $113.0$
  \\ \hline
\end{tabular}
\caption{\footnotesize This table shows the phenomenological input
  used in this work. All values are given in $\MeV$. The charged meson
  masses are electromagnetically corrected.}
\label{tab1}
\end{center}
\end{table}
Table \ref{tab2} shows the four different combinations of the
parameters $\om_m v$, $Z_p/Z_m$ and $\om_m/\ol{\om}_m$ used for our
numerical analysis.
\begin{table}
\begin{center}
\begin{tabular}{|c||c|c|c|c|} \hline
   &
  $\om_m v$ & 
  $Z_p/Z_m$ &
  $\om_m/\ol{\om}_m$ &
  $\ol{\omega}_m v$
  \\[0.5mm] \hline\hline
  (a) &
  $-0.22$ &
  $1.0$ &
  $1.0$ &
  $-0.22$ 
  \\ \hline
  (b) &
  $-0.20$ &
  $0.9$ &
  $1.0$ &
  $-0.20$ 
  \\ \hline\hline
  (c) &
  $-0.22$ &
  $0.9$ &
  $0.9$ &
  $-0.24$ 
  \\ \hline
  (d) &
  $-0.17$ &
  $0.9$ &
  $0.95$ &
  $-0.18$ 
  \\ \hline
\end{tabular}
\caption{\footnotesize The table shows the four different combinations
  of the parameters $\omega_m v$, $Z_p/Z_m$ and $\om_m/\ol{\om}_m$
  used in sect.~\ref{Results}. Line $(d)$ corresponds to the leading
  mixing approximation for which $\omega_m v$ and
  $\omega_m/\ol{\omega}_m$ are computed and do therefore not play the
  role of free input parameters.}
\label{tab2}
\end{center}
\end{table}
The numbers given in the text of this paper correspond to line $(b)$
which may be considered as our best values to first order in the
derivative expansion. 

The first step is to solve (\ref{NWFRC}) for the $Z$--factors and to
extract the values of the zero momentum parameters $\ol{M}_i^2$,
$\ol{f}_i$. The corresponding relations are
(\ref{DifferencesOfDecayConstants}), (\ref{RenormalizedCouplings}),
(\ref{BBB0}), (\ref{BBB1}) and the results are found in tables
\ref{tab3} and \ref{tab4}.
\begin{table}
\begin{center}
\begin{tabular}{|c||c|c|c||c|c|c|} \hline
  &
  $Z_\pi$ & 
  $Z_{K^\pm}$ &
  $Z_{K^\pm}-Z_{K^0}$ &
  $\frac{\stackrel{ }{\ol{M}_{\pi^\pm}}}{\MeV}$ &
  $\frac{\ol{M}_{K^\pm}}{\MeV}$ &
  $\frac{\ol{M}_{K^0}}{\MeV}$ 
  \\[0.5mm] \hline\hline
  (a) &
  1.22 &
  0.89 &
  0.0063 &
  $149.2$ &
  $465.3$ &
  $468.7$ 
  \\ \hline
  (b) &
  1.20 &
  0.90 &
  0.0058 &
  $148.0$ &
  $467.9$ &
  $471.4$ 
  \\ \hline\hline
  (c) &
  1.24 &
  0.88 &
  0.0068 &
  $150.7$ &
  $462.2$ &
  $465.4$ 
  \\ \hline
  (d) &
  1.18 &
  0.91 &
  0.0052 &
  146.5 &
  470.9 &
  474.6
  \\ \hline
\end{tabular}
\caption{\footnotesize Values for wave function renormalizations $Z_i$
  and zero momentum mass parameters $\ol{M}_i$.}
\label{tab3}
\end{center}
\end{table}
\begin{table}
\begin{center}
\begin{tabular}{|c||c|c||c|c|c|} \hline
  &
  $\frac{\ol{f}_\pi}{\MeV}$ &
  $\frac{\ol{f}_{K^\pm}}{\MeV}$ &
  $\frac{\sigma_0}{\MeV}$ &
  $\left(\frac{Z_m}{Z_h}\right)^{1/2}\hspace{-3mm}\frac{v}{\MeV}$ &
  $\left(\frac{Z_m}{Z_h}\right)^{1/2}\hspace{-3mm}\frac{w}{\MeV}$
  \\[0.5mm] \hline\hline
  (a) &
  83.7 &
  119.6 &
  53.9 &
  24.2 &
  $-0.69$
  \\ \hline
  (b) &
  84.3 &
  118.9 &
  53.8 &
  23.3 &
  $-0.67$ 
  \\ \hline\hline
  (c) &
  82.8 &
  120.4 &
  54.1 &
  25.3 &
  $-0.71$ 
  \\ \hline
  (d) &
  85.2 &
  118.2 &
  53.7 &
  22.2 &
  $-0.66$ 
  \\ \hline
\end{tabular}
\caption{\footnotesize Expectation values of scalar singlet and octets.}
\label{tab4}
\end{center}
\end{table}
We should point out the sizeable differences between $M_i^2$ and
$\ol{M}_i^2$ and similarly for the decay constants. They are due to
the large value of $|\om_m v|$.  In particular, the difference
$\ol{f}_K-\ol{f}_\pi$ is almost twice the value of $f_K-f_\pi$!  In
the next step we determine in table \ref{tab5} the parameters of the
linear meson model from (\ref{BBB2})--(\ref{NNN1}),
(\ref{RenormalizedCouplings}), (\ref{A5}) and
(\ref{FlavoredMesonMasses}), (\ref{BBB3}).
\begin{table}
\begin{center}
\begin{tabular}{|c||c|c|c|c||c|c|c|c|} \hline
  &
  $\frac{m_g^2}{\MeV^2}$ &
  $\frac{\nu}{\MeV}$ &
  $\lambda_2$ &
  $\hat{\om}$ &
  $\left(\frac{Z_h}{Z_m}\right)^{1/2}\hspace{-3mm}
   \frac{\gamma_2}{\MeV}$ &
  $\frac{\gamma_2 v}{\MeV^2}$ &
  $\frac{\gamma_2 w}{\MeV^2}$ &
  $\frac{\gamma_3 v}{\MeV^2}$ 
  \\[0.5mm] \hline\hline
  (a) &
  $(390.9)^2$ &
  6814 &
  17.0 &
  0.35 &
  4780 &
  $(340.1)^2$ &
  $-(57.4)^2$ &
  $-(80.9)^2$ 
  \\ \hline
  (b) &
  $(392.9)^2$ &
  6447 &
  21.3 &
  0.30 &
  4942 &
  $(339.3)^2$ &
  $-(57.7)^2$ &
  $(46.1)^2$
  \\ \hline\hline
  (c) &
  $(388.6)^2$ &
  5964 &
  17.5 &
  0.33 &
  4405 &
  $(333.8)^2$ &
  $-(55.7)^2$ &
  $-(38.6)^2$ 
  \\ \hline
  (d) &
  $(395.2)^2$ &
  6011 &
  26.8 &
  0.25 &
  5167 &
  $(338.8)^2$ &
  $-(58.2)^2$ &
  $(112.3)^2$ 
  \\ \hline
\end{tabular}
\caption{\footnotesize Parameters of the linear meson model.}
\label{tab5}
\end{center}
\end{table}
We observe that the infered value of the cubic coupling $\nu$ depends
only moderately on the details of the effective meson model. In
contrast, the uncertainty for the quartic coupling $\lambda_2$ remains
substantial. Table \ref{tab5} also contains information about the
cubic coupling $\gamma_2$ between two pseudoscalar octets and the
scalar octet as well as for the coupling $\gamma_3$ between the
pseudoscalar octet, the pseudoscalar singlet and the scalar
octet. Even though the sign of $\gamma_3$ remains undetermined we find
$|\gamma_3/\gamma_2|\lta0.1$. The coupling $\gamma_2$ therefore
dominates the decay of the $0^{++}$ mesons.

We are now ready to compute the mass matrix elements of
the $\eta$--$\eta^\prime$ sector using (\ref{NNN1}). The eigenvalues
$M_\eta$, $M_{\eta^\prime}$ follow from (\ref{BBB7}) the mixing angles
{}from (\ref{YYY5}) and the relations for $f_\eta$ and $f_{\eta^\prime}$
are given by (\ref{BBB8}) and (\ref{BBB9}). These are our main
``predictions'' for observable quantities. They are displayed in table
\ref{tab6}.
\begin{table}
\begin{center}
\begin{tabular}{|c||c|c|c||c|c|c|c|} \hline
  &
  $\frac{M_\eta}{\MeV}$ &
  $\frac{f_\eta}{\MeV}$ &
  $\frac{f_{\eta^\prime}}{\MeV}$ &
  $\frac{f_{\eta}}{f_\pi}$ &
  $\frac{f_{\eta^\prime}}{f_\pi}$ &
  $\frac{f_{\eta}}{f_\eta^{\rm exp}}$ &
  $\frac{f_{\eta^\prime}}{f_{\eta^\prime}^{\rm exp}}$
  \\[0.5mm] \hline\hline
  (a) &
  550.8 &
  106.6 &
  93.5 &
  1.15 &
  1.01 &
  1.09 &
  1.25
  \\ \hline
  (b) &
  546.9 &
  113.4 &
  83.8 &
  1.23 &
  0.91 &
  1.16 &
  1.12
  \\ \hline\hline
  (c) &
  549.1 &
  103.5 &
  88.8 &
  1.12 &
  0.96 &
  1.06 &
  1.19
  \\ \hline
  (d) &
  536.6 &
  111.3 &
  82.9 &
  1.20 &
  0.90 &
  1.14 &
  1.11
  \\ \hline
\end{tabular}
\caption{\footnotesize ``Predictions'' for $M_\eta$, $f_\eta$ and
  $f_{\eta^\prime}$.}
\label{tab6}
\end{center}
\end{table}
We find a very satisfactory agreement with experiment for line $(b)$.
The uncertainty in the ``prediction'' for $f_\eta$ and
$f_{\eta^\prime}$ is reflected in the differences as compared to lines
$(a)$ and $(c)$. Also the leading mixing approximation $(d)$ is not
too far from experiment, even though contributions beyond this
approximation need to be included. The general tendency of the higher
derivative contributions in the $\eta$--$\eta^\prime$ sector is an
enhancement of $f_{\eta^\prime}$ and a decrease in $M_\eta$ and
$f_\eta$ (cf.~fig.~\ref{Plot5}). From table \ref{tab7} we note that
the octet decay constant $f_{\eta8}$ is rather close to $f_K$ whereas
the corresponding singlet decay constant almost equals $f_\pi$.  In
table \ref{tab7} we also show the mixing angles in the
$\eta$--$\eta^\prime$ sector.
\begin{table}
\begin{center}
\begin{tabular}{|c||c|c|c|c||c|c||c|c||c|} \hline
  & $\frac{f_{\eta8}}{\MeV}$ &
  $\frac{f_{\eta^\prime8}}{\MeV}$ &
  $\frac{f_{\eta0}}{\MeV}$ &
  $\frac{f_{\eta^\prime0}}{\MeV}$ &
  $\theta_p(\eta)$ &
  $\theta_p({\eta^\prime})$ &
  $\frac{m_p}{\MeV}$ &
  $\frac{m_p^2}{M_{\eta^\prime}^2}$ &
  $\frac{f_{K^0}-f_{K^\pm}}{\MeV}$
  \\[0.5mm] \hline\hline
  (a) &
  116.6 &
  132.1 &
  95.3 &
  107.9 &
  $-15.6$ &
  $-31.2$ &
  839.1 &
  0.77 &
  0.25
  \\ \hline
  (b) &
  117.1 &
  124.2 &
  96.3 &
  102.1 &
  $-13.7$ &
  $-28.0$ &
  865.9 &
  0.82 &
  0.28
  \\ \hline\hline
  (c) &
  116.0 &
  126.6 &
  94.0 &
  102.6 &
  $-16.3$ &
  $-31.1$ &
  839.8 &
  0.77 &
  0.22
  \\ \hline
  (d) &
  117.7 &
  120.4 &
  97.5 &
  99.8 &
  $-14.7$ &
  $-28.6$ &
  843.5 &
  0.78 &
  0.30
  \\ \hline
\end{tabular}
\caption{\footnotesize Mixing angles in the $\eta$--$\eta^\prime$
  system and isospin violation in the decay constants.}
\label{tab7}
\end{center}
\end{table}
We find a large mixing between $\eta$ and $\eta^\prime$ with an
important dependence on the momentum. The mixing for
$q^2=-M_{\eta^\prime}^2$ is substantially larger than that for
$q^2=-M_\eta^2$.  The mixing corresponding to
$\theta_p(\eta)\simeq-13.7^o$ is somewhat smaller than earlier
estimates from chiral perturbation theory \cite{GL82-1} where the size
of the mixing angle was extracted indirectly from the requirement of a
realistic value for $M_\eta$. On the other hand, the mixing
corresponding to $\theta_p(\eta^\prime)$ is larger.  We should point
out that our direct method of computing all elements of the matrix for
the inverse propagator in the $\eta$--$\eta^\prime$ system is quite
different from the indirect consistency requirement for $M_\eta$.  We
furthermore see in table \ref{tab7} a large deviation of
$m_p^2/M_{\eta^\prime}^2$ from one despite the fact that this
difference is formally only a quadratic term in the quark mass
expansion.  We also present in table \ref{tab7} the isospin violation
in the decay constants (\ref{NNN29}), $f_{K^0}-f_{K^\pm}$. It is
reduced significantly as compared to the value obtained to lowest
order in the quark mass expansion (\ref{EstimateW}) or in chiral
perturbation theory \cite{GL82-1}.

Our results in the scalar sector depend in addition on $Z_h/Z_m$,
$\lambda_3\sigma_0^2$, $\ol{\omega}_h v$ and $\zeta_h v^2$. We use
$\lambda_3\sigma_0^2$, $\ol{\omega}_h v$ and $\zeta_h v^2$ as input
parameters together with a fixed value $M_{K_0^*}=1430\MeV$.  From
(\ref{Gamma5}), (\ref{Gamma6}) we determine $m_h^2$ and $\gamma_6 v$
as functions of these three couplings. Here we use our ``optimal
values'' for $\lambda_2$, $\nu$, etc.~corresponding to the second line
$(b)$ in tables \ref{tab2}--\ref{tab7}. We determine $Z_h/Z_m$ as a
function of $\lambda_3$, $\ol{\omega}_h v$ and $\zeta_h v^2$ according
to
\begin{equation}
  \frac{Z_h}{Z_m}=\frac{m_g^2+\sigma_0(3\sigma_0\lambda_2+\nu)+
  \frac{1}{6}(9\sigma_0\lambda_2-\nu+12\sigma_0^3\lambda_3)
  [\ol{f}_K-\ol{f}_\pi]}
  {\left[1+\frac{1}{2}\ol{\omega}_h v\right]
  \hat{M}_{K_o^*}^2}\; .
\end{equation}
The results of this analysis and, in particular, the masses of the
lowest lying $0^{++}$ octet are given for several values of
$\lambda_3$, $\ol{\omega}_h v$ and $\zeta_h v^2$ in table \ref{tab9}.
\begin{table}
\begin{center}
\begin{tabular}{|r|c|c||c|c|c|c|c||c|c|} \hline
  \rule[-2.0mm]{0mm}{6.5mm}
  $\ol{\omega}_h v$ &
  $\zeta_h v^2$ &
  $\lambda_3\sigma_0^2$ &
  $\frac{\hat{M}_{K_o\hskip-8pt\hphantom{x}^*}}{\MeV}$ &
  $\frac{Z_h}{Z_m}$ &
  $\frac{m_h}{\MeV}$ &
  $\frac{\gamma_6 v}{\MeV^2}$ &
  $\frac{\gamma_2}{\MeV}$ &
  $\frac{M_{a_o}}{\MeV}$ &
  $\frac{M_{f_8}}{\MeV}$
  \\[0.5mm] \hline\hline
  0.0 &
  0.00 &
  $0.0$ &
  1430 &
  0.35 &
  1407 &
  $(254.7)^2$ &
  8394 &
  1360 &
  1453
  \\ \hline
  $-0.2$ &
  0.00 &
  $0.0$ &
  1430 &
  0.39 &
  1335 &
  $(241.6)^2$ &
  7963 &
  1178 &
  1541
  \\ \hline
  0.0 &
  0.00 &
  $9.4$ &
  1430 &
  0.35 &
  1378 &
  $(382.6)^2$ &
  7973 &
  1267 &
  1480
  \\ \hline
  $-0.2$ &
  0.00 &
  $9.4$ &
  1430 &
  0.39 &
  1307 &
  $(363.0)^2$ &
  7564 &
  1097 &
  1570
  \\ \hline
  0.0 &
  0.02 &
  $0.0$ &
  1295 &
  0.42 &
  1274 &
  $(230.6)^2$ &
  7601 &
  1232 &
  1315
  \\ \hline
  0.0 &
  0.04 &
  $0.0$ &
  1144 &
  0.54 &
  1126 &
  $(203.7)^2$ &
  6715 &
  1088 &
  1162
  \\ \hline
\end{tabular}
\caption{\footnotesize Masses and parameters of the scalar octet.}
\label{tab9}
\end{center}
\end{table}
Here $\lambda_3\sigma_0^2=9.4$ corresponds to the ``maximal value''
$\lambda_2/2$ compatible with a convergent expansion in $\sigma_0$
(cf.~sect.~\ref{ExpansionInTheChiralCondensate}). From an estimate of
$\zeta_h v^2$ due to the exchange of the vector field
$\prl_\mu\rho_V^\mu$ (see appendix \ref{VectorMesonContributions}) we
learn that the two last lines in table \ref{tab9} are preferred.
Additional large mixing effects (see section
\ref{MassRelationsForTheScalarOctet} and appendix
\ref{MixingWithOtherStates}) may further lower $M_{a_o}$ and lead to a
mass of the isotriplet $a_0$ consistent with the observed resonance
$a_0(980)$. The scalar partner of the $\eta$ appears to be associated
with the broad resonance $f_0(1300)$. We finally emphasize the large
deviation of $Z_h/Z_m$ from one which underlines the importance of
nonminimal kinetic terms in the linear meson model.

\sect{Conclusions}
\label{Conclusions}

In this paper we have investigated the effective action for the linear
meson model.  Including the discussion of the vector and axialvector
fields from appendix \ref{VectorMesonContributions} a fairly simple
picture emerges. Expressed in terms of scalar fields $\Phi$ and vector
fields $\rho_V^\mu$, $\rho_A^\mu$ the quark mass expansion seems to
converge rather well for the three light flavors. The same holds for
the derivative expansion, leading to an approximate momentum
dependence of propagators $\sim(Z q^2+\ol{M}^2)^{-1}$. The only
exception from this picture seems to be the scalar isotriplet
$a_0(980)$ which can be explained (see appendix
\ref{MixingWithOtherStates}) by a large contribution of two--kaon or
four--quark states.

The divergence $\prl_\mu\rho_A^\mu$ has the same quantum numbers as
the pseudoscalar octet plus singlet. We have estimated the resulting
mixing effect or, equivalently, the terms induced in the effective
action for $\Phi$ from integrating out $\prl_\mu\rho_A^\mu$ (``partial
Higgs effect''). We find a large non--minimal kinetic term which
induces substantial quark mass corrections to the kinetic terms for
the pseudoscalars ($\omega_m^{(\rho)}v\simeq-0.15$). This effect
remains compatible with a converging quark mass expansion for the
flavored pseudoscalars $\pi$ and $K$. On the other hand, an
investigation of the masses $M_\eta$ and $M_{\eta^\prime}$ as well as
the decay constants $f_\eta$ and $f_{\eta^\prime}$ shows that the
non--minimal kinetic term induces in turn a large momentum dependent
mixing in the $\eta$--$\eta^\prime$ sector. Here we find that
contributions which are of third or higher order in a formal quark
mass expansion are comparable in size to the contributions arising to
second order. It is the nonlinearity generated by this large mixing
which leads to ``predictions'' for $M_\eta$, $f_\eta$ and
$f_{\eta^\prime}$ which are compatible with experimental observations.
This explains why the measured value of $f_\eta$ is quite far away
{}from its value for zero quark masses\footnote{We emphasize once again
  that the nonlinearities in $\Dt$ or the poor convergence of an
  expansion in $\Dt$ appear only in the eigenvalues $M_\eta$ and
  $M_{\eta^\prime}$, the mixing angles $\theta_p(\eta)$,
  $\theta_p(\eta^\prime)$ and the decay constants $f_\eta$,
  $f_{\eta^\prime}$. The matrix elements of the $\eta$--$\eta^\prime$
  propagator converge satisfactorily.}. It is amazing to see that the
nonlinearities in $M_\eta$, $f_\eta$ and $f_{\eta^\prime}$ as
functions of the variable $\omega_m v$ all conspire such that a common
value of $\omega_m v$ can explain simultaneously these three
quantities. Even more, our phenomenological estimate $\omega_m
v\simeq-0.2$ is quite close to the estimate from the partial Higgs
effect, $\omega_m^{(\rho)}v\simeq-0.15$. The latter involves
completely different quantities like the $\rho\pi\pi$ coupling and the
masses of the axialvectors!

It will be very interesting to see if our ``phenomenological'' picture
of the effective linear meson model can be obtained from the solution
of a flow equation similar to \cite{JW95-1}. It would be highly
nontrivial if the couplings would come out in such an approach in a
range consistent with the analysis of the present paper. This
concerns, in particular, the quartic coupling $\lambda_2$ which was
found in \cite{JW95-1} to be essentially determined by an infrared
fixed point behavior. Furthermore, a more systematic analysis of the
(axial)vector meson sector including quark mass effects should lead to
a quantitative estimate of several effective cubic and quartic
vertices relevant for the decay properties of these mesons. Beyond a
successful explanation of the observed values for $f_\eta$ and
$f_{\eta^\prime}$ our results constitute the ``phenomenological
basis'' for interesting further developments. They also shed light on
the important question of the convergence of the quark mass expansion.

\vspace{1.5cm}

\appendix

\noindent
{\Huge\bf Appendices}
\nopagebreak

\sect{Meson decay constants}
\label{MesonDecayConstants}

In this appendix we discuss the meson decay constants within the
linear sigma model. Most results displayed here are well known from
current algebra and are simply rephrased in a somewhat different
language. The only slightly delicate issue concerns the choice of the
normalization of fields. This determines the appropriate definition of
wave function renormalization constants. A careful treatment of these
constants is relevant for quantitative relations between decay
constants and meson masses as discussed in the main text. We adopt
here definitions of $f_\pi$, $f_K$, $f_\eta$, etc.~which are directly
related to measured partial decay widths of the corresponding mesons.

The weak leptonic decay of the charged pion, $\pi^-\lra\mu^-+\olnu_\mu$,
involves the effective three point vertex ($\olgm$ being the
Euclidean analog of $\gm_5$)
\begin{equation}\begin{array}{rcl}
 \ds{\Gm_{\pi\mu\olnu}} &=& \ds{i\int
 \frac{d^4p_\pi}{(2\pi)^4}
 \frac{d^4p_\mu}{(2\pi)^4}
 \frac{d^4p_{\olnu}}{(2\pi)^4}\Bigg[
 \frac{g^2\cos\vth_c}{4\sqrt{2}M_W^2}
 F_\pi^\rho(p_\pi,p_\mu,-p_{\olnu})}\nnn
 &\times& \ds{
 \pi^-(p_\pi)\olmu(p_\mu)\gm_\rho(1+\olgm)
 \nu_\mu(-p_{\olnu})(2\pi)^4 \dt(p_\pi-p_\mu-p_{\olnu})
 +{\rm h.c.}\Bigg]}\; .
 \label{AA1}
\end{array}\end{equation}
Here we have projected onto the leptonic $V-A$ structure following
{}from virtual $W$--exchange with $M_W$ the $W$--boson mass, $g$ the
weak gauge coupling and $\vth_c$ the Cabibbo angle. 
(Analogously, the effective vertex for the charged kaon decay is
obtained from (\ref{AA1}) by the replacements $F_\pi^\rho\ra F_K^\rho$,
$\pi^-\ra K^-$ and $\cos\vth_c\ra\sin\vth_c$.)
The vertex
function $F_\pi^\rho$ can depend only on two independent momenta (e.g.,
$p_\pi$ and $p_{\olnu}$) and the leptonic pion decay involves its
value for on--shell momenta, $p_\pi^2=-M_{\pi^\pm}^2$, $p_\mu^2=-M_\mu^2$,
$p_{\olnu}^2=0$. In the present context we neglect the dependence of
$F_\pi^\rho$ on the leptonic momenta and use the parameterization\footnote{A
  term $\sim p_{\olnu}^\rho$ would not contribute to the pion decay
  anyhow, since on shell we have $\slash{p}_{\olnu}\nu(-p_{\olnu})=0$.}
\begin{equation}
 F_\pi^\rho=p_\pi^\rho F_\pi(p_\pi^2)\; .
\end{equation}
Our task is therefore the evaluation of the pion decay constant
\begin{equation}
 f_\pi=F_\pi(p_\pi^2=-M_{\pi^\pm}^2)
\end{equation}
which is determined experimentally from the leptonic decay
width of the pion (up to electromagnetic corrections)
\begin{equation}
 \Gm_{\pi\ra\mu\olnu}=\frac{G_F^2}{4\pi}m_\mu^2 M_{\pi^\pm} f_\pi^2
 \left(1-\frac{m_\mu^2}{M_{\pi^\pm}^2}\right)^2
 \cos^2\vth_c
\end{equation}
derived from (\ref{AA1}) with
$G_F=\sqrt{2}g^2/(8M_W^2)$.

In order to compute $F_\pi$ we consider the linear $\si$--model
coupled to external currents. For this purpose we replace all
derivatives acting on $\Phi$ by covariant ones
\begin{equation}
 D^\mu\Phi=\prl^\mu\Phi-
 \frac{i}{2}\la_z R_z^\mu\Phi+
 \frac{i}{2}\Phi\la_z L_z^\mu =
 \prl^\mu\Phi-
 \frac{i}{2}V_z^\mu\left[\la_z,\Phi\right]-
 \frac{i}{2}A_z^\mu\left\{\la_z,\Phi\right\}\; .
 \label{CovariantDerivative}
\end{equation}
Here the vector-- and axialvector currents $V_z^\mu$ and $A_z^\mu$ are
related to the left and right handed currents $L_z^\mu$ and $R_z^\mu$,
respectively, by
\begin{equation}\begin{array}{rcl}
 \ds{L_z^\mu} &=& \ds{V_z^\mu-A_z^\mu}\nnn
 \ds{R_z^\mu} &=& \ds{V_z^\mu+A_z^\mu}\; .
\end{array}\end{equation}
By this replacement $\Gamma_{\rm kin}=\int d^4 x\Lc_{\rm kin}$
(\ref{Lkin}) becomes a functional of the (local) background fields
$V^\mu$ and $A^\mu$. Current conservation is automatically embodied in
this construction.  The coupling of mesons to $W$--bosons can be
infered via the identification
\begin{equation}\begin{array}{rcl}
 \ds{L_{1,2}^\mu} &=& \ds{g\cos\vth_c W_{1,2}^\mu}\nnn
 \ds{L_{4,5}^\mu} &=& \ds{g\sin\vth_c W_{1,2}^\mu}\; .
\end{array}\end{equation}
Once the couplings of mesons to $W$--bosons are known the couplings to
lepton pairs follow by insertion of the field equation
\begin{equation}
 W_i^\mu(p)=-\frac{g}{4}\frac{1}{p^2+M_W^2}
 \int\frac{d^4q}{(2\pi)^4}
 \olpsi(q-p)\tau_i\gm^\mu(1+\olgm)\psi(q)\; .
\end{equation}
Here $\psi$ stands for the lepton doublets and we will neglect $p^2$
as compared to $M_W^2$. Extracting the coefficient linear in
$L_{1,2}^\mu$
\begin{equation}\begin{array}{rcl}
 \ds{\Gamma_{L}} &=& \ds{-\int\frac{d^4p}{(2\pi)^4}
 K_{L,z}^\mu(p)L_{z,\mu}(p)}\nnn
 &=& \ds{-\frac{i}{2\sqrt{2}}
 \int\frac{d^4p}{(2\pi)^4}
 \pi^-(p)F_\pi^\mu(p)
 \left[L_{1,\mu}(p)-i L_{2,\mu}(p)\right]+\ldots}
\end{array}\end{equation}
we can relate $F_\pi$ to the part of $K_{L,1}^\mu$ which is linear in
$\pi_1$
\begin{equation}
 K_{L,1}^\mu(p)=\frac{i}{2}p^\mu
 F_\pi(p)\pi_1(p)+\ldots\; .
\end{equation} 
The discussion of leptonic decays of charged kaons is analogous with
\begin{equation}
 K_{L,4}^\mu(p)=\frac{i}{2}p^\mu
 F_K(p)K_4(p)+\ldots\; .
\end{equation}
Here $\pi_{1,2}$ and $K_{4,5}$ are the pion and kaon fields,
$\pi^-=\frac{1}{\sqrt{2}}(\pi_1+i\pi_2)$,
$K^-=\frac{1}{\sqrt{2}}(K_4+i K_5)$, with standard normalization of
their kinetic terms such that their inverse propagator (two point
function) in the vicinity of its zero at timelike momenta is given by
$p^2+M_{\pi^\pm}^2(M_{K^\pm}^2)$. 

For an evaluation of the terms linear in the currents $V_z^\mu$,
$A_z^\mu$ and linear in the meson fields one covariant derivative
should be replaced by
\begin{equation}
 D^\mu\Phi\lra -\frac{i}{2}V_z^\mu [\la_z,\VEV{\Phi}]-
 \frac{i}{2}A_z^\mu\{\la_z,\VEV{\Phi})\}
 \label{A11}
\end{equation}
with $\VEV{\Phi}={\rm diag}(\vph_u,\vph_d,\vph_s)$ the expectation value of
$\Phi$. All other covariant derivatives have to act as simple
derivatives on the meson fields. Consider first the limit of vanishing
quark masses where $\vph_u=\vph_d=\vph_s=\olsi_0$. The term linear in
the vector current $V_z^\mu$ vanishes whereas the axial current
couples linearly to the pion field as a consequence of spontaneous
chiral symmetry breaking, i.e. $D_\mu\Phi\ra-i\olsi_0 A_z^\mu\la_z$. The
relevant term in $\Lc_{\rm kin}$ (\ref{Lkin}) reads
\begin{equation}
 \Lc_{\rm kin}\lra -\frac{1}{2}\left(
 Z_\vph+X_\vph^-\olsi_0^2+U_\vph\olsi_0\right) Z_m^{-1}
 \si_0 A_z^\mu \Tr\left\{\la_z,\prl_\mu m\right\}
\end{equation}
and, with $Z_m=Z_\vph+X_\vph^-\olsi_0^2+U_\vph\olsi_0$
(cf.~sect.~\ref{KineticTermsInTheLinearSigmaModel}), one infers
\begin{equation}
 K_{L,z}^\mu=-\frac{1}{4}\si_0
 \Tr\left\{\la_z,\prl_\mu m\right\} \; .
\end{equation}
In the limit of vanishing quark masses the fields $m_z$ are already
properly normalized and we can identify $\pi_1=m_1$, $K_1=m_4$. This
yields 
\begin{equation}
 f_\pi=2\si_0\; .
\end{equation}
The proportionality between $f_\pi$ and $\si_0$ is no surprise. It is
well known that in the limit of vanishing quark masses the pion decay
is related to the non--conservation of the axial current induced by
chiral symmetry breaking.

Going beyond the limit of vanishing quark masses the expectation
values of $\vph_u$, $\vph_d$ and $\vph_s$ are different. Nevertheless,
there is no term linear both in the vector current $V_z^\mu$ and a
pseudoscalar meson. This follows from the discrete symmetries $\Cc$,
$\Pc$ and is related to the observation that the first term on the
right hand side of (\ref{A11}) is hermitean whereas the second is
anti--hermitean. On the other hand, the vector current couples
linearly to the scalars as, for example (see eqs.~(\ref{Phi}),
(\ref{Identification})) 
\begin{equation}\begin{array}{rcl}
 \ds{\Tr\left( D^\mu\Phi\right)^\dagger D_\mu\Phi}
 &\lra& \ds{ -\hal A_z^\mu\Tr\left[\{\la_z,\VEV{\Phi}\}
 \left( Z_m^{-\hal}\prl_\mu m+\frac{2}{\sqrt{6}}Z_p^{-\hal}\prl_\mu p
 \right)\right] }\nnn
 && \ds{ -\frac{i}{2} V_z^\mu\Tr\left[ [\la_z,\VEV{\Phi}]
 \left( Z_h^{-\hal}\prl_\mu h+\frac{2}{\sqrt{6}}Z_s^{-\hal}\prl_\mu s
 \right)\right] }\; .
\end{array}\end{equation}
For our
computation of $f_\pi$ we can omit the term linear in $V_z^\mu$.

Let us consider first the approximation where the kinetic term
(\ref{Lkin}) is truncated to $Z_\vph\Tr\left( D^\mu\Phi\right)^\dagger
\left( D_\mu\Phi\right)$. In this limit $m$ describes already the
  properly normalized pion and kaon fields. One finds
\begin{equation}
 K_{L,1}^\mu=\frac{i}{2}p^\mu\pi_1(p)Z_\vph Z_m^{-\hal}
 \left(\vph_u+\vph_d\right)
\end{equation}
or
\begin{equation}
 f_\pi=Z_\vph Z_m^{-\hal}\left(\vph_u+\vph_d\right)\; .
\end{equation}
Similarly, we find for the leptonic decay of the charged kaons
\begin{equation}
 f_{K^\pm}=Z_\vph Z_m^{-\hal}\left(\vph_u+\vph_s\right)
\end{equation}
and define
\begin{equation}
 f_{K^0}=Z_\vph Z_m^{-\hal}\left(\vph_d+\vph_s\right)\; .
\end{equation}
Noting
$\olsi_0=\frac{1}{3}\left(\vph_u+\vph_d+\vph_s\right)$ and observing
that in this approximation $Z_m=Z_\vp$ one arrives at
(\ref{RenormalizedCouplings})
\begin{equation}
 f_\pi+f_{K^\pm}+f_{K^0}=6\olsi_0 Z_m^{\hal}=6\si_0\; .
\end{equation}
With the definitions $\Dt_u=Z_m^{\hal}\vph_u-\si_0$ etc.~and
$\VEV{h}=\sqrt{2}Z_h^{\hal}\VEV{\Phi_s}=2Z_h^{\hal}
\left(\VEV{\Phi}-\olsi_0\right)=2\left(Z_h/Z_m\right)^{\hal}{\rm
    diag}(\Dt_u,\Dt_d,\Dt_s)$ we obtain the relations
  (\ref{DifferencesOfDecayConstants}).

Next we consider the more general kinetic term (\ref{Lkin}). The first
effect is a nontrivial wave function renormalization between the
fields $m_1$ and $\pi_1$, i.e.
\begin{equation}
 m_1=Z_\pi^{-\hal}\pi_1\; ,\;\;
 m_4=Z_{K^{\pm}}^{-\hal}K_4\; .
\end{equation}
This effect multiplies $f_\pi$ by a factor $Z_\pi^{-\hal}$ and
similarly for $f_{K^\pm}$, $f_{K^0}$. Here we note that $Z_\pi$,
$Z_{K^\pm}$ and $Z_{K^0}$ should be defined at the corresponding poles
such that the inverse renormalized two--point function is approximated
in the vicinity of the pole by $q^2+M_{\pi^\pm}^2$ with
$M_{\pi^\pm}\simeq135.1\MeV$ the physical pion mass (after subtraction
of electromagnetic effects). The second effect reflects the
modification of the general kinetic term into which
(\ref{CovariantDerivative}) is inserted. Since the axialvector field
$A_\mu(q)$ is needed for on--shell momenta, we conclude that all
kinetic terms must be evaluated at the poles. Expanding the inverse
propagators around $q^2=-M^2$, knowledge of the coefficient linear in
$q^2$ suffices for a computation of the meson decay constants. For the
pions this is given by $Z_\pi$ and the evaluation of the full kinetic
term gives a factor $Z_\pi$ in the formula for $f_\pi$. This can be
seen directly by inserting (\ref{CovariantDerivative}) into the
contributions $\sim X_\vph^-$, $U_\vph$ (\ref{ZZZ2}), (\ref{ZZZ1}) and
using the relations (\ref{A3}). In summary, the total effect of the
generalized kinetic term is a multiplication of $f_\pi$ with
$Z_\pi^{1/2}$. Similarly, $f_{K^\pm}$ and $f_{K^0}$ are proportional
to $Z_{K^\pm}^{1/2}$ and $Z_{K^0}^{1/2}$, respectively. This explains
the relations (\ref{A69}). We emphasize that $Z_\pi$, $Z_{K^\pm}$ and
$Z_{K^0}$ should be evaluated from the coefficient linear in $q^2$ in
an expansion of the inverse propagator around
$q^2=-M_{\pi^\pm}^2,-M_{K^\pm}^2,-M_{K^0}^2$, respectively. More
precisely, they are defined by (\ref{DefinitionOfZi}) with $q_0^2$
replaced by $-M_i^2$.

We finally extend the discussion to the decay constants of the
non--flavored pseudoscalars $f_{\pi^0}$, $f_\eta$ and $f_\etap$. By
eqs.~(\ref{BBB9}) and (\ref{BBB9a}) they are related to the couplings
of these mesons to the corresponding components of the axialvector
currents, or in a different language, the expectation value of the
corresponding current between the vacuum and the meson state. For
instance, $f_{\eta8}$ parameterizes the coupling of $\eta$ to the
current $A_8$ whereas $f_{\etap0}$ corresponds to the coupling of the
$\etap$ to the singlet current. For a comparison of $f_{\eta8}$ with
$f_\pi$ we therefore have to replace
$\Tr\left(\{\la_1,\VEV{\Phi}\}\la_1\right)Z_\pi^{1/2}$ by
$\Tr\left(\{\la_8,\VEV{\Phi}\}\la_8\right)Z_8^{1/2}$. The ratios of
$f_{\pi^0}$, $f_{\eta8}$ and $f_{\etap0}$ to $f_\pi$ are then easily
computed
\begin{eqnarray}
 \label{BBB8}
 \ds{\frac{f_{\pi^0}}{f_\pi}} &=& \ds{
 \left(\frac{Z_{\pi^0}}{Z_\pi}\right)^\hal}\\[2mm]
 \label{TTT4}
 \ds{\frac{f_{\eta8}}{f_\pi}} &=& \ds{
 \left(\frac{Z_{8}}{Z_\pi}\right)^\hal
 \frac{\vph_u+\vph_d+4\vph_s}{3(\vph_u+\vph_d)}=
 \left(\frac{Z_{8}}{Z_\pi}\right)^\hal
 \frac{2\olf_{K^\pm}+2\olf_{K^0}-\olf_\pi}{3\olf_\pi}}\\[2mm]
 \label{TTT15}
 \ds{\frac{f_{\eta0}}{f_\pi}} &=& \ds{
 \left(\frac{Z_{8}}{Z_\pi}\right)^\hal
 \frac{2(\vph_u+\vph_d+\vph_s)}{3(\vph_u+\vph_d)}=
 \left(\frac{Z_{8}}{Z_\pi}\right)^\hal
 \frac{\olf_{K^\pm}+\olf_{K^0}+\olf_\pi}{3\olf_\pi}}\; .
\end{eqnarray}
Similarly, the octet and singlet decay constants for the normalization
of the $\etap$, $f_{\etap8}$ and $f_{\eta^\prime0}$, are given by
\begin{equation}
 \label{TTT5}
 \frac{f_{\etap8}}{f_{\eta8}}=\frac{f_{\etap0}}{f_{\eta0}}=
 \left(\frac{Z_p}{Z_m Z_8}\right)^\hal
 \tilde{z}_p(-M_{\eta^\prime}^2)^{1/2}\; .
\end{equation}
Here $\tilde{z}_p(-M_{\eta^\prime}^2)$ accounts for higher derivative
effects and is normalized for $q^2=-M_\eta^2$ according to
$\tilde{z}_p(-M_{\eta}^2)=1$.  If the higher derivative effects are
omitted one also has $\tilde{z}_p(-M_{\eta^\prime}^2)=1$. Furthermore,
if we neglect the mixing effects (or for $z_8(q^2)=z_p(q^2)$) we can
identify $\tilde{z}_p(q^2)$ with $z_p(q^2)$ appearing in (\ref{ZZZ7}).
We note that (\ref{TTT5}) is appropriate for a definition of $Z_p$ at
$q_0^2=-M_\eta^2$. If one instead would define $Z_p$ at
$q_0^2=-M_{\eta^\prime}^2$ the factor $z_p(-M_{\eta^\prime}^2)^{1/2}$
would be absorbed by this alternative definition.

\sect{Vector mesons}
\label{VectorMesonContributions}

In this section we briefly discuss\footnote{See also \cite{Mei88-1}
  and references therein.} the vector and pseudovector
fields and their interactions with scalars and
pseudoscalars. This will permit us to estimate the part of the
effective interactions involving four (pseudo)scalars which is
induced by the exchange of vector fields. We introduce the fields
$\rho_L^\mu$ and $\rho_R^\mu$ as hermitian $3\times3$ matrices which
transform as ${\bf 8}\oplus{\bf 1}$ under $SU_L(3)$ and $SU_R(3)$,
respectively, being neutral with respect to the other part of the
flavor group. The interaction with (constituent) quarks
\begin{equation}
 \Lc_{\rho\olq q}=\frac{1}{\sqrt{2}}
 Z_q g_{\rho\olq q}
 \left(\olq_L\gm_\mu\rho_L^\mu q_L+
 \olq_R\gm_\mu\rho_R^\mu q_R\right)
 \label{CubicQuarkVectorInteraction}
\end{equation}
respects $\cst$ and is also consistent with left--right symmetry
($\rho_L\leftrightarrow\rho_R$) and charge conjugation
($\rho_L\ra-\rho_R^T$, $\rho_R\ra-\rho_L^T$). The invariant kinetic
and mass terms read
\begin{equation}\begin{array}{rcl}
 \ds{\Lc_{\rho,2}} &=& \ds{\frac{Z_\rho}{8}\Tr
 \left(\prl_\mu\rho_{L\nu}-\prl_\nu\rho_{L\mu}\right)
 \left(\prl^\mu\rho_L^{\nu}-\prl^\nu\rho_L^{\mu}\right)+
 \frac{Z_\rho}{4\al_\rho}\Tr\left(\prl_\mu\trho_L^\mu\right)^2}\nnn
 &+& \ds{\frac{Z_\rho}{12\al_\rho^\prime}
 \left(\prl_\mu\Tr\rho_L^\mu\right)^2+
 \frac{1}{4}m_\rho^2\Tr\trho_{L\mu}\trho_L^\mu+
 \frac{1}{12}m_\rho^{\prime 2}\Tr\rho_L^\mu\Tr\rho_{L\mu}+
 \left( L\ra R\right)}
 \label{VectorMassTerms}
\end{array}\end{equation}
where
\begin{equation}
 \trho_{L,R}^\mu=\rho_{L,R}^\mu-
 \frac{1}{3}\Tr\rho_{L,R}^\mu\equiv
 \trho_{L,R}^{z,\mu}\la_z
\end{equation}
denotes the octets and $\frac{1}{\sqrt{6}}\Tr\rho_{L,R}^\mu$
represents the singlets. In general, the field $\rho^\mu$ can
describe spin--one and spin--zero ($\sim\prl_\mu\rho^\mu$) particles.
For $\al_\rho,\al_\rho^\prime\ra0$ the spin--zero components decouple
and the fields $\rho^\mu$ only describes spin--one bosons.
($\prl_\mu\rho^\mu=0$). In the opposite limit
$\alpha_\rho,\alpha_\rho^\prime\ra\infty$ there remains a kinetic term
only for the spin--one bosons whereas $\prl_\mu\rho^\mu$ can be
determined algebraically from the field equations.

There is only one possible cubic coupling involving two scalars or
pseudoscalars and the vector octet. To lowest order in a derivative
expansion it reads
\begin{equation}
 \Lc_{\Phi^2\rho}=\frac{i}{\sqrt{2}}
 g_{\rho\pi\pi}\hat{Z}\Tr\left[\left(
 \prl_\mu\Phid\Phi-\Phid\prl_\mu\Phi\right)\trho_L^\mu+
 \left(\prl_\mu\Phi\Phid-\Phi\prl_\mu\Phid\right)\trho_R^\mu\right]\; .
 \label{CubicVectorOctetCoupling}
\end{equation}
The appropriate value of $\hat{Z}$ will be determined later such that
$g_{\rho\pi\pi}$ appears in the width of the decay $\rho\ra\pi\pi$
according to
\begin{equation}
  \label{AAA0}
  \Gamma(\rho\ra\pi\pi)=
  \frac{g_{\rho\pi\pi}^2}{48\pi}
  \frac{\left(M_\rho^2-4M_{\pi}^2\right)^{\frac{3}{2}}}
  {M_\rho^2}\; .
\end{equation}
Using the experimental values $\Gamma(\rho\ra\pi\pi)\simeq150\MeV$ and
$M_\rho\simeq770\MeV$ this yields
\begin{equation}
  \label{AAA1}
  g_{\rho\pi\pi}\simeq6.0\; .
\end{equation}
A second coupling appears for the singlets
\begin{equation}
 \Lc_{\Phi^2\rho}^\prime=\frac{i}{3}
 \frac{g_{\rho\pi\pi}^\prime}{\sqrt{2}}
 \hat{Z}\Tr\left(\prl_\mu\Phid\Phi-\Phid\prl_\mu\Phi\right)
 \left(\Tr\rho_L^\mu-\Tr\rho_R^\mu\right)\; .
 \label{CubicVectorSingletCoupling}
\end{equation}
We note that similarly the couplings of the quarks to the vector and
pseudovector singlets could be different from the octet couplings
leading to a modification of (\ref{CubicQuarkVectorInteraction}). In
the following we will neglect for simplicity the differences between
the singlets and octets (i.e., $g_{\rho\pi\pi}^\prime=g_{\rho\pi\pi}$,
$m_\rho^{\prime 2}=m_\rho^2$,
$\al_\rho^\prime=\al_\rho$). Accordingly, we give for the quartic
couplings involving two (pseudo)scalars and two (pseudo)vectors
only those appearing for the octets and extend them to the
singlets. In contrast to the cubic coupling
(\ref{CubicVectorOctetCoupling}) the lowest order term does not involve
derivatives: 
\begin{eqnarray}
 \label{QuarticVectorScalarCoupling}
 \ds{\Lc_{\Phi^2\rho^2}} &=& \ds{
 \frac{1}{2}\hat{Z}
 \left(g_{\rho\pi\pi}^2+f_1\right)
 \left(\Tr\Phid\Phi\rho_L^\mu\rho_{L\mu}+
 \Tr\Phi\Phid\rho_R^\mu\rho_{R\mu}\right)}\\[2mm]
 &-& \ds{
 \hat{Z}\left( g_{\rho\pi\pi}^2+f_2\right)
 \Tr\Phid\rho_R^\mu\Phi\rho_{L\mu}+
 \hat{Z}f_3\left(\Tr\Phid\Phi-\rho_0\right)
 \Tr\left(\rho_L^\mu\rho_{L\mu}+\rho_R^\mu\rho_{R\mu}\right)
 \nonumber}\; .
\end{eqnarray}
Our conventions are such that for $f_1=f_2=f_3=0$ the couplings
(\ref{CubicVectorOctetCoupling}), (\ref{CubicVectorSingletCoupling}),
(\ref{QuarticVectorScalarCoupling}) and an appropriate scalar kinetic
term can be written in terms of a gauge covariant derivative
\begin{equation}
 D_\mu\Phi=\prl_\mu\Phi-
 \frac{i}{\sqrt{2}}g_{\rho\pi\pi}\rho_{R\mu}\Phi+
 \frac{i}{\sqrt{2}}g_{\rho\pi\pi}\Phi\rho_{L\mu}
\end{equation}
as $\hat{Z}\Tr\left( D^\mu\Phi\right)^\dagger\left( D_\mu\Phi\right)$.
For $g_{\rho\olq q}=g_{\rho\pi\pi}$ this also extends to the couplings
to quarks.

Chiral symmetry breaking by the expectation value of $\Phi$ leads to a
mixing between $\rho_L^\mu$ and $\rho_R^\mu$. In the absence of quark
masses the $\rho$--mass matrix reads
\begin{equation}
 \ol{M}_{VA}^2=\left(\ba{ccc}
 m_\rho^2+2\left( g_{\rho\pi\pi}^2+f_1\right)\hat{Z}\olsi_0^2 &,&
 -2\left( g_{\rho\pi\pi}^2+f_2\right)\hat{Z}\olsi_0^2 \\
 -2\left( g_{\rho\pi\pi}^2+f_2\right)\hat{Z}\olsi_0^2 &,&
 m_\rho^2+2\left( g_{\rho\pi\pi}^2+f_1\right)\hat{Z}\olsi_0^2
 \ea\right)\; .
\end{equation}
The mass eigenstates are the vector and pseudovector mesons (and
associated scalars)
\begin{equation}\begin{array}{rcl}
 \ds{\rho_V^\mu} &=& \ds{
 \frac{1}{\sqrt{2}}\left(\rho_R^\mu+\rho_L^\mu\right)}\nnn
 \ds{\rho_A^\mu} &=& \ds{
 \frac{1}{\sqrt{2}}\left(\rho_R^\mu-\rho_L^\mu\right)}\; .
\end{array}\end{equation}
They transform under charge conjugation as
\begin{equation}\begin{array}{rcl}
 \ds{\rho_V} &\stackrel{C}{\lra}& \ds{-\rho_V^T}\nnn
 \ds{\rho_A} &\stackrel{C}{\lra}& \ds{\rho_A^T}
\end{array}\end{equation}
and we conclude that the transversal parts of $\rho_V^\mu$ and
$\rho_A^\mu$ describe the $1^{--}$ and $1^{++}$ octets and
singlets of the light meson spectrum. The mass of the vector mesons is
given by
\begin{equation}
 M_V^2=m_\rho^2+2\left( f_1-f_2\right)\hat{Z}\olsi_0^2
\end{equation}
whereas the pseudovector mass reads
\begin{equation}
 \ol{M}_A^2=
 m_\rho^2+2\left(2g_{\rho\pi\pi}^2+f_1+f_2\right)
 \hat{Z}\olsi_0^2\; .
 \label{A8}
\end{equation}
We have put here a bar on $\ol{M}_A^2$ in order to indicate that the
relation with physical axialvector masses involves an additional wave
function renormalization
\begin{equation}
  \label{AAA2}
  M_A^2=\frac{\ol{M}_A^2}{Z_A}\; .
\end{equation}
In fact, chiral symmetry breaking induces also a difference in the
kinetic terms for $\rho_V$ and $\rho_A$ by invariants of the type
$\Tr\Phi^\dagger F_R^{\mu\nu}\Phi F_{L\mu\nu}$ with
$F_{L,R}^{\mu\nu}=\prl^\mu\rho_{L,R}^\nu-\prl^\nu\rho_{L,R}^\mu$. 
Combining the most general term involving up to two powers of $\Phi$
\begin{equation}
  \label{OPE17}
  \Delta\Lc_{\rm kin,\rho}=
  \frac{\beta_1}{4}\Tr\Phi^\dagger
  F_R^{\mu\nu}\Phi F_{L\mu\nu}+
  \frac{\beta_2}{8}\Tr\left(
  \Phi^\dagger\Phi
  F_L^{\mu\nu}F_{L\mu\nu}+
  \Phi\Phi^\dagger
  F_R^{\mu\nu}F_{R\mu\nu}\right)
\end{equation}
with (\ref{VectorMassTerms}) one finds for the kinetic term relevant
for the spin--one bosons in the limit of vanishing quark masses
\begin{equation}
  \label{OPE18}
  \Lc_{\rm kin,\rho}=
  \frac{Z_V}{8}\Tr
  F_V^{\mu\nu}F_{V\mu\nu}+
  \frac{Z_A}{8}\Tr
  F_A^{\mu\nu}F_{A\mu\nu}
\end{equation}
with
\begin{eqnarray}
  \label{OPE19}
  \ds{Z_V} &=& \ds{
  Z_\rho+(\beta_2+\beta_1)\ol{\sigma}_0^2}\nnn
  \ds{Z_A} &=& \ds{
  Z_\rho+(\beta_2-\beta_1)\ol{\sigma}_0^2}\; .
\end{eqnarray}
The wave function renormalization $Z_\rho$ can be fixed by convention.
If we choose the normalization of $\rho_{L,R}^\mu$ such that the
kinetic term for $\rho_V^\mu$ has the standard form ($Z_V=1$) there
remains nevertheless an additional parameter $Z_A$ multiplying the
kinetic term of $\rho_A^\mu$ which typically differs from one. Similar
effects can influence the effective kinetic terms for the spin--zero
components $\sim\prl_\mu\rho^\mu$.

We observe that for $g_{\rho\pi\pi}^2+f_2>0$ the vector octet is
indeed lighter than the pseudovector octet. For $\abs{f_2}\ll
g_{\rho\pi\pi}^2$ the mass splitting can be related to the
$\rho\pi\pi$ coupling and therefore to the $\rho$ lifetime by
\begin{equation}
 \ol{M}_A^2-M_V^2=
 4\left( g_{\rho\pi\pi}^2+f_2\right)
 \hat{Z}\olsi_0^2=
 \frac{1}{9}\left(2\ol{f}_K+
 \ol{f}_\pi\right)^2 g_{\rho\pi\pi}^2
 \frac{\hat{Z}}{Z_m}x_\rho
 \label{A6}
\end{equation}
with
\begin{equation}
  \label{AAA3}
  x_\rho=1+\frac{f_2}{g_{\rho\pi\pi}^2}\; .
\end{equation}
The mass splitting within the octet because of nonvanishing quark
masses can also be understood from the interactions
(\ref{QuarticVectorScalarCoupling}) and (\ref{OPE17}) inserting
$\Phi=\olsi_0+\hal(w\lambda_3-\sqrt{3}v\lambda_8)Z_h^{-1/2}$. (This
can be used to determine the parameters appearing in these
expressions.)

It is instructive to write the cubic coupling
(\ref{CubicVectorOctetCoupling}) in terms of mass eigenstates by using
(\ref{Phi}):
\begin{equation}\begin{array}{rcl}
 \ds{\Lc_{\Phi^2\rho}} &=& \ds{
 \frac{i}{2}g_{\rho\pi\pi}\hat{Z}\Tr\Bigg\{
 \left[\prl_\mu\Phi_p\Phi_p-\Phi_p\prl_\mu\Phi_p+
 \prl_\mu\Phi_s\Phi_s-\Phi_s\prl_\mu\Phi_s\right]\rho_V^\mu}\nnn
 &+& \ds{i\Big[
 \prl_\mu\Phi_p\Phi_s-\Phi_p\prl_\mu\Phi_s+\Phi_s\prl_\mu\Phi_p-
 \prl_\mu\Phi_s\Phi_p+
 \frac{2}{3}\left(\chi_s\prl_\mu\chi_p-\prl_\mu\chi_s\chi_p\right)}\nnn
 &+& \ds{\frac{2}{\sqrt{3}}
 \left(\chi_s\prl_\mu\Phi_p-\prl_\mu\chi_s\Phi_p-
 \chi_p\prl_\mu\Phi_s+\prl_\mu\chi_p\Phi_s\right)\Big]\rho_A^\mu\Bigg\}}\nnn
 &+& \ds{
 \sqrt{2}g_{\rho\pi\pi}\hat{Z}\olsi_0\Tr\left\{
 \left(\Phi_p+\frac{1}{\sqrt{3}}\chi_p\right)
 \prl_\mu\rho_A^\mu\right\} }\; .
 \label{A1}
\end{array}\end{equation}
We note a mixing of the longitudinal component $\prl_\mu\rho_A^\mu$
with the pseudoscalar mesons $\Phi_p$, $\chi_p$ for $\olsi_0>0$. This
is possible since $\prl_\mu\rho_A^\mu$ represents a $0^{-+}$ state. In
contrast, $\prl_\mu\rho_V^\mu$ transforms as $0^{+-}$ and a mixing
with $0^{++}$ states $\Phi_s$ is possible only for the charged
scalars. It is induced by a nonvanishing expectation value
$\VEV{\Phi_s}$.  As required by $\Cc$ and $\Pc$ invariance the vector
mesons have cubic couplings to two pseudoscalars only if those are
distinct, e.g.~there is a $\rho^0\pi^+\pi^-$ but no $\rho^0\pi^0\pi^0$
coupling. Typical decays of pseudovectors involve the coupling of
$\rho_A$ to one pseudoscalar and one scalar.  Additional cubic
couplings involving two (pseudo)vectors and one (pseudo)scalar are
generated by (\ref{QuarticVectorScalarCoupling}) if $\olsi_0$ is
inserted for one of the fields $\Phi$.

For an estimate of the effective $\Phi$--interactions induced by the
exchange of $\rho$--fields we have to solve the field equations for
$\rho_V^\mu$ and $\rho_A^\mu$ as functions of $\Phi$. The result is
then inserted into the action thus eliminating $\rho^\mu$ and
replacing it by functions of $\Phi$. We will do so keeping only terms
linear and quadratic in $\Phi$. This is sufficient for invariants
containing up to four powers of $\Phi$. For the contributions from the
quartic term (\ref{QuarticVectorScalarCoupling}) we keep only the
lowest order $\Phi={\rm diag}(\olsi_0)$, i.e.~we neglect the mass
splitting within the octets. The resulting field equations
for $\rho_V^\mu$ and $\rho_A^\mu$ are
\begin{equation}\begin{array}{rcl}
 \ds{\left(-\prl^\nu\prl_\nu\dt_\mu^\si+
 \left(1-\frac{1}{\al_\rho}\right)\prl^\si\prl_\mu+
 M_V^2\dt_\mu^\si\right)\rho_{V,\si}^{ab} }
 &=& \ds{-ig_{\rho\pi\pi}\hat{Z}\Big(
 \prl_\mu\Phid\Phi-\Phid\prl_\mu\Phi}\nnn
 &+& \ds{
 \prl_\mu\Phi\Phid-\Phi\prl_\mu\Phid\Big)^{ab} }\nnn
 \ds{\left(-\prl^\nu\prl_\nu\dt_\mu^\si+
 \left(1-\frac{1}{\al_\rho}\right)\prl^\si\prl_\mu+
 \ol{M}_A^2\dt_\mu^\si\right)
 \rho_{A,\si}^{ab} }
 &=& \ds{ig_{\rho\pi\pi}\hat{Z}\Big(
 \prl_\mu\Phid\Phi-\Phid\prl_\mu\Phi}\nnn
 &-& \ds{
 \prl_\mu\Phi\Phid+\Phi\prl_\mu\Phid\Big)^{ab} }
 \label{VectorEOM}
\end{array}\end{equation}
If we omit first effects from chiral symmetry breaking one has
$M_V^2=\ol{M}_A^2=m_\rho^2$. Inserting (\ref{VectorEOM}) into
(\ref{VectorMassTerms}) and (\ref{CubicVectorOctetCoupling}) and
keeping only terms involving up to two derivatives, the vector and
pseudovector mesons contribute to the kinetic term (\ref{Lkin}) only a
structure
\begin{equation}
 X_\vp^{-(\rho)}=-\frac{4g_{\rho\pi\pi}^2
 \hat{Z}^2}{m_\rho^2}\; .
 \label{A2}
\end{equation}

For $\olsi_0>0$ the $\cst$ symmetry is spontaneously broken. The
effective interactions mediated by $\rho_V$ and $\rho_A$ still
preserve the vector--like $SU(3)$ symmetry if quark mass effects are
neglected. In this approximation it is most convenient to give
directly the contribution from $\rho$--exchange to the wave function
renormalization constants $Z_m$, $Z_p$, $Z_h$ and $Z_s$ as well as
$\omega_m$, $\omega_{pm}$ and $\omega_h$. They can be read off from
(\ref{A1})
\begin{equation}\begin{array}{rcl}
 \ds{Z_m^{(\rho)}} &=& \ds{Z_p^{(\rho)}=
 -\frac{4g_{\rho\pi\pi}^2}
 {\ol{M}_A^2}\hat{Z}^2\olsi_0^2}\nnn
 \ds{Z_h^{(\rho)}} &=& \ds{Z_s^{(\rho)}=
 \omega_h^{(\rho)}=0}\nnn
 \ds{\om_m^{(\rho)}} &=& \ds{
 -\frac{4g_{\rho\pi\pi}^2}{\ol{M}_A^2}
 \frac{\hat{Z}^2}{Z_h^\hal Z_m}\olsi_0}\nnn
 \ds{\om_{pm}^{(\rho)}} &=& \ds{
 -\frac{8}{\sqrt{6}}\frac{g_{\rho\pi\pi}^2}
 {\ol{M}_A^2}
 \frac{\hat{Z}^2}{(Z_h Z_m Z_p)^\hal}\olsi_0}\; .
 \label{A4}
\end{array}\end{equation}
This yields the same expressions as inserting (\ref{A2}) into
(\ref{A3}) if $m_\rho^2$ is replaced by $\ol{M}_A^2$. Indeed, to
linear order in the quark masses there is no contribution to the wave
function renormalization constants from the exchange of the vector
field $\rho_V^\mu$. Only the exchange of the spin--zero component
$\prl_\mu\rho_A^\mu$ induces the corrections (\ref{A4}). This can
easily be understood from the structure of the interactions
(\ref{A1}). Possible contributions to the quadratic terms defining
$Z_m$, $Z_h$, $\om_m$ etc.~can only arise through terms in the field
equations for $\rho$ which are linear in $\Phi-\VEV{\Phi}$. By
Lorentz--invariance such terms must be $\sim\int d^4
x\prl_\mu\Phi\rho^\mu\sim\int d^4 x(\Phi-\VEV{\Phi})\prl_\mu\rho^\mu$.
The discrete symmetries $\Cc$ and $\Pc$ allow to this order only a
term $\sim\Phi_p\prl_\mu\rho_A^\mu$ which corresponds to the mixing of
the $0^{-+}$ states and a structure
$\sim\left[\Phi_s-\VEV{\Phi_s},\VEV{\Phi_s}\right]\prl_\mu\rho_V^\mu$
for the mixing in the scalar sector. The mixing in the pseudoscalar
sector receives contributions $\sim\sigma_0,v$ and therefore
contributes to the terms in (\ref{A4}). In contrast, the mixing in the
scalar sector vanishes for $v=0$ and gives therefore only a correction
(\ref{OPQ1}) to the scalar kinetic terms which is quadratic in $v$
\begin{equation}
  \label{OPQ3}
  \zeta_h^{(\rho)}=\frac{1}{4}
  \frac{\hat{Z}^2}{Z_h^2}
  \frac{g_{\rho\pi\pi}^2}{M_V^2}\; .
\end{equation}

The contribution to $Z_m^{(\rho)}$ (\ref{A4}) is known as the
``partial Higgs effect''. Inserting (\ref{A8}) for $\ol{M}_A^2$ one
finds for $f_1=f_2=0$ and $\hat{Z}=Z_\vph$
\begin{equation}
 \label{AAA4}
 Z_m=Z_\vph+Z_m^{(\rho)}=
 Z_\vph\frac{m_\rho^2}{\ol{M}_A^2}
\end{equation}
and observes that $Z_m$ vanishes for $m_\rho^2\ra0$. In this limit the
symmetry $\cst$ becomes an exact local gauge symmetry. As a result of
the Higgs effect the pseudoscalar octet disappears from the spectrum.
What remains are massive pseudovector mesons which acquire their mass
through spontaneous symmetry breaking of the axial $SU_A(3)$ by
$\olsi_0\neq0$. In the real world, however, the local gauge symmetry
is explicitly broken by the mass term $m_\rho^2$ and by the deviation
of $f_1$ and $f_2$ from zero. Below we will take $\hat{Z}$ different
{}from $Z_\vph$ and this again violates local $SU_L(3)\times SU_R(3)$
symmetry and modifies (\ref{AAA4}). The partial Higgs effect
corresponds to the mixing between the two $0^{-+}$ octet states
contained in $\Phi_p$ and $\prl_\mu\rho_A^\mu$. We may further improve
the estimate (\ref{A4}) by taking into account the contributions $\sim
q^2$ on the left hand side of (\ref{VectorEOM}).  The propagator of
the divergence of $\rho_A^\mu$ differs from the one for the
pseudovector mesons, since the kinetic term is given by the ``gauge
fixing'' term $\sim\frac{1}{\al_\rho}$ in (\ref{VectorMassTerms}).
(There are similar differences in higher order in the momentum $q^2$
and in contributions to the kinetic terms from chiral symmetry
breaking.) The expression for $\ol{M}_A^2$ appropriate in (\ref{A4})
should involve the inverse propagator for the longitudinal component
of $\rho_A^\mu$, $Z_P q^2+\ol{M}_A^2$, where to lowest order
$Z_P=\alpha_\rho^{-1}$. It should be taken at a momentum scale
corresponding to the light pseudoscalar octet masses, $q_0^2=-m_m^2$.
We also define a renormalized mass parameter
\begin{equation}
  \label{L07}
  M_P^2=\ol{M}_A^2/Z_P\; .
\end{equation}
Inserting this into (\ref{A4}) yields the relation
\begin{eqnarray}
 \ds{
 \left(\frac{Z_h}{Z_m}\right)^\hal\om_m^{(\rho)}
 \left(2\ol{f}_K+
 \ol{f}_\pi\right)}
 &=& \ds{-\frac{2}{3}
 \frac{g_{\rho\pi\pi}^2}
 {\ol{M}_{A}^2-Z_P m_m^2}
 \left(2\ol{f}_K+
 \ol{f}_\pi\right)^2
 \frac{\hat{Z}^2}{Z_m^2}}\nnn
 &=& \ds{
 -\frac{6}{x_\rho}
 \frac{\hat{Z}}{Z_m}
 \frac{M_A^2-M_V^2/Z_A}{M_A^2-m_m^2 Z_P/Z_A} }
\end{eqnarray}
where we used (\ref{A6}) for the last equation. We note that unless
$Z_P/Z_A=M_A^2/M_P^2$ is very large the precise value of $Z_P$ has
only little influence. In fact, we will see that $Z_P/Z_A\lta0.35$
(cf.~table \ref{TTA1}). The average masses of the vectors and
pseudovectors are given by ($M_\rho=770\MeV$, $M_{K^*}=892\MeV$,
$M_{a_1}=1230\MeV$, $M_{K_1}=1340\MeV$)
\begin{equation}
  \label{L05}
  \begin{array}{rcl}
  \ds{M_V^2} &=& \ds{
  \frac{2}{3}M_{K^*}^2+
  \frac{1}{3}M_\rho^2\simeq(853\MeV)^2}\nnn
  \ds{M_A^2} &=& \ds{
  \frac{2}{3}M_{K_1}^2+
  \frac{1}{3}M_{a_1}^2\simeq(1300\MeV)^2}\; .
  \end{array}
\end{equation}
This leads to a quantitative estimate for the $\rho$--exchange
contribution to $\om_m v$
\begin{eqnarray}
  \label{L06}
  \ds{\om_m^{(\rho)}v} &=& \ds{
  -\frac{4}{x_\rho}\frac{\hat{Z}}{Z_m}
  \frac{(\ol{f}_K-\ol{f}_\pi)}
  {(2\ol{f}_K+\ol{f}_\pi)}
  \frac{\left(1-M_V^2/M_A^2\right)}
  {\left(1-m_m^2/M_P^2\right)}
  \left[1-
  \frac{M_V^2}{M_A^2-M_V^2}
  \left(\frac{1}{Z_A}-1\right)\right]}\nnn
  &\simeq& \ds{
  -\frac{0.28}{x_\rho}
  \frac{\hat{Z}}{Z_m}
  \left[1-0.75\left(
  \frac{1}{Z_A}-1\right)\right]}
\end{eqnarray}
where in the last step we have neglected the weak dependence on
$M_P^2$.  At this stage we see already that a typical order of
magnitude for $\omega_m^{(\rho)}v$ is around $-0.2$ which is
quite close to what is needed to explain $f_\eta$ and
$f_{\eta^\prime}$ (see sect.~\ref{DecayConstantsOfEtaAndEtap}).

For a more detailed estimate we have to determine the appropriate
choice of $\hat{Z}$. After elimination of $\rho_A^\mu$ by solving its
fields equation the relevant terms in the effective interaction for
$\Phi$ and $\rho_V$ are
\begin{eqnarray}
  \label{AAA6}
  \ds{{\cal L}_{\Phi V}} &=& \ds{
  Z_m\Tr(\prl^\mu\Phi)^\dagger
  \prl_\mu\Phi}\nnn
  &+& \ds{
  \frac{i}{2}g_{\rho\pi\pi}\hat{Z}
  \Tr\left[\left(
  \prl_\mu\Phi^\dagger\Phi-\Phi^\dagger\prl_\mu\Phi+
  \prl_\mu\Phi\Phi^\dagger-\Phi\prl_\mu\Phi^\dagger
  \right)\rho_V^\mu\right]}\; .
\end{eqnarray}
As a result of the partial Higgs effect we notice (\ref{A4}) the
appearance of
\begin{equation}
  \label{AAA7}
  Z_m=Z_\vph-
  4g_{\rho\pi\pi}^2
  \frac{\ol{\sigma}_0^2}{\ol{M}_A^2-Z_P m_m^2}
  \hat{Z}^2
\end{equation}
instead of\footnote{More precisely, $Z_\vph$ stands in this appendix
  for $Z_\vph+U_\vph\ol{\sigma}_0+\tilde{X}_\vph^-\ol{\sigma}_0^2$
  where $\tilde{X}_\vph^-$ is the part of $X_\vph^-$ which is
  unrelated to ``$\rho$--exchange''.} $Z_\vph$ in front of the kinetic
term for $\Phi$. Within the lowest order derivative approximation
employed here we therefore have to identify
\begin{equation}
  \label{AAA8}
  \hat{Z}=Z_m
\end{equation}
in order to get the standard coupling of $\rho_V$ to the renormalized
scalar field. Indeed, the lowest order $\rho\pi\pi$ interaction now
takes the form
\begin{equation}
  \label{AAA9}
  {\cal L}_{\rho\pi\pi}=
  g_{\rho\pi\pi}\epsilon_{ijk}
  \pi_j\prl_\mu\pi_k\rho_i^\mu=
  -2ig_{\rho\pi\pi}\rho_3^\mu
  \pi^+\prl_\mu\pi^- +\ldots
\end{equation}
as can be seen by inserting $\rho_V^\mu=\rho_i^\mu\tau_i$,
$\Phi=\frac{1}{2}Z_m^{-1/2}
f_\pi\exp\left(i\frac{\pi_k\tau_k}{f_\pi}\right)$,
$i,j,k=1,2,3$. With the choice (\ref{AAA8}) the interactions
(\ref{CubicVectorOctetCoupling}), (\ref{CubicVectorSingletCoupling})
can be combined for $f_1=f_2=0$ into a covariant kinetic term for
$\Phi$
\begin{eqnarray}
  \label{AAA10}
  \ds{{\cal L}_{\Phi V}^\prime} &=& \ds{
  Z_m\Tr\left( D^\mu\Phi\right)^\dagger
  \left( D_\mu\Phi\right)}\nnn
  \ds{D_\mu\Phi} &=& \ds{
  \prl_\mu\Phi-
  \frac{i}{2}g_{\rho\pi\pi}
  \left[\rho_{V\mu},\Phi\right]}\; .
\end{eqnarray}
We observe in passing that because of the violation of gauge symmetry
by the mass term $\sim m_\rho^2\rho_A^\mu\rho_{A\mu}$ one cannot have
simultaneously local gauge invariance of the interaction terms with
respect to $SU_L(3)\times SU_R(3)$ (corresponding to $\hat{Z}=Z_\vph$)
and the vector--like subgroup $SU_V(3)$ after spontaneous symmetry
breaking (corresponding to $\hat{Z}=Z_m$). We could, of course,
replace in (\ref{CubicVectorOctetCoupling})
$g_{\rho\pi\pi}Z_m=\hat{g}_{\rho\pi\pi}Z_\vph$. The quartic
interactions (\ref{QuarticVectorScalarCoupling}), however, would then
be proportional to $Z_m g_{\rho\pi\pi}^2$ $=$
$Z_\vph\hat{g}_{\rho\pi\pi}^2 \left(1+\frac{Z_\vph-Z_m}{Z_m}\right)$.
In this normalization (\ref{AAA10}) corresponds to nonvanishing
$\hat{f}_1=\hat{f}_2=\frac{Z_\vph-Z_m}{Z_m}$ and a different
normalization of the gauge coupling.  We will consider here the case
of approximate $SU_V(3)$ gauge symmetry (\ref{AAA10}) in contrast to
part of the literature where the choice $\hat{Z}=Z_\vph$ is adopted
while $\hat{f}_1$ and $\hat{f}_2$ are neglected. In our case, the
interaction terms are only invariant with respect to global chiral
$SU_L(3)\times SU_R(3)$ transformations. Local $SU_V(3)$ invariance
arises effectively only after chiral symmetry breaking if the symmetry
breaking terms $m_\rho^2$, $f_i$, etc.~are neglected.

The wave function renormalization $Z_A$ can be evaluated from the
average mass of the pseudovector mesons (\ref{A6})
\begin{equation}
  \label{TTT27}
  Z_A=\frac{M_V^2+4g^2_{\rho\pi\pi}\sigma_0^2x_\rho}
  {M_A^2}\; .
\end{equation}
If we neglect the weak dependence of $\omega_m^{(\rho)}v$ on $Z_P$,
(\ref{L06}), (\ref{AAA8}) and (\ref{TTT27}) are already sufficient to
compute $\omega_m^{(\rho)}v$ as a function of $x_\rho$. For a more
accurate estimate we have to determine $Z_P$ or $M_P$ which will be
done in appendix \ref{MixingWithOtherStates} (cf.~(\ref{TTT28})). This
allows to determine $Z_A$, $\ol{M}_A$, $Z_P$, $M_P$ and
$\omega_m^{(\rho)} v$ for given $x_\rho$. The results are displayed in
table \ref{TTA1} for different values of $x_\rho$ and $M_0$
(cf.~appendix \ref{MixingWithOtherStates}).
\begin{table}
\begin{center}
\begin{tabular}{|c||c|c|c||c|c|c||c|c|c|} \hline
  $x_\rho$ &
  $\frac{Z_p}{Z_m}$ &
  $Z_A$ &
  $\frac{\ol{M}_A}{\MeV}$ &
  $Z_P$ &
  $\frac{M_P}{\MeV}$ &
  $\frac{M_0}{\MeV}$ &
  $\omega_m^{(\rho)} v$ &
  $Z_m/Z_\vph$ &
  $\delta_\omega^{(\rho)}$
  \\[0.5mm] \hline\hline
  1.0 &
  1.00 &
  0.67 &
  1068 &
  0.00 &
  $-$ &
  $-$ &
  $-0.14$ &
  0.73 &
  0.00
  \\ \hline
  0.8 &
  0.99 &
  0.62 &
  1030 &
  0.22 &
  2196 &
  1852 &
  $-0.17$ &
  0.71 &
  0.08
  \\ \hline
  1.0 &
  0.99 &
  0.67 &
  1069 &
  0.22 &
  2278 &
  1944 &
  $-0.15$ &
  0.73 &
  0.07
  \\ \hline
  1.2 &
  0.99 &
  0.72 &
  1106 &
  0.22 &
  2358 &
  2031 &
  $-0.14$ &
  0.74 &
  0.07
  \\ \hline\hline
  1.0 &
  0.9 &
  0.67 &
  1069 &
  0.16 &
  2673 &
  2282 &
  $-0.15$ &
  0.73 &
  0.05
  \\ \hline
\end{tabular}
\caption{\footnotesize The table gives the ``leading mixing'' results
  for various parameters related to $\eta$--$\eta^\prime$ mixing. Only
  in the last line we use a fixed value $Z_p/Z_m=0.9$ corresponding to
  a nonvanishing $U_\vph$ in (\ref{A3}).}
\label{TTA1}
\end{center}
\end{table}
The first five lines are evaluated in the leading mixing approximation
as described in the end of sect.~\ref{HigherDerivativeContributions}.
The last line assumes in addition $U_\vph\neq0$ in (\ref{A3}) and
corresponds to line $(d)$ in sect.~\ref{Results}.  The main
uncertainties in $\omega_m^{(\rho)}v$ arise from $x_\rho\neq1$ and
{}from possible higher derivative contributions to the decay
$\rho\ra\pi\pi$ which would modify the value of $g_{\rho\pi\pi}$. In
addition, there are also corrections from the mass splitting in the
vector-- and pseudovector octets. The latter are suppressed by an
additional power of quark masses. We believe that the values for
$Z_m/Z_\vph$ in table \ref{TTA1} give a rather realistic estimate for
the quantitative importance of the ``partial Higgs effect''. For the
values in table \ref{TTA1} we have assumed
$\omega_m=\omega_m^{(\rho)}$ for the computation of
$\ol{f}_K-\ol{f}_\pi$, $\sigma_0$, etc. If we use instead the values
of line $(b)$ in table \ref{tab4}, $\omega_m^{(\rho)}v$ typically
decreases by $0.02$ thus reaching values close to $-0.2$. In
conclusion, typical values for $\omega_m^{(\rho)}v$ are in the range
between $-0.14$ and $-0.17$. This coincides more or less with the
values needed for an explanation of the observed decay constants
$f_\eta$ and $f_{\eta^\prime}$! We also note that the partial Higgs
effect reflected in the deviation of $Z_m/Z_\vph$ from one leads
typically to a $30\%$ correction in the kinetic terms.

Let us finally turn to the size of the mixing effects in the scalar
sector parameterized by $\zeta_h$ (\ref{OPQ3}). Despite their
suppression by a factor $\sim v^2$ they may be quantitatively relevant
because the propagator of the exchanged quantum now involves the mass
of the vector instead of the axialvector mesons. Since this mixing
concerns only the $K_0^*$ mesons (cf.~(\ref{OPQ1})) we use for a
quantitative estimate of $\zeta_h$ the mass $M_{K^*}=892\MeV$ and
replace in (\ref{OPQ3}) $M_V^2$ by $M_{K^*}^2+\tilde{Z}_P
q_0^2=M_{K^*}^2-\tilde{Z}_P M_{K_o^*}^2$. (Here $\tilde{Z}_P$ differs
{}from $Z_P$ by contribution $\sim\ol{\sigma}_0^2$ similar to
(\ref{OPE17})).  This implies
\begin{equation}
  \label{OPQ5}
  \zeta_h^{(\rho)}v^2=\frac{1}{9}\frac{Z_m}{Z_h}
  \left(\frac{g_{\rho\pi\pi}
  (\ol{f}_K-\ol{f}_\pi)}{M_{K^*}}\right)^2
  \left(1-\tilde{Z}_P\frac{M_{K_o^*}^2}
  {M_{K^*}^2}\right)^{-1}\simeq
  \frac{0.0185}{1-2.57\tilde{Z}_P}
\end{equation}
and we note that this formula is not applicable for $\tilde{Z}_P$
around $0.4$.  There $M_{K^*}^2/\tilde{Z}_P$ is of the same size as
$M_{K_o^*}^2$ and the mixing of states cannot be described by a
derivative expansion anymore (cf.~appendix
\ref{MixingWithOtherStates}). Also the imaginary part of the
propagators of unstable particles have to be taken into account.
Because of the uncertainty in the value of $\tilde{Z}_P$ it is
difficult to give a precise quantitative estimate of
$\zeta_h^{(\rho)}v^2$. We only know that this quantity is positive and
exceeds the value for $Z_P=0$. This leads to the estimate $\zeta_h
v^2\gta0.02$.

We conclude that the ``classical'' exchange of spin one mesons only
contributes to the effective quartic (and higher) interactions of the
pseudoscalars but does not modify their effective propagators. They
are therefore not relevant for the investigations of the present work
which concentrate on masses and mixings. The only contributions from
higher states to the propagators concern the mixing with higher
$0^{-+}$ states for $\Phi_p$ and $\chi_p$ and additional scalar states
for $\Phi_s$. Such states are contained in $\prl_\mu\rho_A^\mu$ and
$\prl_\mu\rho_V^\mu$.  These mixings contribute to quantities like
$X_\vph^-$, $Z_h/Z_m$ etc. A brief general discussion of mixing
effects is given in appendix \ref{MixingWithOtherStates}.

\sect{Mixing with other states}
\label{MixingWithOtherStates}

In QCD the pseudoscalar and scalar mesons described by the field
$\Phi$ are only part of a rich spectrum of quark--antiquark states
plus glueballs and possibly also $\ol{q}q\ol{q}q$ states. There are
strong couplings between the various states and their physics
therefore influences the behavior of the $0^{-+}$ and $0^{++}$
particles described in this work. On the level of effective
propagators which are the main subject of this paper we have to
consider mixing effects with other $0^{-+}$ or $0^{++}$ states in the
spectrum. For the pseudoscalar octet this concerns only the higher
mass $0^{-+}$ octets, whereas for the scalar octet we also have to
consider the possible mixing with $\ol{q}q\ol{q}q$ states in the
$0^{++}$ channel. We mention that it is not crucial in this context if
the two--meson or four--quark states (with $\ol{q}q\ol{q}q$ quantum
numbers) correspond to ``particles'' like the $a_0(980)$, $f_0(980)$
or not --- the composite fields describing the $\ol{q}q\ol{q}q$ states
may also have ``propagators'' without a pole.  Finally, for the
pseudoscalar $\etap$ we also have to include a mixing with
pseudoscalar glueballs. We collectively denote these additional
resonances as ``higher states''.

The general method for dealing with the higher states is to integrate
them out and to compute an effective theory for $\Phi$ alone. The
investigations of this paper should be understood in this context.
There are various methods for integrating out the higher states. One
consists in computing first the effective action including additional
fields for the higher states. In a second step the field equations for
these states are solved for arbitrary values of $\Phi$.  The resulting
``classical fields'' are functionals of $\Phi$ and can be reinserted
into the effective action, thus leading to an effective action which
depends only on $\Phi$. The discussion in appendix
\ref{VectorMesonContributions} can serve as an example. For those
results of the present paper which are only based on symmetries it is
actually not necessary to perform the integration of additional fields
in practice.  Nevertheless, some insight in the origin of some of the
constants of the effective action, like $X_\vph^-$, $\tV_\vph$ etc.,
can be gained by considering the possible form of the effective action
including additional states. We should point out that we neglect
throughout the imaginary part of the two--point functions which is due
to the decay of unstable resonances. This approximation may become
invalid in the immediate vicinity of poles in the propagators.

We have already encountered the mixing of $0^{-+}$ states in the
discussion of the longitudinal component of $\rho_A^\mu$ in appendix
\ref{VectorMesonContributions}. Let us rephrase this with a somewhat
different perspective by introducing a field
\begin{equation}
  \label{CanonicalDimenison}
  \tau_P=\prl_\nu\rho_A^\nu-
  \frac{1}{3}\Tr\prl_\nu\rho_A^\nu
\end{equation}
for the additional $0^{-+}$ state. With this normalization the
propagator for $\tau_P$ can be approximated by $q^2G_P(q^2)$ with
$G_P^{-1}=Z_P q^2+ \ol{M}_A^2$ (cf.~appendix
\ref{VectorMesonContributions}). The inverse propagator for the
coupled system of $\sqrt{2}\Phi_p$ and $\tau_P$ contains off--diagonal
terms
\begin{equation}
  \label{OO1}
  G^{-1}(q)=\left(
  \begin{array}{rcl}
  G_\vph^{-1}(q) &,& G_{\vph P}^{-1}(q)\nnn
  G_{P\vph}^{-1}(q) &,& G_P^{-1}(q)/q^2
  \end{array}
  \right)
\end{equation}
which are responsible for the mixing. From (\ref{A1}) one finds
\begin{equation}
  \label{OO2}
  G_{\vph P}^{-1}(q)=G_{P\vph}^{-1}(q)=
  2g_{\rho\pi\pi}Z_m\olsi_0\; .
\end{equation}
A similar mixing occurs between $\chi_p$ and the singlet
$\tau_P^\prime$ contained in $\Tr\prl_\mu\rho_A^\mu$.  It is no
accident that the mixing vanishes for $\olsi_0=0$ or $q^2=0$: In the
limit of unbroken chiral symmetry ($\olsi_0=0$) the fields $\Phi$ and
$\rho$ belong to different representations of $\cst$ and cannot mix.
Also for $\olsi_0\neq0$ and vanishing quark masses $\Phi_p$ describes
Goldstone bosons which can only have derivative couplings. By
construction the quark mass terms only appear as source terms for
$\Phi$.

\begin{figure}
\unitlength1.0cm
\begin{picture}(8.,4.)
\put(2.8,2.0){$\sigma_0,v$}
\put(5.5,2.0){$v$}
\put(10.1,2.0){$v$}
\put(12.4,2.0){$\sigma_0$}

\put(2.2,0.0){$m$}
\put(4.2,0.0){$\tau_P$}
\put(6.2,0.0){$p$}

\put(9.1,0.0){\bf $m$}
\put(11.1,0.0){\bf $\tau_P^\prime$}
\put(13.1,0.0){\bf $p$}

\put(1.2,0.0){
\epsfysize=2.0cm
\epsffile{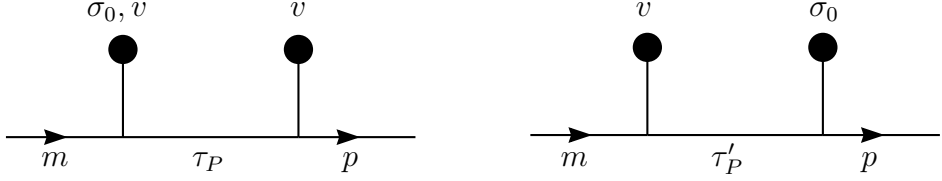}
}
\end{picture}
\caption{\footnotesize Feynman diagrams contributing to
  $\eta$--$\eta^\prime$ mixing due to exchange of
  $\prl_\mu\rho_A^\mu$.}
\label{Feyn1} 
\end{figure}

It is equivalent to diagonalize the matrix (\ref{OO1}) or to eliminate
$\tau_P$ by solving its field equations for $\tau_P[\Phi]$ which is
more adapted to our purpose. 
The elimination of $\tau_P$ gives an additional contribution to the
effective inverse propagator $G_\vph^{-1}(q)+\Delta G_\vph^{-1}(q)$,
namely
\begin{equation}
  \label{OO3}
  \Delta G_\vph^{-1}(q)=
  -G_{\vph P}^{-2}(q)
  G_P(q)q^2\; .
\end{equation}
There are also contributions to the off--diagonal $m-p$ kinetic term
related to $\eta$--$\eta^\prime$ mixing which are represented
graphically in fig.~\ref{Feyn1}.
{}From an investigation of $G_\vph^{-1}+\Delta G_\vph^{-1}$ we can
obtain some general insight in the structure of mixing effects. First,
we observe that for $G_\vph^{-1}=Z_\vph q^2+\ol{M}^2$ the effective
propagator $\ol{G}=(G_\vph^{-1}+\Delta G_\vph^{-1})^{-1}$ typically
has two poles corresponding to the values of $q^2$ for which the
determinant of $G^{-1}(q)$ vanishes. In the vicinity of the lower mass
pole $G_\vph^{-1}+\Delta G_\vph^{-1}$ can be approximated by a typical
one particle two point function. The propagator vanishes at the value
$q^2=-M_P^2$ where $G_P^{-1}(q)$ has a zero. One observes that due to
the particular factor of $(q^2)^{-1}$ in the inverse $\tau_P$
propagator (\ref{OO1}) the value of $M_P$ is always larger than both
masses corresponding to the location of the poles\footnote{Without the
  $(q^2)^{-1}$ factor $M_P^2$ would be inbetween the two poles.}. In
addition, the residue of $\ol{G}$ at the pole with the higher value of
$-q^2$ has the opposite sign as for the lower mass pole. (The higher
mass pole does not correspond to a stable particle even within our
approximations.) If we denote by $q^2=-M_0^2$ the location of the
higher pole one finds with (\ref{OO2}) and $\ol{M}^2=m_m^2 Z_m$
\begin{equation}
  \label{TTT28}
  M_P^2=\ol{M}_A^2/Z_P=
  M_0^2\left[1+
  4g_{\rho\pi\pi}^2\sigma_0^2
  \frac{1}{Z_P}\frac{Z_m}{Z_\vph}
  \left( M_0^2-
  m_m^2\frac{Z_m}{Z_\vph}\right)^{-1}
  \right]\; .
\end{equation}
An estimate of $M_0$ is not obvious and subject to large
uncertainties. One can then determine $Z_\vph/Z_m$ from (\ref{AAA7}),
(\ref{AAA8}) and solve the resulting system of equations in dependence
on $Z_P$.  Results are displayed in table \ref{TTA1} for values of
$Z_P$ in a range for which $M_\eta$ comes out with a reasonable size
in the leading mixing approximation (fig.~\ref{Plot5}). We note that
much larger values of $Z_P$ lead to a very large mixing in the
$\eta$--$\eta^\prime$ sector and completely destroy any reasonable
picture. We also show in table \ref{TTA1} three different values of
$x_\rho$.

Second, it is clear that for real octet or singlet fields the
off--diagonal elements must be real and equal.  Integrating out the
additional fields gives a negative contribution to the coefficient
$\sim q^2$ in the quadratic term for
$\Phi-\left\langle\Phi\right\rangle$ as long as $G_P(q)$ stays
positive. The mixing gives therefore a negative contribution to $Z_m$
(\ref{A3}) and explains why $X_\vph^-$ is negative. Third, from
(\ref{OO3}) we learn that the mixing effects are proportional to the
propagator $G_P$. This suggests that mixing effects with light
additional states are particularly important. Fourth, the mixing
effects also contribute to higher derivative terms in the effective
action for $\Phi$. Expanding $G_P(q)$ around $q_0^2$ gives
\begin{equation}
  \label{NNN23}
  \Dt G_\vph^{-1}=
  -4\frac{g_{\rho\pi\pi}^2Z_m^2\ol{\si}_0^2}
  {\ol{M}_A^2+
  Z_P q_0^2}q^2
  \left[1-
  \frac{Z_P\left( q^2-q_0^2\right)}
  {\ol{M}_A^2+
  Z_P q_0^2}\right]
\end{equation}
where the first term corresponds to $Z_m^{(\rho)}q^2$ as given by
(\ref{A4}). Comparing with (\ref{NNN24}) and using (\ref{A4}) we find
a contribution
\begin{equation}
  \label{NNN19}
  \ol{H}_m^{(\rho)}=
  -\frac{1}{4}\om_m^{(\rho)}v
  \frac{2\ol{f}_K+\ol{f}_\pi}
  {\ol{f}_K-\ol{f}_\pi}
  \frac{1}{M_P^2-m_m^2}\; .
\end{equation}

The momentum dependence from the effective propagator $G_P$ may be
particularly important for the $\eta^\prime$ since its mass is closest
to $M_P^2$. The resulting $q^2$--dependence of the quantities
appearing in (\ref{ZZZ7}) has therefore to be treated with care. We
remark that typical masses in the neutral sector are higher than in
the flavored one and therefore guess $M_P$ around $2000\MeV$ with a
large uncertainty.  In terms of the unrenormalized fields the inverse
propagator in the flavor neutral sector takes on the form
\begin{equation}
 \left(\ba{ccc}
 \left[Z_\vph+X_\vph^-(q^2)\ol{\sigma}_0^2\right] q^2+
 M_p^2 Z_p &,& 
 \left[\hat{\omega}(q^2)q^2+
 M_{8p}^2\right] Z_m^{1/2}Z_p^{1/2}\\
 \left[\hat{\omega}(q^2)q^2+
 M_{8p}^2\right] Z_m^{1/2}Z_p^{1/2} &,& 
 \left[Z_\vph+X_\vph^-(q^2)\ol{\sigma}_0^2\right] q^2+
 \ol{M}_8^2
 \ea\right)\; .
 \label{TTT11}
\end{equation}
Here we assume that the nontrivial momentum dependence beyond the
approximation linear in $q^2$ arises dominantly from propagator
effects contained in $X_\vph^-(q^2)$ and $\hat{\omega}(q^2)$ according
to
\begin{eqnarray}
  \label{TTT12}
  \ds{X_\vph^-(q^2)} &=& \ds{
  X_\vph^-(-m_m^2)
  \frac{\left( M_P^2-m_m^2\right)}
  {\left( M_P^2+q^2\right)}}\nnn
  \ds{\hat{\omega}(q^2)} &=& \ds{
  \hat{\omega}(-M_\eta^2)
  f_\omega(q^2)\; ,\;\;\;\;
  f_\omega(q^2)=
  \frac{M_P^2-M_\eta^2}
  {M_P^2+q^2}}\; .
\end{eqnarray}
Our conventions imply
\begin{eqnarray}
  \label{TTT13}
  \ds{\hat{\omega}} &=& \ds{
  \hat{\omega}(-M_\eta^2)}\nnn
  \ds{X_\vph^-} &=& \ds{
  X_\vph^-(-m_m^2)}
\end{eqnarray}
and we remind that we have defined both, $Z_p$ and $Z_8$, for
$q^2=-M_\eta^2$. The leading mixing approximation implies
\begin{eqnarray}
  \label{TTT14}
  \ds{Z_m} &=& \ds{
  Z_\vph+X_\vph^-(-m_m^2)
  \ol{\sigma}_0^2}\nnn
  \ds{Z_m Z_8} &=& \ds{
  Z_p=Z_\vph+X_\vph^-(-M_\eta^2)
  \ol{\sigma}_0^2}\; .
\end{eqnarray}
With this definition the only higher derivative effect in the matrix
(\ref{ZZZ7}) for $q^2=-M_\eta^2$ appears in the factor
$f_\omega^{-1}(-m_m^2)=(M_P^2-m_m^2)/(M_P^2-M_\eta^2)=1.02$,
(\ref{ZZZ4}). Here we have used the estimate $M_P=2670\MeV$
{}from table \ref{TTA1}. Also the resulting deviation of $Z_p/Z_m$
{}from unity is small
\begin{eqnarray}
  \label{TTT16}
  \ds{\frac{Z_p}{Z_m}} &=& \ds{
  Z_8=
  \frac{Z_\vph+X_\vph^-(-M_\eta^2)
  \ol{\sigma}_0^2}
  {Z_\vph+X_\vph^-(-m_m^2)
  \ol{\sigma}_0^2}=
  1+\left[ X_\vph^-(-M_\eta^2)-X_\vph^-\right]
  \sigma_0^2}\nnn
  &=& \ds{
  1+\frac{1}{4}\omega_m v
  \frac{(2\ol{f}_K+\ol{f}_\pi)}
  {(\ol{f}_K-\ol{f}_\pi)}
  \frac{(M_\eta^2-m_m^2)}
  {(M_P^2-M_\eta^2)}\simeq1+0.05\omega_m v}
\end{eqnarray}
and compatible with the linearization for $Z_8$ according to
(\ref{NNN19}). (The numerical value for $\omega_m v=-0.20$ is
$Z_p/Z_m=0.99$.) There is no modification of the decay constants
$f_{\eta8}$ and $f_{\eta0}$, (\ref{TTT4}), (\ref{TTT15}). On the other
hand, the propagator effect in $X_\vph^-(q^2)$ and $\hat{\omega}(q^2)$
could lead to substantial effects for $q^2=-M_{\eta^\prime}^2$
depending on the value of $M_P$: In the diagonal elements of the
inverse propagator (\ref{ZZZ7}) one has to insert
\begin{eqnarray}
  \label{TTT17}
  \ds{z_p(q^2)} &=& \ds{
  z_8(q^2)=\frac{Z_\vph+X_\vph^-(q^2)
  \ol{\sigma}_0^2}
  {Z_\vph+X_\vph^-(-M_\eta^2)
  \ol{\sigma}_0^2}}\nnn
  &=& \ds{
  \frac{\left(1-\frac{1}{4}\omega_m v
  \frac{(2\ol{f}_K+\ol{f}_\pi)}
  {(\ol{f}_K-\ol{f}_\pi)}
  \frac{(q^2+m_m^2)}
  {(q^2+M_P^2)}
  \right)}
  {\left(1+\frac{1}{4}\omega_m v
  \frac{(2\ol{f}_K+\ol{f}_\pi)}
  {(\ol{f}_K-\ol{f}_\pi)}
  \frac{(M_\eta^2-m_m^2)}
  {(M_P^2-M_\eta^2)}
  \right)}}
\end{eqnarray}
whereas $\hat{\omega}$ is replaced by $\hat{\omega}(q^2)$
(\ref{ZZZ4}), (\ref{TTT12}). Furthermore, the correct definition of
the decay constants $f_{\eta^\prime0}$ and $f_{\eta^\prime8}$ involves
now the factor (\ref{TTT5})
\begin{equation}
  \label{TTT20}
  \tilde{z}_p(-M_{\eta^\prime}^2)=
  z_p(-M_{\eta^\prime}^2)\; .
\end{equation}

For the scalar octet an interesting possibility of mixing concerns
states in the two--meson channels. In fact, the four--point function
for $\Phi_p$ may develop resonance--like structures in the momentum
range corresponding to the sum of two pseudoscalar meson masses.
(These momentum dependent structures are not accounted for by the
four--point function at zero external momenta described by the
effective potential $U$.) Such resonance structures can be replaced by
effective interactions with a composite $0^{++}$ field $\tau_S$. The
effective two--point functions for $\tau_S$ obtained in this way do
not necessarily correspond to a propagating particle or resonance,
since their real part may be strictly positive and bounded for all
values of $q^2$ on the real axis. Consider a generic structure for the
mixing between $\sqrt{2}\Phi_s$ and $\tau_S$
\begin{equation}
  \label{NNN22}
  G^{-1}(q)=\left(
  \begin{array}{ccc}
  Z q^2+\ol{M}^2 & , & b(q^2)\\
  b(q^2) & , & c(q^2)
  \end{array}
  \right)\; .
\end{equation}
If $c(q^2)$ has a zero for $\sqrt{-q^2}$ in the vicinity of the sum of
two pseudoscalar masses one finds two values of $q^2$ for which an
eigenvalue of $G^{-1}$ vanishes. In the case of the isospin triplet
they could be associated with $a_0(980)$ and $a_0(1320)$.  On the
other hand, the two--particle threshold could also be reflected by a
finite enhancement of $c(q)^{-1}$ or $b(q)$ without a zero of $c(q)$.
If $c(q^2)$ dips in this momentum region to values smaller than
$b^2(q^2)/(\ol{M}^2+Z q^2)$ the zero eigenvalue of $G^{-1}$ will occur
precisely in the threshold region, namely for $q^2=-M_0^2$ as
determined by $c(-M_0^2)=b^2(-M_0^2)(\ol{M}^2-Z M_0^2)^{-1}$.  After
solving for $\tau_S[\Phi_s]$ the location of the single pole of
$[G_\vph^{-1}(q)-b^2(q^2)c^{-1}(q^2)]^{-1}$ would then necessarily be
found at $-M_0^2$ in the threshold region.  For $\ol{M}^2/Z$ not too
far from the two--particle threshold this effect could explain
naturally why the isotriplet in $\Phi_s$ is found precisely at the
$2K$ threshold! Mixing effects from $\Delta
G_\vph^{-1}=-b^2(q^2)c^{-1}(q^2)$ are large in this case. Since this
mechanism requires a critical strength for $b^2 c^{-1}$ not all
members of the scalar octet have to be in the vicinity of
two--particle thresholds. More precisely, the phenomenon of
``threshold mass shifting'' which induces mesons masses near a
two--particle threshold occurs whenever $\ol{M}^2/Z$ is above the
threshold and the quantity $b^2(q)/(c(q)Z(q))$ makes a strong enough
jump in the threshold region. An alternative way of looking at this
``threshold mass shifting'' notes that the loop contribution to the
two--point function for the $a_0$ becomes important if the mass is
close to the sum of the masses of the two pseudoscalars circulating in
the loop.

Even without a detailed discussion of the complicated analyticity
properties we conclude that for both alternatives the effective
inverse propagator for the isotriplet in $\Phi_s$ should have a zero
at the observed $a_0(980)$ resonance. In this momentum region the
mixing effects with two--kaon states are expected to be very strong.
No detailed understanding of the properties of the $a_0(980)$ seems
possible without incorporating the two--kaon channel. In addition, the
effective inverse propagator may (or may not) have a second zero
corresponding to the possible resonance $a_0(1320)$. In this momentum
region the mixing effects should be much smaller because of the larger
value of $Z q^2+\ol{M}^2-c(q)$. We note that both resonances are
described by the same value of\footnote{It is convenient to choose
  composite fields such that $c(0)=0$.} $\ol{M}^2$, but different
effective $Z_h$ and $\omega_h$. The two different associations
$a_0(980)$ vs.~$a_0(1320)$ in the main text become in this case only
two facets of the same story. If the $a_0(1320)$ exists the values
characterizing the potential, like $m_h^2$, $\lambda_2$, $\lambda_3$,
etc.~should be independent of the identification of the isotriplet.
The actual differences in these values are then a measure for the
influence of neglected terms. We find that these differences can
indeed be small if the mixing is large enough for the $a_0(980)$.
This is compatible with the existence of the $a_0(1320)$ as a real
resonance.

Finally, we turn to the mixing of a pseudoscalar $0^{-+}$ glueball
$g$ with the $\etap$. This is particularly interesting in view of a
possible experimental detection of $g$. Because of the anomaly the
$\etap$ is not a Goldstone boson for vanishing quark masses and the
off--diagonal element in the mixing matrix may not vanish for zero
momentum. We introduce for the glueball a pseudoscalar singlet field
$g$ with an effective action
\begin{eqnarray}
  \label{GlueballAction}
  \ds{\Lc_{\rm g}} &=& \ds{
  \hal\prl^\mu g\prl_\mu g+
  \hal m_{gl}^2 g^2+h_{gl}\omega g}\nnn
  \ds{\omega} &=& \ds{
  i\left(\det\Phi-\det\Phid\right)}\; .
\end{eqnarray}
The coupling between $\Phi$ and $g$ conserves all symmetries
($\Pc(g)=-g$, $\Cc(g)=g$). The real coupling $h_{gl}$ may depend on the
momentum of $g$. Expanding $h_{gl}(q)$ around $q_0^2=-m_m^2$ the mixing
with the glueball contributes to the effective potential for $\Phi$
(cf.~\ref{OmegaSquared})
\begin{equation}
  \label{GlueballPotential}
  U^{(g)}[\Phi]=-
  \hal h_{gl}^2(q_0)
  \left[m_{gl}^2+q_0^2\right]^{-1}
  \omega^2
\end{equation}
whereas for the kinetic term (\ref{Lkin}) it induces a coupling
\begin{equation}
  \label{GlueCoupling}
  \tV_\vph^{(g)}=
  h_{gl}^2(q_0)\left[m_{gl}^2+q_0^2\right]^{-2}-
  \frac{\prl h_{gl}^2}{\prl q^2}(q_0)
  \left[m_{gl}^2+q_0^2\right]^{-1}\; .
\end{equation}
It is probably difficult to disentangle (\ref{GlueballPotential}) from
other contributions to the potential. On the other hand, a
determination of the size of the parameter $\tV_\vph$ will put
restrictions on $h_{gl}$. Since $h_{gl}$ is directly related to the mixing
between $\etap$ and $g$ one may obtain from it interesting information
on the decay of the pseudoscalar glueball into mesons or photons. To
lowest order we can use
\begin{equation}
  \label{OO5}
  \omega=-\sqrt{6}\olsi_0^2 Z_p^{-\hal} p
\end{equation}
and obtain the inverse propagator for the $\etap$--glueball system as
\begin{equation}
  \label{EtapGlueball}
  G^{-1}(q)\simeq\left(
  \begin{array}{{ccc}}
  q^2+m_p^2 &,&
  -\frac{\sqrt{6}}{2}Z_p^{-\hal}Z_m^{-1}\si_0^2 h_{gl}
  \nnn
  -\frac{\sqrt{6}}{2}Z_p^{-\hal}Z_m^{-1}\si_0^2 h_{gl}
  &,& q^2+m_{gl}^2
  \end{array}
  \right)\; .
\end{equation}
For not too large $h_{gl}$ the mixing angle between $g$ and $\etap$ is
suppressed by the small ratio $\si_0/m_{gl}$
\begin{equation}
  \label{MixingAngleGlueballEtap}
  -\vth_{gl}\simeq
  h_{gl}\frac{\si_0^2}{m_{gl}^2-m_p^2}
  \simeq10^{-3}\, h_{gl}\; .
\end{equation}


\end{document}